\documentclass[amsmath,amssymb,aps,10pt,prd,letterpaper,nofootinbib,balancelastpage,notitlepage,superscriptaddress,twocolumn,floatfix]{revtex4-2}
\usepackage{graphicx}	% Standard figures 
\usepackage{epstopdf}	% Convert eps to pdf
\usepackage{dcolumn}	% Column align
\usepackage{ragged2e}	
\usepackage{bm}		    % Bold math
\usepackage{enumitem}   % Enumeration
\usepackage{rotating}
\usepackage[sort&compress]{natbib}	 	% Standard reference package
	\setcitestyle{square,numbers,comma}	% Set citation style
\usepackage[colorlinks=true,urlcolor=blue,linkcolor=black,citecolor=red]{hyperref}	
\newcommand{\orcid}[1]{\href{https://orcid.org/#1}{\,\includegraphics[width=8px]{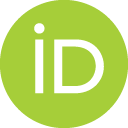}}}
\newcommand{\up}[1]{\textsuperscript{#1}}				% Raised text

\newcommand{\bparagraph}[1]{\textbf{#1}---\ignorespaces}

\newcommand{\Alice}{{314159~Alice}}
\newcommand{\Bob}{{271828~Bob}}
\newcommand{\tabref}[2][]{Tab{#1}.~\ref{#2}}		% Table reference
\newcommand{\figref}[2][]{Fig{#1}.~\ref{#2}}		% Figure reference
\newcommand{\subfigref}[3][]{Fig{#1}.~\hyperref[#2]{\ref{#2}(#3)}}		% Figure reference
\newcommand{\sectref}[2][]{Sec{#1}.~\ref{#2}}		% Section reference
\newcommand{\appref}[2][x]{Appendi{#1}~\ref{#2}}	% Appendix reference
\renewcommand{\eqref}[2][]{Eq{#1}.~(\ref{#2})}		% Equation reference
\newcommand{\eqrefRange}[2]{Eqs.~(\ref{#1})--(\ref{#2})}		% Equation range reference
\newcommand{\citeR}[2][]{Ref{#1}.~\cite{#2}}			% Ref. Citation
\renewcommand{\Re}{\ensuremath\,\mathrm{Re}}
\renewcommand{\mod}{\ensuremath\,\mathrm{mod}\,}
\newcommand{\lb}{\ensuremath{\left}}					% Left Brackets
\newcommand{\rb}{\ensuremath{\right}}					% Right Brackets
\newcommand{\nl}{\nonumber \\ & \quad }					% in-array new line

\newcommand{\alice}{\ensuremath{\circledast}}           % command for \alice symbol
\newcommand{\order}[1]{\ensuremath{\mathcal{O}(#1)}}    % O(x) notation
\newcommand{\Hz}[1]{\ensuremath{\,\mathrm{{#1}Hz}}}

\begin{document}
%%%%%%%%%%%%%%%%%%%%%%%%%%%%%%%%%%%%%%%%%%%%%%%%%%%%%%%%%%%%%%%%%%%%%%%%%%%%%%%%%%%%%%%%%%	
%%%%%%%%%%%%%%%%%%%%%%%%%%%%%%%%%%%%%%%%%%%%%%%%%%%%%%%%%%%%%%%%%%%%%%%%%%%%%%%%%%%%%%%%%%	
%%%%%%%%%%%%%%%%%%%%%%%%%%%%%%%%%%%%%%%%%%%%%%%%%%%%%%%%%%%%%%%%%%%%%%%%%%%%%%%%%%%%%%%%%%	
%%%%%%%%%%%%%%%%%%%%%%%%%%%%%%%%%%%%%%%%%%%%%%%%%%%%%%%%%%%%%%%%%%%%%%%%%%%%%%%%%%%%%%%%%%	

%%%%%%%%%%%%%%%%%%%%%%%%%%%%%%%%%%%%%%%%%%%%%%%%%%%%%%%%%%%%%%%%%%%%%%%%%%%%%%%%%%%%%%%%%%	
% Title, Author and Affiliation
\title{\texorpdfstring{Asteroids for $\bm{\mu}$Hz gravitational-wave detection}{Asteroids for microhertz gravitational-wave detection}}
\date{\today}
%%%%%%%%%%%%%%%%%%%%%%%%%%%%%%%%%%%%%%%%%%%%%%%%%%%%%%%%%%%%%%%%%%%%%%%%%%%%%%%%%%%%%%%%%%	

%%%%%%%%%%%%%%%%%%%%%%%%%%%%%%%%%%%%%%%%%%%%%%%%%%%%%%%%%%%%
\author{Michael A.~Fedderke\orcid{0000-0002-1319-1622}}
\email{mfedderke@jhu.edu}
\affiliation{The William H.~Miller III Department of Physics and Astronomy, The Johns Hopkins University, Baltimore, MD  21218, USA}
%%%%%%%%%%%%%%%%%%%%%%%%%%%%%%%%%%%%%%%%%%%%%%%%%%%%%%%%%%%%
\author{Peter W.~Graham\orcid{0000-0002-1600-1601}}
\email{pwgraham@stanford.edu}
\affiliation{Stanford Institute for Theoretical Physics and Kavli Institute for Particle Astrophysics \& Cosmology, Department of Physics, Stanford University, Stanford, CA 94305, USA}
%%%%%%%%%%%%%%%%%%%%%%%%%%%%%%%%%%%%%%%%%%%%%%%%%%%%%%%%%%%%
\author{Surjeet Rajendran\orcid{0000-0001-9915-3573}\,}
\email{srajend4@jhu.edu}
\affiliation{The William H.~Miller III Department of Physics and Astronomy, The Johns Hopkins University, Baltimore, MD  21218, USA}
%%%%%%%%%%%%%%%%%%%%%%%%%%%%%%%%%%%%%%%%%%%%%%%%%%%%%%%%%%%%

%%%%%%%%%%%%%%%%%%%%%%%%%%%%%%%%%%%%%%%%%%%%%%%%%%%%%%%%%%%%%%%%%%%%%%%%%%%%%%%%%%%%%%%%%%
% Abstract
\begin{abstract}
%%%%%%%%%%%%%%%%%%%%%%%%%%%%%%%%%%%%%%%%%%
A major challenge for gravitational-wave (GW) detection in the $\mu$Hz band is engineering a test mass (TM) with sufficiently low acceleration noise.
We propose a GW detection concept using asteroids located in the inner Solar System as TMs.
Our main purpose is to evaluate the acceleration noise of asteroids in the $\mu$Hz band.
We show that a wide variety of environmental perturbations are small enough to enable an appropriate class of $\sim 10\,\text{km}$-diameter asteroids to be employed as TMs.
This would allow a sensitive GW detector in the band $\text{(few)} \times 10^{-7}\,\text{Hz} \lesssim f_{\textsc{gw}} \lesssim \text{(few)} \times 10^{-5}\,\text{Hz}$, reaching strain $h_c \sim 10^{-19}$ around $f_{\textsc{gw}} \sim 10\,\mu$Hz, sufficient to detect a wide variety of sources.
To exploit these asteroid TMs, human-engineered base stations could be deployed on multiple asteroids, each equipped with an electromagnetic transmitter/receiver to permit measurement of variations in the distance between them.
We discuss a potential conceptual design with two base stations, each with a space-qualified optical atomic clock measuring the round-trip electromagnetic pulse travel time via laser ranging. 
Tradespace exists to optimize multiple aspects of this mission: for example, using a radio-ranging or interferometric link system instead of laser ranging.  
This motivates future dedicated technical design study.
This mission concept holds exceptional promise for accessing this GW frequency band.
\end{abstract}
%%%%%%%%%%%%%%%%%%%%%%%%%%%%%%%%%%%%%%%%%%%%%%%%%%%%%%%%%%%%%%%%%%%%%%%%%%%%%%%%%%%%%%%%%%

\maketitle

\tableofcontents

%%%%%%%%%%%%%%%%%%%%%%%%%%%%%%%%%%%%%%%%%%%%%%%%%%%%%%%%%%%%%%%%%%%%%%%%%%%%%%%%%%%%%%%%%%
\section{Introduction}
\label{sect:idea}
%%%%%%%%%%%%%%%%%%%%%%%%%%%%%%%%%%%%%%%%%%%%%%%%%%%%%%%%%%%%%%%%%%%%%%%%%%%%%%%%%%%%%%%%%%
The direct discovery by LIGO/Virgo~\cite{PhysRevLett.116.061102,PhysRevLett.119.161101,PhysRevX.9.031040,Abbott:2020niy,LIGOScientific:2021djp} of gravitational waves~(GWs) in the few Hz to kHz range, generated by the final inspiral of compact objects in the few-to-hundred solar-mass class, has heralded a new era of observation of the Universe.
The science case for a broad coverage of the gravitational-wave frequency spectrum is exceptionally strong~\cite{Sesana:2019vho,Baibhav:2019rsa,Sedda:2019uro,Baker:2019pnp}.
Indeed, there is already broad existing or planned coverage for gravitational-wave observations over much of the frequency range from nHz to kHz: the continuing ground-based observational program by LIGO/Virgo/KAGRA~\cite{Akutsu:2020zlw}; ongoing pulsar timing array measurements in the 1--300\Hz{n} range~\cite{Kramer_2013,Babak:2015lua,Shannon1522,2016MNRAS.458.1267V,Aggarwal:2018mgp,Kerr_2020,Arzoumanian:2020vkk}, including interesting recent hints from NanoGRAV~\cite{Arzoumanian:2020vkk}; the planned space-based observational program by LISA in the 0.1\Hz{m}--100\Hz{m} range~\cite{Baker:2019nia,LISA_Sci_Req,LISA_L3,PhysRevLett.120.061101} and TianQin around 0.01\Hz{}--1\Hz{}~\cite{Luo:2015ght,Milyukov:2020fm}; and various `mid-band' detectors based on atomic interferometry~\cite{Dimopoulos:2007cj,Dimopoulos:2008sv,Hogan:2010fz,Graham:2017pmn,Coleman:2018ozp,Canuel:2018fq, Abe:2021ksx,Tino:2019tkb,Badurina:2019hst,Zhan_2019,AEDGE:2019nxb,Badurina:2021rgt} or atomic clock techniques~\cite{Kolkowitz:2016wyg} that aim for the band between LISA and the ground-based laser-interferometric detectors.
Moreover, there are many future proposals for sensitivity improvements over much of this range; see, e.g., \citeR[s]{BBO,Crowder:2005nr,Seto:2001qf,Kawamura:2018esd,kawamura2020current,Maggiore:2019uih,LIGOVoyagerCE,CosmicExplorer,Ackley:2020atn}.
Proposals also exist to extend exploration up to frequencies as high as the MHz--GHz range~\cite{Aggarwal:2020olq}.

However, the frequency band between pulsar timing arrays (PTAs) and LISA, roughly $0.1$--$100\Hz{\mu}$ suffers from a dearth of existing or proposed coverage at interesting levels of strain sensitivity; this is known as the `$\mu$Hz gap'.
This band is home to many interesting astrophysical and cosmological GW sources~\cite{Sesana:2019vho} and its exploration is well motivated~\cite{Spallicci:2011nr}. 
For instance, the extensive study in \citeR{Sesana:2019vho} (see, e.g., their Fig.~1) indicates that a detector sufficiently sensitive in this band would have as promising targets inspiraling massive black-hole binaries out to redshift $z\sim 10$, merging massive black-hole binaries out to redshift $z\sim 20$, resolved galactic black-hole binaries, stars merging with the massive black hole at the center of our own galaxy some time out from merger, and cosmologically distant (redshift $z\sim 7$) intermediate mass-ratio inspiral (IMRI) events, as well as being able to eventually reach and characterize unresolved galactic and cosmological gravitational-wave backgrounds. 
Additionally, other surprises may await detection in this band, as GW detectors operating in this band may also have access to other, non-GW new physics such as various dark-matter candidates (see, e.g., \citeR[s]{Seto:2004zu,Adams:2004pk,Graham:2015ifn,Grabowska:2018lnd}).

GW detection in this band is challenging, and existing technologies struggle to access this band from either `above' or `below' (i.e., moving into the band from higher or lower frequencies, respectively).
One of the few local-test-mass-based proposals that has attempted to outline the technical requirements to probe interesting levels of strain sensitivity in this band is the $\mu$Ares proposal~\cite{Sesana:2019vho}, a mission concept similar to LISA.
That study indicated that a mission would require interferometer arm lengths significantly (around 200 times) larger than those proposed for LISA, as well as greatly improved low-frequency test-mass (TM)%
%%%%%%%%%%%%
\footnote{\label{ftnt:proofMass}%
    Test masses are also sometimes referred to as `proof masses'.
    } %
%%%%%%%%%%%%
isolation.%
%%%%%%%%%%%%
\footnote{\label{ftnt:lowFreqTMmuAres}%
    The $\mu$Ares `strawman mission concept'~\cite{Sesana:2019vho} projected strain sensitivity assumes that a TM acceleration noise amplitude spectral density (ASD) slightly exceeding the best absolute levels attained in the LISA Pathfinder mission~\cite{PhysRevLett.120.061101} around $f\sim1$--$10\Hz{m}$ can be maintained \emph{without} degradation (i.e., flat in frequency) down to $f\sim 10^{-7}$\Hz{}; whereas the LISA Pathfinder acceleration noise ASD is already rising approximately as $\sqrt{S_a} \propto f^{-1}$ below $f \sim (\text{few})\times 10^{-4}\,$Hz.} %
%%%%%%%%%%%%
Likewise, accessing this band with PTAs is challenging, as PTAs lose sensitivity with increasing frequency~\cite{Moore:2014eua}, and their high-frequency sensitivity is limited by the Nyquist sampling frequency of the array (typically, a few $\mu\text{Hz}$~\cite{Kramer_2013,Babak:2015lua,Shannon1522,2016MNRAS.458.1267V,Aggarwal:2018mgp,Kerr_2020,Arzoumanian:2020vkk}).
Proposals utilizing astrometric techniques on large-scale survey data (e.g., \emph{Gaia}~\cite{GaiaOverview} and Roman Space Telescope~\cite{RomanWebsite} surveys) are also able to access this band (see, e.g., \citeR[s]{Pyne:1995iy,schutz_2009,Book:2010pf,Klioner:2017asb,Moore:2017ity,Park:2019ies,Wang:2020pmf}), but existing projections indicate that the levels of strain sensitivity attainable are somewhat modest~\cite{Wang:2020pmf}; such approaches are however able to overcome some important noise sources~\cite{Fedderke:2020yfy} that limit all local-TM-based techniques operating in the inner Solar System below $\sim\mu$Hz frequencies.
Recent work has also studied how orbital perturbations (of, e.g., binary millisecond pulsars, or the Moon) that arise specifically from a broadband stochastic GW background could access this band~\cite{Blas:2021mqw,Blas:2021mpc}.
A recent study has also considered how timing perturbations, arising from GW in this band, to higher-frequency continuous-wave GW sources could allow some sensitivity to the former~\cite{Bustamante-Rosell:2021daj}.

In this paper, we revisit the $\mu$Hz gap and propose an alternative technique to access this band.
Our study is motivated by the following observations: existing approaches accessing this band from above suffer from worsening acceleration noise on small human-engineered TMs; these can however be tracked exceedingly well.
On the other hand, approaches accessing this band from below suffer from an inability to track (really, time) excellent astrophysically massive TMs to the requisite sensitivity levels.
This raises a natural question: is it possible to marry the favorable acceleration noise characteristics of astrophysically massive natural TMs with the sensitive tracking approaches that are more characteristic of missions using human-engineered TMs? 
Tracking approaches with sufficient sensitivity however require the deployment of human-engineered apparatus at the TM locations; realistically, this limits the consideration of available TM to those that occur naturally in the (inner) Solar System.
The question is therefore sharpened: are there natural, astrophysically massive bodies existing in the (inner) Solar System that behave as sufficient good TMs for us to use them in a GW detector that can access the $\mu$Hz gap?

We demonstrate that the answer is yes: a few carefully selected inner Solar System asteroids in the 10\,km-class are sufficiently massive to make them attractive as TM for a ranging-type GW detector. Considering in turn solar intensity, solar wind, thermal cycling, collisional, electromagnetic, and other relevant perturbations to these asteroids, we show that both center-of-mass (CoM) and rotational perturbations from these sources appear to be small enough that strain sensitivities that are comparable to those projected for $\mu$Ares could be achievable in the $\mu$Hz band, roughly $h_c \sim 10^{-19}$ at $f_{\textsc{gw}}\sim 10\Hz{\mu}$. 

While our main point is to demonstrate that asteroids serve as surprisingly good TMs despite prevailing ambient environmental conditions, good test masses alone do not constitute a gravitational-wave detector.
We complete the picture by sketching out a mission concept to link together asteroids using simple asteroid-to-asteroid laser ranging measurements between deployed base stations. 
We point out that state-of-the-art terrestrial atomic clocks possess the metrological capability to perform the required measurements. 
We also discuss other link options.

Our work is distinguished from \citeR[s]{Blas:2021mqw,Blas:2021mpc}, that considered looking for resonant orbital perturbations of binary systems---including the Earth--Moon system, via Lunar Laser Ranging~\cite{Murphy:2013qya}---in order to detect a broadband GW background, both in the technique proposed and because our proposal would be sensitive to narrowband signals at any frequency in our band.%
%%%%%%%%%%
\footnote{\label{ftnt:BlasNarrowband}%
		Of course, as noted in \citeR[s]{Blas:2021mqw,Blas:2021mpc}, narrowband GW signals that happen to accidentally co-incide with the resonant frequencies of a binary system would also be visible in their approach, but this provides only limited narrowband coverage.
	} %
%%%%%%%%%%
Our proposal is also very different in nature to the type of approach considered in \citeR{Tsai:2021irw} (which we note is phrased as a fifth-force, and not a GW, search), in that we propose to deploy active base stations on a small number of asteroids themselves to perform direct asteroid-to-asteroid ranging, rather than to track the perihelion precession of asteroid orbits, or use the full astrometric data for a larger collection of asteroid orbits.
It is also clearly distinguished from other existing studies using objects in the Solar System that aim to (1) use bodies such as the Moon themselves as GW detectors~\cite{LSGreport,PAIK2009167,PhysRevD.90.102001,LGW-ESA,LGWA:2020mma,LSGA} by viewing them as large resonant-type detectors~\cite{PhysRev.117.306,PhysRevLett.18.498}, (2) deploy an interferometer setup on the Moon~\cite{Jani:2020gnz,LSGA}, as it is seismically quieter than Earth, or (3) track human-engineered spacecraft on long interplanetary missions~\cite{Soyuer:2021qqp}.

The remainder of this paper is structured as follows:
in \sectref{sect:concept} we provide more details on the overall mission concept.
We give a detailed accounting for all the important test-mass acceleration, torque, thermal, and seismic noise sources that we have identified in \sectref{sect:noise}.
In \sectref{sect:linkNoise}, we discuss possible links, clocks, and associated noise sources.
We highlight a number of the more detailed mission design considerations for this concept in \sectref{sect:missionDesign}, and also propose concrete developmental goals.
Our final projected strain sensitivity is shown in \figref[s]{fig:results_combo} and \ref{fig:results} in \sectref{sect:sensitivity}.
We conclude in \sectref{sect:conclusion}.
\appref{app:crossPowerTerms} gives a technical derivation of a claim made in the text of \sectref{sect:TSI} regarding cross-power terms.

%%%%%%%%%%%%%%%%%%%%%%%%%%%%%%%%%%%%%%%%%%%%%%%%%%%%%%%%%%%%%%%%%%%%%%%%%%%%%%%%%%%%%%%%%%
\section{Mission concept}
\label{sect:concept}
%%%%%%%%%%%%%%%%%%%%%%%%%%%%%%%%%%%%%%%%%%%%%%%%%%%%%%%%%%%%%%%%%%%%%%%%%%%%%%%%%%%%%%%%%%

%%%%%%%%%%%%%%%%%%%%%%%%%%%%%%%%%%%%%%%%%%%%%%%%%%%%%%%%%%%%%%%%%%%%%%%%%%%%%%%%
\begin{figure*}[!t]
    \centering
    \includegraphics[width=\textwidth]{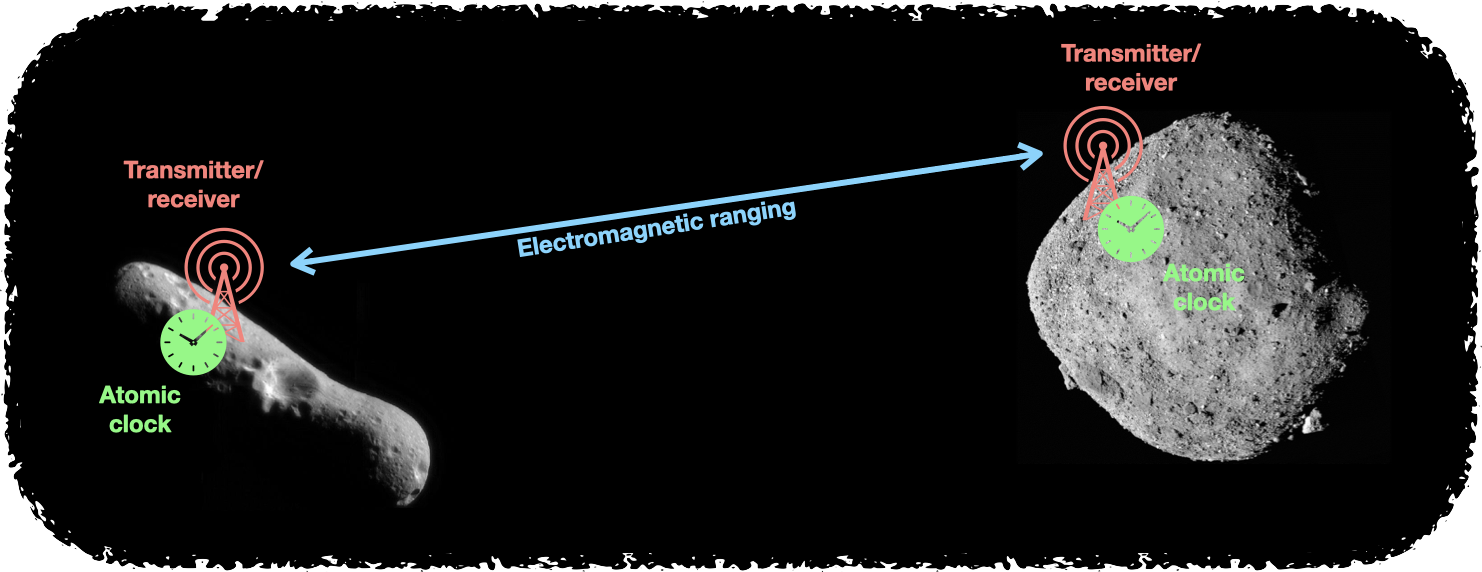}
    \caption{\label{fig:sketch}%
            Sketch of the mission concept outlined in \sectref{sect:concept} (not to scale). 
            Two $10$\,km-class asteroids act as inertial test masses.
            Base stations deployed on the asteroids exchange electromagnetic pulses via a transmitter/receiver link system.
            The round-trip travel time for the pulses is recorded by referencing a local space-qualified optical atomic clock.
            This creates a ranging-type gravitational-wave detector.
            A more sophisticated setup, with the base stations held in orbit around the asteroids but continually referenced to reflectors deployed on the asteroid surfaces, is also discussed in \sectref{sect:concept}.
            The projected strain sensitivity for this concept is shown in \figref{fig:results}.
            Image credits~\cite{NASAmediaGuidelines}: NASA/JPL (433~Eros)~\cite{ErosImage}, NASA/Goddard/University of Arizona (101955~Bennu)~\cite{BennuImage}.
    }
\end{figure*}
%%%%%%%%%%%%%%%%%%%%%%%%%%%%%%%%%%%%%%%%%%%%%%%%%%%%%%%%%%%%%%%%%%%%%%%%%%%%%%%%

A gravitational wave (with $+$ polarization) can be described by the metric perturbation (in transverse-traceless gauge): 
%%%%%%
\begin{align}
    ds^2 = - dt^2 + &\big[1 + h_{+}^{(0)} \sin\big(\omega_{\textsc{gw}}\! \left(t - z\right)\! \big)\big] dx^2 \nonumber \\ + &\big[1 - h_{+}^{(0)} \sin\big(\omega_{\textsc{gw}}\! \left(t - z\right)\! \big)\big] dy^2 + dz^2.
    \label{eq:GWmetric}
\end{align}
%%%%%%
This metric describes a gravitational wave with amplitude $h_{+}^{(0)}$ and angular frequency $\omega_{\textsc{gw}}=2\pi f_{\textsc{gw}}$ propagating in the $+\bm{\hat{z}}$ direction. 
The physical effect of this wave is to induce strain along the $x$--$y$ plane. 
As a result, two test masses located in this plane that are separated by a distance $L$ will  experience a relative acceleration $\Delta a_{\textsc{gw}} \sim h L \omega_{\textsc{gw}}^2$ (when $\omega_{\textsc{gw}} L \ll 1$). 
This acceleration occurs at the angular frequency $\omega_{\textsc{gw}}$ of the gravitational wave. 
The wave can be detected by measuring this acceleration. 

The challenge in gravitational-wave detection lies in the fact that the amplitudes $h$ of expected gravitational-wave sources are extremely small.
The measurement of the induced relative acceleration requires precision accelerometry.
This task is best accomplished by measuring the distance between the test masses as a function of time.  
Over the period $T_{\textsc{gw}}\sim 2\pi/\omega_{\textsc{gw}}$ of the gravitational wave, the relative acceleration $\Delta a_{\textsc{gw}}$ gives rise to a distance change $\Delta L \sim h L$. 
To measure this distance, a laser (or possibly radio) ranging system can be used.

On the metrology side, gravitational-wave detection requires a time standard (i.e., a clock) that is sufficiently accurate over the period of the gravitational wave. 
A second, and equally important requirement is that the distance between the well-identified test-mass locations is perturbed dominantly by the gravitational wave at the measurement frequency. 
These two requirements set the noise curve of typical gravitational-wave detectors. 
The astrophysics of gravitational-wave sources is such that the precision necessary in measuring the distance between the test masses is greater at high frequencies. 
For a sensor with a fixed precision, this determines the high frequency end of the detector's sensitivity. 
At low frequencies, even small environmental accelerations can lead to large displacements between the test masses. 
This leads to a rapid rise in the noise at low frequencies, limiting the reach of the detector. 

At $\mu$Hz frequencies, strains $h \sim 10^{-17}$--$10^{-18}$ are expected from astrophysical sources~\cite{Sesana:2019vho}.
Such a strain will cause the distance between two test masses separated by $L \sim 1$\,AU to fluctuate by $\Delta L \sim h L \sim 1$--$0.1 \, \mu \text{m}$, respectively. 
While distance measurements with this level of precision over such a long baseline are undoubtedly challenging, they are within the capabilities of current metrological technologies. 
However, the stability of test masses at the $0.1$--$1 \, \mu \text{m}$ level in the $\mu$Hz frequency band has not been demonstrated. 
For example, this stability is about 2--3 orders of magnitude more stringent than the stability demonstrated by the LISA Pathfinder mission~\cite{PhysRevLett.120.061101} (extrapolated as necessary into this frequency band). 

One way to tackle the problem of stability might be to use a large test mass, since the center-of-mass position of a large body is likely to be less sensitive to environmental perturbations. 
CoM stability alone is however not enough; extended-body rigid stability (e.g., a lack of seismic activity) at the same level is also necessary since the distance between the test masses is actually measured from the surfaces of the masses.

In this paper, we show that asteroids with diameters $\sim$~few km are a natural class of massive bodies to consider as test masses for a gravitational-wave detector. 
These are large enough to have a sufficiently stable center of mass. 
On the other hand, they are small enough to have lost their heat of formation, eliminating a major cause of seismic activity.
Moreover, for a $\sim 10$\,km asteroid, the fundamental frequency of seismic waves is $\sim 10$\,mHz~\cite{WALKER2006142}. 
This implies that at frequencies below $\sim 10$\,mHz, the seismic response of the system to lower frequency perturbations will be suppressed by this fundamental harmonic. 
These asteroids are also not expected to possess their own atmosphere, eliminating another potential source of surface fluctuations (or perturbations to metrology systems).
These arguments, and others we give later in the paper, suggest that it is reasonable to expect the surface of such an asteroid to be stable enough to potentially use asteroids as test masses for gravitational-wave detection in the $\mu$Hz band.

%%%%%%%%%%%%%%%%%%%%%%%%%%%%%%%%%%%%%%%%%%%%%%%%%%%%%%%%%%%%%%%%%%%%%%%%%%%%%%%%
\begin{figure*}[p]
    \centering
    \includegraphics[width=0.95\textwidth]{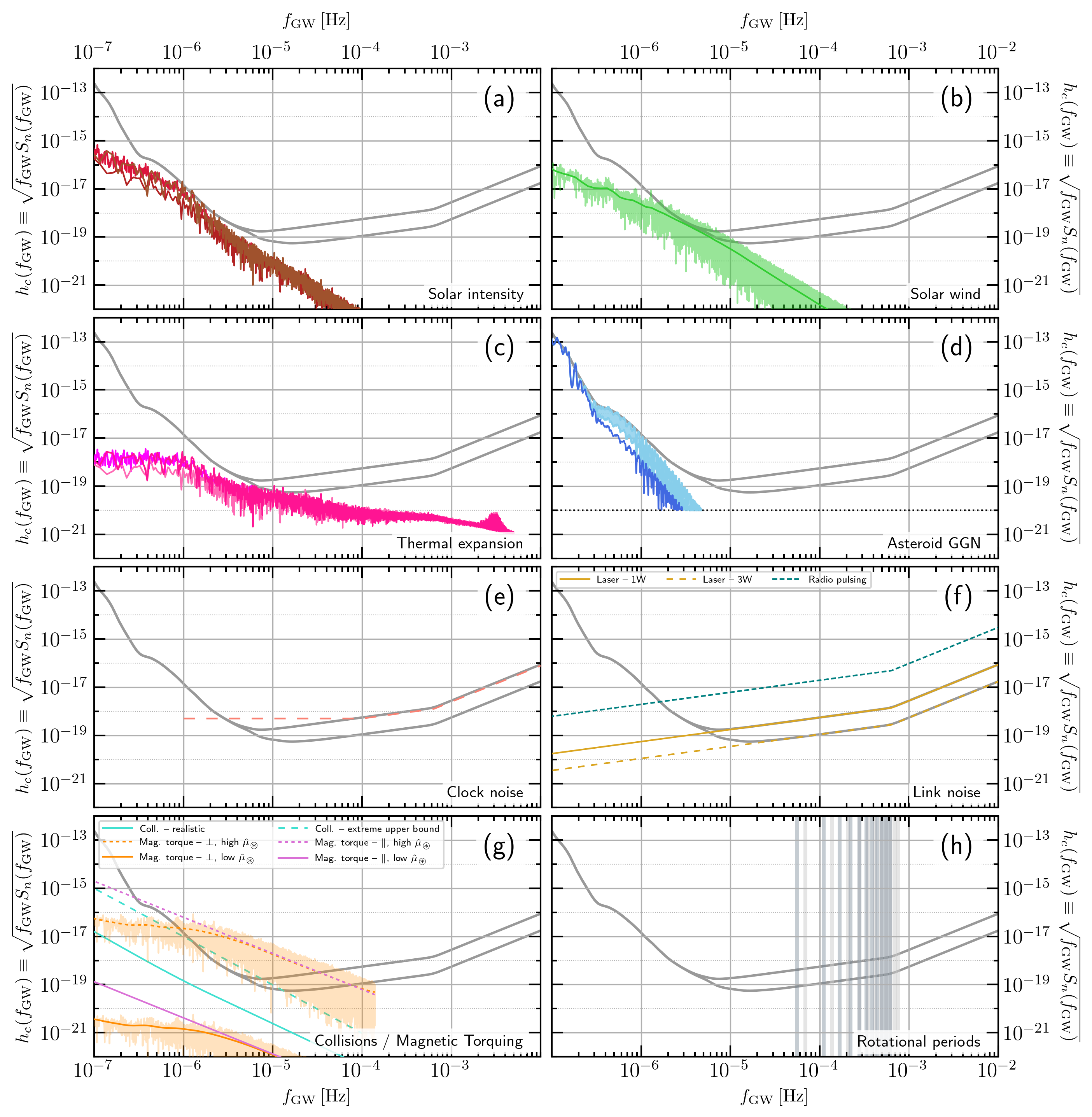}
    \caption{\label{fig:noise_contributions}%
            Individual noise contributions to overall characteristic strain sensitivity around the \Hz{\mu} band.
            In every panel (a)--(h), the solid thick grey lines denote two smoothed estimates of our combined sensitivity reach [\sectref{sect:sensitivity}]; see \figref{fig:results}.
            Each panel shows one or more of the noise estimates detailed in \sectref[s]{sect:noise} and \ref{sect:linkNoise}: (a) solar intensity CoM noise [\sectref{sect:TSI}; see also \sectref{sect:TSIandWindTorque}]; (b) solar wind CoM noise (green band, with smoothed result as a solid green line) [\sectref{sect:solarWind}; see also \sectref{sect:TSIandWindTorque}]; (c) thermal expansion [\sectref{sect:thermal}]; (d) asteroid GGN~\cite{Fedderke:2020yfy} simulation (solid dark blue) and close-pass estimate (shaded light blue), in both cases plotted only where $h_c \geq 10^{-20}$ [\sectref{sect:GGN}]; (e) clock noise (long-dashed salmon) (not included in enveloped noise curve; see text) [\sectref{sect:clocknoise}]; (f) link noise for two different laser powers in a laser pulsing setup (solid and dashed gold being 1\,W and 3\,W of laser power transmitted, respectively) [\sectref{sect:laserPulsing}], and in a radio pulsing setup (dotted teal) [\sectref{sect:radioPulsing}];
            (g) collisional CoM noise using in-quadrature (realistic; solid turquoise) or linear (worst-case; dashed turquoise) summation [\sectref{sect:collisions}; see also \sectref{sect:collisionsTorque}], 
            and a variety of magnetic torquing estimates using both high (dotted lines) and low (433~Eros-like; solid lines) values for the asteroid specific magnetic moment, for torques both parallel (purple lines) and perpendicular (orange bands with smoothed orange lines) to the angular momentum axis of the asteroid [\sectref{sect:EMforcesTorque}]; and (h) expected excluded bands (taken to be 5\% wide) around 4\,hr (light gray) and 5\,hr (darker blue-gray) asteroid rotational periods, and their first 10 harmonics [\sectref{sect:rotationalMotion}].
            In panels (a) and (c), the different colored curves denote the results of using different measured solar intensity fluctuation power spectral densities; the different lines are detailed in the relevant sections of \sectref{sect:externalAccnNoise}.
            These results assume asteroid TMs with 8\,km radii and densities of 2.5\,g/cm\up{3} (mass of $5.4\times 10^{15}$\,kg), located 1.5\,AU from the Sun, with a fixed 1\,AU baseline; see discussion in \sectref{sect:TSI}.
            A combined plot is given in \figref{fig:results_combo}; an enveloped noise curve is given in \figref{fig:results}.
            }
\end{figure*}
%%%%%%%%%%%%%%%%%%%%%%%%%%%%%%%%%%%%%%%%%%%%%%%%%%%%%%%%%%%%%%%%%%%%%%%%%%%%%%%%

The primary mission concept that we consider is described in \figref{fig:sketch}. 
Imagine deploying base stations on the surfaces of two asteroids separated by $\sim \text{AU}$.%
%%%%%%%%%%%%
    \footnote{See, e.g.,~\citeR[s]{Yeomans2085,NearShoemakerWebsite,Baker1327,HayabusaWebsite,TSUDA202042,Hayabusa2Website,Lauretta:2017erq,Smith672,OsirisRex,OririsRexWebsite1,OririsRexWebsite2,Bibring493,Taylor387,RosettaWebsite,RAYMAN2006605,DawnWebsite} for a variety of missions that have successfully rendezvoused with asteroids or comets, as orbiters (excluding non-orbital flybys), soft landers, sample collectors, and/or `rover' deployers.
    See also \citeR{SiddiqiNASA} for a broad historical overview.
    Additionally, DART~\cite{DARTwebsite} is a recently launched mission aiming to modify a binary asteroid orbital system (parent: 65803~Didymos) by impacting with the smaller body in the binary, as a planetary defense technology demonstrator.
    } %
%%%%%%%%%%%
Each base station contains a laser (or possibly radio) ranging system consisting of a receiver and transmitter.
The base station also contains a local space-qualified optical atomic clock that serves as a time standard.%
%%%%%%%%%%%%
    \footnote{\label{ftnt:clocksInSpace}%
    For discussions of on-orbit atomic clocks, see generally \citeR[s]{Tino:2007az,Kolkowitz:2016wyg,Su:2017kng,Liu:2018km,Ebisuzaki:2018ujm,10.1093/nsr/nwaa215,Voyage2050ESA,Burt:2021iad,DSACwebsite,GONZALEZ2000335,PHARAOwebsite,HighPerformanceClocksInSpace,ColdAtomsInSpace,SALOMON20011313,10.1117/12.2308164}.
    } %
%%%%%%%%%%%%
The ranging scheme works as follows.
A base station sends a pulse of light (or, possibly, radio waves) to the other station at pre-determined times. 
The fidelity of the time of transmission is maintained by the local atomic clock at this base station.
When this pulse arrives at the distant base station, it is received, amplified, and sent back to the original station; the amplification of the pulse must maintain its chronal properties.
This returned pulse is received finally at the initial base station, which compares the time of arrival of the return pulse to the time at which the original pulse was sent. 
This comparison is made by referencing the local atomic clock on the initial base station. 
These comparisons permit the measurement of the light-travel time (i.e., the proper distance) between these base stations as a function of time.  

This above scheme has the potential disadvantage that the detector can operate only when there is a line of sight between the base stations. 
Since asteroids rotate, this could lead to an $\mathcal{O}\left(1\right)$ loss in duty cycle if the base station were mounted rigidly to the asteroid.
A more complex setup could in principle recover much of this duty cycle. 
For example, one may consider maintaining a parent satellite in orbit at a suitable distance from the asteroid surface.
In addition, one or more probes would be deployed by this parent satellite to place an array of reflectors (e.g., retroreflectors) at various locations on the asteroid surface (such a setup has been envisaged for, e.g., improving asteroid rotational-state measurements~\cite{FRENCH2020113537}).
The distance between the satellite and the landed reflectors on the asteroid would be continuously measured (e.g., by sending signals from the satellite to the reflectors and measuring the time of arrival of the passively reflected signals); the satellite thus inherits the stability of the asteroid's position. 
The orbiting satellites at the locations of each of the asteroids may now measure the distance between themselves via a laser (or possibly radio) ranging system that can now be housed in the orbiting satellite, and is thus easier to stabilize and point in the desired direction.
In this protocol, the atomic clocks are also housed in the satellites, and they can be used as the time standard both for anchoring the location of the satellite to the appropriate (local) asteroid and for measuring the ranging distance to the other (distant) satellite.  

%%%%%%%%%%%%%%%%%%%%%%%%%%%%%%%%%%%%%%%%%%%%%%%%%%%%%%%%%%%%%%%%%%%%%%%%%%%%%%%%%%%%%%%%%%
\section{Test-mass noise sources}
\label{sect:noise}
%%%%%%%%%%%%%%%%%%%%%%%%%%%%%%%%%%%%%%%%%%%%%%%%%%%%%%%%%%%%%%%%%%%%%%%%%%%%%%%%%%%%%%%%%%

Broadly speaking, the consideration of whether asteroids constitute sufficiently good TM for our proposed mission concept in the face of various external environmental perturbations can be broken down into a four categories: (1) forces acting on the asteroid CoM; (2) torques acting on the rotational state of the asteroid; (3) rigid-body asteroid kinematic considerations (rotational and orbital) that could limit sensitivity; and (4) the excitation of internal degrees of freedom, such as thermal expansion and seismic noise.

In this section, we consider each of these four categories in turn, and discuss and estimate the impacts of both dominant and sub-dominant perturbations on the asteroids.
We will have frequent occasion to refer to \figref{fig:noise_contributions}, which presents most of the dominant or otherwise especially relevant noise estimates.

Throughout this section, we perform estimates for noise sources assuming the existence of a (fictitious) fiducial asteroid that we call `\Alice' (shorthand symbol `$\alice$').
We will generally take \Alice\ to be exactly spherical, with a radius of $R_{\alice}= 8$\,km and a uniform mass-density of $\rho_{\alice} = 2.5\,\text{g/cm}^3$, resulting in a mass of $M_{\alice}=5.4\times 10^{15}\,$kg.
Moreover, we will assume that \Alice\ is located at a fixed distance of $r_{\alice} = 1.5\,$AU from the Sun, is a perfectly uniform blackbody absorber, and has a rotational period of $T_{\alice} = 5$\,hrs.
Where necessary we will also assume the existence of a second such fictitious asteroid that we call `\Bob'; we assume \Bob\ to have the same physical and orbital characteristics as \Alice, except that it has a rotational period of $T = 4\,$hrs.
\Alice\ and \Bob\ will be assumed to be separated by a (fixed) baseline distance of $L = 1\,$AU for our estimates (this is conservative; cf.~\figref{fig:baselineASDs} later in the paper).
Many of these physical and orbital parameters are close to those of one of the largest real near-Earth asteroids: 433~Eros~\cite{JPL-SBD,Yeomans2085,1998AJ....116.2023M,THOMAS200218,Veverka2088} (see \tabref{tab:asteroids}).
However, in some cases, some of these fiducial asteroid assumptions will prove too na\"ive for certain noise estimates to be made correctly; in those cases we will relax the relevant simplifying assumption(s) in order to estimate noise sources that would only arise from non-sphericity, partially or non-uniformly reflective surfaces, and/or the elliptical and inclined orbital motion of real asteroids.

%%%%%%%%%%%%%%%%%%%%%%%%%%%%%%%%%%%%%%%%%%%%%%%%%%%%%%
%%%%%%%%%%%%%%%%%%%%%%%%%%%%%%%%%%%%%%%%%%%%%%%%%%%%%%
\subsection{Fluctuating forces: center-of-mass motion}
\label{sect:externalAccnNoise}
%%%%%%%%%%%%%%%%%%%%%%%%%%%%%%%%%%%%%%%%%%%%%%%%%%%%%%
%%%%%%%%%%%%%%%%%%%%%%%%%%%%%%%%%%%%%%%%%%%%%%%%%%%%%%
The most obvious source of perturbations to asteroids as test masses are those that directly act on the asteroid center of mass: external forces.

In this subsection, we consider in turn the forces that arise from \hyperref[sect:TSI]{(1)} the (fluctuating) solar radiation pressure, \hyperref[sect:solarWind]{(2)} the (fluctuating) solar wind, \hyperref[sect:GGN]{(3)} gravitational perturbations from other asteroids in the inner Solar System, \hyperref[sect:collisions]{(4)} collisions with dust and particles that permeate the inner Solar System, and \hyperref[sect:EMforces]{(5)} electromagnetic forces arising from both magnetic field gradients in the Solar System and electrical charging of the asteroid.

%%%%%%%%%%%%%%%%%%%%%%%%%%%%%%%%%%%%%%%%%%%%%%%%%%%
\begin{table*}[t]
    \begin{ruledtabular}
    %%%%%%%%%%%%%%%%%%%%%%%%%%%%%%%%%%%%%
    \caption{\label{tab:asteroids}%
        %%%%%%%%%%%%%%%%%%%%%%%%%%%%%%%%%%%%%
        Selected inner Solar System asteroids in the 10\,km class, and their physical and orbital properties (where available, else `---'): mass $M$~[kg], volume $V$~[km${}^{3}$], surface area $A$~[km${}^{2}$], mean radius $\bar{R}$~[km], extent~[km$\times$km$\times$km], mass density $\rho$~[g/cm${}^{3}$], rotational period $T_{\text{rot}}$~[hrs]; semimajor axis $\mathfrak{a}$~[AU], orbital eccentricity $\mathfrak{e}$, orbital inclination $\mathfrak{i}$~[${}^\circ$], and orbital period $T_{\text{orb}}$~[yrs].
        For ease of comparison, $\text{hr}^{-1} \sim 3\times 10^{-4}\Hz{}$ and $\text{yr}^{-1}\sim 3\times 10^{-8}\Hz{}$.
        `Extent' is approximately the size of a rectangular box into which the asteroid would fit snugly, whose volume in general differs by an $\order{1}$ factor from the volume actually occupied by the asteroid material.
        We do not necessarily intend to suggest that all of these would make good candidates for our mission concept; we present these data merely to demonstrate the existence of a handful of asteroids in the appropriate size, orbit, and rotational-period class, for which further study may be warranted.}
    %%%%%%%%%%%%%%%%%%%%%%%%%%%%%%%%%%%%%
    \begin{tabular}{rcccccccccccl}
    %%%%%%%%%%%%%%%%%%%%%%%%%%%%%%%%%%%%%
        &   \multicolumn{7}{|c|}{\textbf{Physical Parameters}} & \multicolumn{4}{c|}{\textbf{Orbital Parameters}} &  \\ 
        \hline
        %%%%%%%%%%%%%%%%%%%%%%%%%%%%%%%%%%%%%
        Name & $M$ & $V$ & $A$ & $\bar{R}$ & Extent & $\rho$ & $T_{\text{rot}}$ & $\mathfrak{a}$ & $\mathfrak{e}$ & $\mathfrak{i}$ & $T_{\text{orb}}$ & Ref. \\ \hline
        %%%%%%%%%%%%%%%%%%%%%%%%%%%%%%%%%%%%%
        433 Eros  & $6.7\times 10^{15}$ & $2.5\times 10^3$ & $1.1\times 10^3$ & 8.5 & $34\times11\times11$ & 2.7 & 5.3 & 1.46 & 0.22 & 10.8 & 1.8 & \cite{JPL-SBD,Yeomans2085,1998AJ....116.2023M,THOMAS200218,Veverka2088} \\
        %%%%%%%%%%%%%%%%%%%%%%%%%%%%%%%%%%%%%
        1627 Ivar & --- & --- & --- & 4.6 & --- & --- & 4.8 & 1.86 & 0.40 & 8.5 & 2.5 & \cite{JPL-SBD,DELBO2003116} \\
        %%%%%%%%%%%%%%%%%%%%%%%%%%%%%%%%%%%%%
        2064 Thomsen & --- & --- & --- & 6.8 & --- & --- & 4.2 & 2.18 & 0.33 & 5.7 & 3.2 & \cite{JPL-SBD} \\
        %%%%%%%%%%%%%%%%%%%%%%%%%%%%%%%%%%%%%
        6618 Jimsimons & --- & --- & --- & 5.8 & --- & --- & 4.1 & 1.87 & 0.04 & 23.8 & 2.6 & \cite{JPL-SBD} \\
        %%%%%%%%%%%%%%%%%%%%%%%%%%%%%%%%%%%%%
        1866 Sisyphus & --- & --- & --- & 4.2 & --- & --- & 2.4 & 1.89 & 0.54 & 41.2 & 2.6 & \cite{JPL-SBD} \\
        %%%%%%%%%%%%%%%%%%%%%%%%%%%%%%%%%%%%%
        3200 Phaethon & --- & --- & --- & 3.1 & --- & --- & 3.6 & 1.27 & 0.89 & 22.3 & 1.4 & \cite{JPL-SBD} \\
        %%%%%%%%%%%%%%%%%%%%%%%%%%%%%%%%%%%%%
        1036 Ganymed  & --- & --- & --- & 18.8 & --- & --- & 10.3 & 2.67 & 0.53 & 26.7 & 4.4 & \cite{JPL-SBD} \\
        %%%%%%%%%%%%%%%%%%%%%%%%%%%%%%%%%%%%%
        4954 Eric  & --- & --- & --- & 5.4 & --- & --- & 12.1 & 2.00 & 0.45 & 17.4 & 2.8 & \cite{JPL-SBD} \\
        %%%%%%%%%%%%%%%%%%%%%%%%%%%%%%%%%%%%%
    %%%%%%%%%%%%%%%%%%%%%%%%%%%%%%%%%%%%%
    \end{tabular}
    \end{ruledtabular}
\end{table*}
%%%%%%%%%%%%%%%%%%%%%%%%%%%%%%%%%%%%%%%%%%%%%%%%%%%

%%%%%%%%%%%%%%%%%%%%%%%%%%%%%%%%%%%%%%%%%%%%%%%%%%%%%%
\subsubsection{Solar intensity fluctuations}
\label{sect:TSI}
%%%%%%%%%%%%%%%%%%%%%%%%%%%%%%%%%%%%%%%%%%%%%%%%%%%%%%
At the location of Earth, the total solar irradiance~(TSI)---the energy flux density delivered by the Sun---is approximately $\bar{I}_{\odot} \equiv 1.36\,\text{kW/m}^2$ on average~\cite{Froehlich:2004wed}.
Assuming that a proportion $0\leq \epsilon\leq 2$ of the momentum carried by the incoming radiation is transferred to a body (with $1<\epsilon \leq2$ corresponding to partial-to-total retro-reflection of the incident radiation), the corresponding \emph{static} radiation pressure is%
%%%%%%%%%%%%
\footnote{\label{ftnt:naturalunits}%
    Throughout this paper, we write formulae using natural Heaviside--Lorentz units. 
    That is, we assume that $\hbar=c=1$, and that the fine-structure constant is $\alpha \equiv e^2/(4\pi)$.
    We restore SI units in numerical estimates where appropriate.
    } %
%%%%%%%%%%%%
$P_{\odot}(r) = \epsilon \bar{I}_{\odot} ( r_{\oplus} / r )^2 \sim 4.5\,\mu\text{N}/\text{m}^2 \cdot \epsilon \cdot ( r_{\oplus} / r )^2$, where $r$ (respectively, $r_{\oplus}$) is the distance of the body (respectively, Earth) from the Sun.
The average Earth--Sun distance is $r_{\oplus} \equiv 1\,\text{AU}\approx 1.5\times 10^{11}\,$m~\cite{PDG}.

Since \Alice\ is a perfect blackbody ($\epsilon = 1$), this pressure would give rise to a \emph{static} acceleration $a_0 \sim 8\times 10^{-14}\,\text{m/s}^2$.
Were this static acceleration a noise source for our concept, it would be fatally large: over $T \sim 1/(10 \mu\text{Hz}) = 10^5\,$s, it would generate a displacement of order $\Delta x \sim (1/2) a_0 T^2 \sim 0.4\,\text{mm}$ on each asteroid, which even taken over an AU baseline between two such asteroids would generate a strain contribution%
%%%%%%%%%%%%
\footnote{\label{ftnt:ignoreOrientation}%
    Note that throughout this paper, we neglect $\order{1}$ factors arising from orientation effects of the GW source as compared to the detector baseline.
    } %
%%%%%%%%%%%%
$h_c \sim 3\times 10^{-15}$, which would so severely limit the strain sensitivity as to make this concept of no real interest (cf.~\figref{fig:results}).

Of course, that is not actually the relevant noise estimate, precisely because $a_0$ is a \emph{static} (i.e., `DC') acceleration. 
Since we have in mind a search for temporally oscillatory (i.e., `AC') GW signals with our detector concept, we must evaluate instead the relevant \emph{in-band} noise contribution.

Consider again the instantaneous acceleration $\bm{a}(t)$ due to solar radiation pressure acting on a body of mass $M$ at distance $r(t)$ from the Sun, that presents a cross-sectional geometrical area $A(t)$ to the Sun, and which is subject to the (fluctuating) solar output $I_{\odot}(t) \equiv \bar{I}_{\odot} \cdot [ 1 + \delta I_{\odot}(t) ]$, where $\bar{I}_{\odot}$ is the average TSI and $\delta I_{\odot}(t)$ is the time-dependent fractional TSI fluctuation, with the DC piece subtracted off so that the temporal average $\langle \delta I_{\odot}(t) \rangle = 0$.
The TSI fluctuation depends mildly on the solar cycle, and its power spectral density (PSD) $S[\Delta I_{\odot}]$ has been well measured at different epochs~\cite{Froehlich:2004wed}.
$\delta I_{\odot}$ should not however depend strongly on heliocentric distance in the $\sim \text{AU}$ range of distances, and so can be taken as measured near Earth.
The magnitude of this acceleration is of order
%%%%%%
\begin{align}
    a(t) \sim \bar{I}_{\odot} \frac{ A_{\text{eff}}(t) }{ M } \left[ 1 + \delta I_{\odot}(t) \right] \lb( \frac{r_{\oplus}}{ r_{\alice} } \rb)^2,
    \label{eq:accnTimeSeries}
\end{align}
%%%%%%
where we have written an effective area $A_{\text{eff}}(t) \equiv \epsilon(t) A(t)$ to account for both the geometric and albedo variations of the surface of the asteroid presented to the Sun as a function of time.

For \Alice, all of the quantities $r(t)$, $A_{\text{eff}}(t) = \pi R_{\alice}^2$, and $M = 4\pi \rho_{\alice} R_{\alice}^3 / 3$ are time independent; therefore, the temporal fluctuation arises from the fluctuating solar output:
%%%%%%
\begin{align}
    \delta a_{\alice}(t) \sim  \frac{ 3 \bar{I}_{\odot} }{ 4 \rho_{\alice} R_{\alice} } \lb( \frac{r_{\oplus}}{ r_{\alice} } \rb)^2 \delta I_{\odot}(t).
    \label{eq:accnTimeSeriesAlice}
\end{align}
%%%%%%
The amplitude spectral density (ASD)%
%%%%%%%%%%
\footnote{\label{ftnt:ASD}%
		The ASD is defined as the square root of the power spectral density (PSD) $S[a_{\alice}](f)$.
		We follow the Fourier transform and PSD-normalization conventions of Appendix C of \citeR{Fedderke:2020yfy}.
	} %
%%%%%%%%%%
of the acceleration is therefore
%%%%%%
\begin{align}
    \sqrt{ S[a_{\alice}](f) } \sim  \frac{ 3 \bar{I}_{\odot} }{ 4 \rho_{\alice} R_{\alice} } \lb( \frac{r_{\oplus}}{ r_{\alice} } \rb)^2 \sqrt{ S[\delta I_{\odot}](f) }.
    \label{eq:accnASDAlice}
\end{align}
%%%%%%

To convert this analytically to an approximate noise estimate for strain measurements made between \Alice\ and \Bob, we make some further assumptions: 
%%%%%%%%%%%%%%%%%%%%%%
\begin{enumerate}[label=(\alph*)]
%%%%%%%%%%%%%%%%%
\item%
    we will neglect geometrical effects in the instantaneous baseline-projection of the independent (vector) acceleration noises on each asteroid, and assume that the component of the acceleration along the baseline is of the order of \eqref{eq:accnTimeSeriesAlice}.
    This is likely conservative by an $\order{1}$ factor since the radiation pressure is (predominantly) radial from the Sun, while the baseline separation vector will in general not be;
%%%%%
\item%
    we will make the assumption, physically well motivated since we assume ranging between similarly sized asteroids separated by $\sim \text{AU}$ baselines, of uncorrelated accelerations of similar amplitude acting on the asteroids at each end of the baseline, so that the net baseline-projected differential acceleration noise ASD $\sqrt{ S[\Delta a](f)}$ is a factor of $\sim \sqrt{2}$ larger than the single-asteroid estimate at \eqref{eq:accnASDAlice}: $ \sqrt{ S[\Delta a](f)} \sim \sqrt{ 2 S[a_{\alice} ](f)}$; and
%%%%%
\item%
    we take a fixed baseline length of $L = 1\,$AU between \Alice\ and \Bob\ (see detailed discussion below under `Refinements for real asteroids').
%%%%%%%%%%%%%%%%%
\end{enumerate}
%%%%%%%%%%%%%%%%%%%%%%

Under these additional assumptions, we can convert between the single-asteroid acceleration ASD and an estimate for the characteristic strain noise amplitude $h_c$ using%
%%%%%%%%%%%%
\footnote{\label{ftnt:leadingFactorOfTwo}%
    The leading factor of 2 in the numerator in \eqref{eq:hCalice} arises because a monochromatic plane gravitational wave of amplitude $h(t) \sim h_0 \cos(\omega_{\textsc{gw}}t)$ normally incident on the plane containing the orbits of two co-planar-orbiting test masses generates, in the long-wavelength $\omega_{\textsc{gw}}L\ll 1$ limit, a baseline-projected acceleration of the form $\Delta a(t) \sim \frac{1}{2}L \ddot{h}(t)$~\cite{Misner:1974qy,Fedderke:2020yfy}.
    This factor of 2 can also be understood directly from \eqref{eq:GWmetric}, since the prefactors of $dx^2$ and $dy^2$ are $\sim [ 1 + h \sin(\ldots) ]$; the implied proper distance change for fixed co-ordinate location of the TMs (the correct prescription in transverse-traceless gauge) is thus $\sim (h/2) \sin(\ldots)$ in the $h \ll 1$ limit.
    Of course, we have not accounted here for any orientation effects of the orbits or of the GW relative to the orbital planes, so our estimates are all at best accurate to \order{1} factors.
    } %
%%%%%%%%%%%%
%%%%%%
\begin{align}
    h_c(f) &\sim \frac{2\sqrt{ f S[\Delta a](f) }}{(2\pi f)^2 L}     \label{eq:hCalice} \\
        &\sim \frac{3}{4\sqrt{2}\cdot \pi^2}\, \frac{ \bar{I}_{\odot} }{  \rho_{\alice} R_{\alice} f^2 L } \lb( \frac{r_{\oplus}}{ r_{\alice} } \rb)^2 \sqrt{ f S[\delta I_{\odot}](f) }.
    \label{eq:hCaliceSimplified}
\end{align}
%%%%%%

Various fractional TSI PSDs are presented in Fig.~12 of \citeR{Froehlich:2004wed}.
These are based on: (1) a composite of solar intensity measurements from a variety of missions spanning 1978--2002 (PMOD composite~\cite{Froehlich:2004wed}), (2) data from the VIRGO instrument on SOHO during solar maximum (Oct.~2000--Feb.~2002), and (3) data from VIRGO during solar minimum (Feb.~1996--Aug.~1997).
Using each of these three PSDs in turn to make separate estimates, we arrive at the noise estimates shown respectively by the (1) red, (2) brown, and (3) maroon lines in \subfigref{fig:noise_contributions}{a}.

%%%%%%%%%%%%%%%%%%%%%%%%%%%%%%%%%%%%%%%%%%
\bparagraph{Refinements for real asteroids}
%%%%%%%%%%%%%%%%%%%%%%%%%%%%%%%%%%%%%%%%%%
Let us consider first the acceleration noise estimate \eqref{eq:accnTimeSeries}, and restore the time dependence of the effective area and asteroid--Sun distance.
We write $X(t) = \bar{X} ( 1 + \delta X(t) )$ for $X \in \{ A_{\text{eff}}, r \}$.
It is not atypical to have $\bar{A}_{\text{eff}} \sim \order{ \bar{A}_{\text{geom}} }$ with $\bar{A}_{\text{geom}}$ the average geometrical cross-sectional area presented to the Sun,%
%%%%%%%%%%%%
\footnote{\label{ftnt:albedos}%
    Typical asteroid albedos lie in the range $\sim 0.1-0.4$~\cite{JPL-SBD}, implying that $\epsilon \sim \order{1}$.
    } %
%%%%%%%%%%%%
and $|\delta A_{\text{eff}}|\sim \order{1}$ (e.g., 433 Eros, which has a highly non-spherical shape; see \tabref{tab:asteroids}).
However, neglecting small longer-term variations in the surface albedo and/or geometrical area from space weathering or impacts on the asteroid surface, $\delta A_{\text{eff}}$ will only have dominant frequency content at or above the asteroid rotational frequency, $\gtrsim 3\times 10^{-5}\Hz{}$.
Moreover, for a realistic asteroid on an elliptical orbit with semi-major axis $\mathfrak{a}$ and ellipticity $\mathfrak{e}$ (typically small but not vanishingly so for asteroids of possible interest to this concept; see \tabref{tab:asteroids}), we have $r(\theta) = \mathfrak{a}(1-\mathfrak{e}^2)\lb(1+\mathfrak{e}\cos\theta\rb)^{-1}$ with $\theta=\theta(t)$ the angle about the orbit from perihelion (which by Kepler's Third Law of course evolves non-trivially in time for an elliptical orbit; see, e.g., the discussion in Appendix A.4 of \citeR{Fedderke:2020yfy}).
This means that we can take $\bar{r} = \mathfrak{a}$ with $- \mathfrak{e} \leq \delta r(t) \leq +\mathfrak{e}$, and with $\delta r(t)$ having dominant frequency content near and around inner Solar System orbital frequencies $\sim 10^{-8}\,$Hz--$10^{-7}\,$Hz (i.e., periods of $\sim 0.3$--$3$\,yrs).
However, note that the \emph{temporal} average $\langle r^{-2}\rangle_T =  \mathfrak{a}^{-2} \lb[ 1-\mathfrak{e}^2\rb]^{-1/2} > \mathfrak{a}^{-2}$, while  $\mathfrak{a}^{-2}(1+\mathfrak{e})^{-2}< r^{-2} <  \mathfrak{a}^{-2}(1-\mathfrak{e})^{-2}$.

Let us now understand the impacts of these observations.
First, consider how these additional time variations impact the normalization of the acceleration noise arising from the in-band solar fluctuation.
Since the effective area modulates rapidly, over timescales $T \sim f_{\textsc{gw}}^{-1}$, we are justified in keeping in place the approximation $A_{\text{eff}} \sim \bar{A}_{\text{geom}}$.
On the other hand, the orbital modulation is slower than the GW period, so we would expect to see a rising and falling noise level as the asteroids move around their orbits. 
Indeed, the variation in $r^{-2}$ noted above would be expected to cause the instantaneous in-band noise level to rise and fall by a factor of $\mathcal{O}(2)$ in either direction for typical asteroid eccentricities of $\mathfrak{e}\sim0.3$ (for asteroids like 3200 Phaethon with $\mathfrak{e}\sim 0.9$, the difference is clearly closer to an order of magnitude, but that may just be a sign to avoid such asteroids in planning this mission), with an average value slightly larger than $\mathfrak{a}^{-2}$ by maybe a few/tens of percent.
In our estimates, we take $r(t) \rightarrow r_{\alice} = 1.5\,$AU, which is a representative value---if perhaps slightly smaller than average---for $\mathfrak{a}$ for relevant asteroids per \tabref{tab:asteroids}. 
It is possible therefore that our noise estimates are slightly too aggressive by something on the order of a $\mathcal{O}(2)$ factor at the worst possible times around the orbits, but on the other hand they are too conservative at more favorable times.
This degree of uncertainty is clearly within the intended level accuracy of our overall estimations here, and we do not attempt to correct for it.

In addition to causing a change to the normalization of the in-band noise directly from the solar fluctuations, we also have additional frequency contributions, and also potentially cross terms that can move two out-of-band noise sources into the band of interest.
Let us substitute into \eqref{eq:accnTimeSeries}, and assume for the sake of argument that $\mathfrak{e} \ll 1$.
We can then expand:
%%%%%%
\begin{align}
    \frac{a(t)}{a_0} &\equiv 1 + \delta a(t) \\
    &\sim \frac{ \lb( 1 + \delta A_{\text{eff}}(t) \rb) \lb(1 + \delta I_{\odot}(t) \rb)}{ (1+\delta r(t))^2 } \label{eq:aExp} \\
    &\sim 1 + \delta I_{\odot}(t) + \delta A_{\text{eff}}(t) - 2 \delta r(t) \nl 
    + \delta I_{\odot}(t) \cdot \delta A_{\text{eff}}(t) - 2\delta r(t) \cdot \delta A_{\text{eff}}(t)\nl
    - 2 \delta r(t) \cdot \delta I_{\odot}(t) + 3 \delta r(t)\cdot \delta r(t) + \cdots,
\end{align}
%%%%%%
where
%%%%%%
\begin{align}
    a_0  &\equiv \bar{I}_{\odot} \frac{\bar{A}_{\text{geom}}}{M}\lb( \frac{r_{\oplus}}{ \mathfrak{a} } \rb)^2.
\end{align}
%%%%%%

If we now ask what the frequency content of this noise is, we first find the expected DC term which we can neglect.
This is followed by linear terms that will contribute noise at their respective dominant frequencies: the orbital contribution from $\delta r(t)$ is out of band%
%%%%%%%%%%%%
\footnote{\label{ftnt:windowing}%
    With appropriate signal windowing, this noise can be very effectively confined out of the band of interest; see, e.g., \citeR[s]{Harris:1978wdg,Fedderke:2020yfy}.
    Similarly for the noise at the rotational frequency.
    }
%%%%%%%%%%%%
on the low-frequency side and rotational motion from $\delta A_{\text{eff}}(t)$ is out of band (or the limiting noise source) on the high-frequency side.
Neither of those will contribute in-band noise, so the only linear term that contributes in-band noise is the term arising from the direct in-band TSI fluctuations $\delta I_{\odot}$ that we have already estimated.

For the quadratic terms, we recall the multiplication--convolution theorems of Fourier analysis: a multiplication in the time domain is a convolution in the frequency domain.
Consider first the quadratic terms with one power of $\delta r(t) \sim \mathfrak{e} \ll 1$: since $\delta r(t)$ has power only at very low frequencies $\sim 10^{-8}$--$10^{-7}$\,Hz, these quadratic terms will only induce small-amplitude side-bands around the dominant frequencies in $\delta I_{\odot}(t)$ and $\delta A_{\text{eff}}$.
These side-bands are suppressed by $\mathfrak{e}$ and are located within $\sim 10^{-8}$--$10^{-7}$\,Hz of the respective dominant frequencies in $\delta I_{\odot}(t)$ and $\delta A_{\text{eff}}$.
Since $\delta A_{\text{eff}}$ is already out of band on the high-frequency side, there is no serious impact from the $\delta r(t) \cdot \delta A_{\text{eff}}(t)$ term. 
The at-worst impact of the $\delta r(t) \cdot \delta I_{\odot}(t)$ term is thus to add and rearrange some in-band power in $\delta I_{\odot}(t)$ to other nearby in-band frequencies; this does not change our estimates by more than \order{1} factors.
The quadratic terms containing two powers of $\delta r(t)$ [and similar higher powers] will contain power at higher harmonics of the dominant orbital frequencies (and the even powers also contain a zero-frequency term that is part of the re-normalization of the in-band direct solar power contribution that we already discussed above); however, owing to the assumption of $\mathfrak{e} \ll 1$ (and this conclusion also holds for $\mathfrak{e}\lesssim 1$), this power is exponentially truncated moving above the orbital frequency band, and so does not significantly leak into the $\mu$Hz band (see \citeR{Fedderke:2020yfy} for a lengthy discussion of a similar effect, and also \sectref{sect:orbitalKinematics}).
The quadratic terms involving $\delta r(t)$ (and higher powers $[\delta r(t)]^n$) thus do not impact our results beyond the in-band power renormalization we discussed above.

The quadratic term that is potentially worrisome is $\delta I_{\odot}(t) \cdot \delta A_{\text{eff}}(t)$.
Because we expect that $\delta A_{\text{eff}}(t)$ has $\order{1}$ fluctuations at rotational frequencies, where  $\delta I_{\odot}(t)$ still has non-zero power, this cross term can contain in-band power near $\sim \mu$Hz.
That is, frequency components of $\delta I_{\odot}(t)$ that lie very close to the dominant frequency content of $\delta A_{\text{eff}}(t)$ interfere and produce a low-frequency beat.
However, we expect that $\delta A_{\text{eff}}(t)$ has quite sharp frequency features at the rotational period and its harmonics, while the ASD of $\delta I_{\odot}(t)$ falls with increasing frequency above the $\mu$Hz band~\cite{Froehlich:2004wed} (it is at the few per-mille level around the asteroid rotational period).
We show in \appref{app:crossPowerTerms} that the resulting low-frequency beat note is thus not expected to modify our noise estimate significantly (i.e., by more than an \order{1} factor).

In summary, the additional time variations from the rotational and orbital motions of the asteroids introduce uncertainties in our estimate of the in-band acceleration noise from the solar radiation pressure by perhaps \order{1} factors.
They do not however appear to modify that estimate significantly.

Another effect we have neglected so far is the variation of the baseline with time in the translation from the acceleration noise ASD to the strain noise ASD: we simply took $L \sim 1\,$AU to be fixed in the conversion at \eqref{eq:hCalice}.
Of course, as both \Alice\ and \Bob\ move around their orbits, this baseline distance will modulate by perhaps up to an order of magnitude in total for typical asteroid orbital configurations (see also \sectref{sect:orbitalKinematics}); it will also rotate with respect to the vector acceleration noises acting on each asteroid.
While both of these modulation effects will have an impact, they are both low-frequency, as they are associated with orbital motion around $\sim 10^{-8}$--$10^{-7}$\,Hz, below the $\mu$Hz band.
We thus do not expect these effects to have any impact on our in-band solar-intensity-induced strain noise estimation beyond perhaps shuffling some in-band power around in frequency to other in-band frequencies, and again perhaps modifying the normalization of the in-band strain noise. 
Taking these additional effects into account correctly would require simulation of the actual asteroid motion in response to applied accelerations, followed by an extraction of the baseline-projected strain time series; were our goal here detailed mission simulation, this would of course be necessary.
However, our goal in this work is to provide rough estimates of noise magnitudes in order to demonstrate the viability of this mission concept.
As such, we retain the simple analytical conversion between acceleration and strain that we have performed at \eqref{eq:hCalice}, and defer detailed modeling for any of these effects to future work, safe in knowledge that (a) the largest impact these effects have are out of band, and (b) our choice to fix $L=1\,$AU in our noise estimation is actually fairly conservative since the distance between typical orbiting points on typical inner Solar System asteroid orbits is larger than 1\,AU most of the time (see \figref{fig:baselineASDs} later in the paper).

%%%%%%%%%%%%%%%%%%%%%%%%%%%%%%%%%%%%%%%%%%%%%%%%%%%%%%
\subsubsection{Solar wind fluctuations}
\label{sect:solarWind}
%%%%%%%%%%%%%%%%%%%%%%%%%%%%%%%%%%%%%%%%%%%%%%%%%%%%%%
The solar wind~\cite{1958ApJ...128..664P,Schwenn:2006sdf,Verscharen:2019vsd} is a stream of positively charged ions (mostly protons and helium, with trace heavier elements), as well as electrons, that flows outward from the Sun with average proton speeds of $\bar{v}_p \sim 4\times 10^2\,\text{km/s}$~\cite{Ipavich:1998wdg,celias-mtofDATASET} and average proton densities of $\bar{n}_p \sim 3$--$8\,\text{cm}^{-3}$~\cite{Ipavich:1998wdg,celias-mtofDATASET,ISSAUTIER20052141,LeChat2012};%
%%%%%%%%%%%%
\footnote{\label{ftnt:differentwinds}%
    Note that there are actually `fast' and `slow' components of the solar wind that have different speeds and densities~\cite{LeChat2012}; these numbers are nevertheless representative.
    }
%%%%%%%%%%%%
these particles will scatter from the asteroid surface, supplying a force on the asteroid.
The protons in the solar wind are currently monitored in real time by the CELIAS/MTOF proton monitor (PM)~\cite{Ipavich:1998wdg,celias-mtofARTICLE,celias-mtofDATASET,celias-mtofURL} on the SOHO satellite~\cite{Domigo:1995erh} located at the Earth--Sun L1 Lagrange point~\cite{Domigo:1995erh}, which is at a distance of $r_{L1} \approx 0.99\,\text{AU} \sim 1\,$AU from the Sun.%
%%%%%%%%%%%
\footnote{\label{ftnt:annualNotCorrected}%
    Of course, the L1 location varies at the percent level on an annual cycle owing to Earth's slightly eccentric orbit ($\mathfrak{e}_{\oplus} \approx 0.017$~\cite{EarthEccen}); we do not correct for this as our estimates are not intended to be accurate at the percent level.
} %
%%%%%%%%%%%
It is also currently monitored by Parker Solar Probe~\cite{ParkerSolarProbeWebsite}, the PlasMag instrument on the DSCOVR satellite~\cite{DSCOVRwebsite}, and a number of instruments on the ACE satellite~\cite{Stone:1998dga,ACEwebsite}.
The wind has previously been studied by a host of other spacecraft missions such as Helios~\cite{1977JGZG...42..561R}, Wind~\cite{Lin1995,Ogilvie1995,Bougeret1995,ISSAUTIER20052141} and Ulysses~\cite{1992A&AS...92..237B,Issautier:2008wqe,LeChat2012}. 

The CELIAS/MTOF PM supplies $\sim 25$\,yrs of data on the instantaneous ($30\,$s sampling resolution) number density $n_p(t)$ and velocity $v_p(t)$ of the proton flux in the solar wind.
These measurements show that the densities and speeds of the wind have temporal fluctuations that will give rise to an in-band noise source for our GW detection proposal, in much the same way as the fluctuating TSI.

We estimate the impact on the asteroid CoM of this fluctuating solar wind proton flux as follows.
Consider a gas of protons of density $n_p(t)$ streaming approximately radially outward from the Sun at speed $v_p(t)$ [note that the proton temperature is roughly a factor of 10 lower~\cite{DSCOVRwebsite} than the kinetic energy associated with the bulk outflow; random proton motion with respect to the bulk flow can thus reasonably be neglected]. 
Each proton carries momentum $p_p(t) = m_p v_p(t)$, and the radial momentum flux carried by the gas is 
%%%%%%
\begin{align}
    \frac{d^2p_p(t)}{dtdA_{\perp}} \approx  m_p v_p(t) \times n_p(t) v_p(t) &\equiv m_p \Omega_p(t),
    \label{eq:solarWindMomentumFlux} \\
    \Omega_p(t) &\equiv n_p(t) [v_p(t)]^2,
    \label{eq:solarWindOmega}
\end{align}
%%%%%%
where we defined a quantity $\Omega_p(t)$ that depends solely on SOHO measurements; $A_{\perp}$ is the cross-sectional area presented by the asteroid to the incoming wind.
Over a broad range of heliocentric distances $0.5\,\text{AU}\lesssim r \lesssim 80\,\text{AU}$ (i.e., out to the edge of the heliosphere), the solar wind radial velocity has been directly measured~\cite{Schwartz:1983sfa,Richardson:1995wrg,Richardson2013,Provornikova_2014,Nakanotani_2020} and found to be largely independent of~$r$, while the number density of solar wind particles is measured to fall as roughly $n_p \propto r^{-2}$~\cite{Schwartz:1983sfa,Richardson:1995wrg,Richardson2013,Provornikova_2014,Nakanotani_2020}; this is also consistent (up to logarithmic corrections) with an isothermal Parker wind model outside the sonic radius $r > r_s$, where $r_s \ll \text{AU} $~\cite{1958ApJ...128..664P}.
In particular, this means that this outward momentum flux applies a time-dependent total force to \Alice\ of
%%%%%%
\begin{align}
    F_{\alice}(t) \approx \epsilon_p(t) \pi m_p R_{\alice}^2 \Omega_p(t) \cdot \lb( \frac{r_{\oplus}}{r_{\alice}} \rb)^{2},
    \label{eq:solarWindForce}
\end{align}
%%%%%%
where $1 \leq \epsilon_p \leq 2$ is an $\order{1}$ parameter characterising the proton--asteroid collisions and surface geometry: the lower (respectively, upper) bound on $\epsilon_p$ is saturated when \Alice\ is a perfect absorber (respectively, retroreflector) of the momentum flux.
The value $\epsilon_p=1$ is also attained if the reflection of protons from a perfectly spherical asteroid surface is exactly specular.
We will thus take $\epsilon_p\sim 1$.

Writing $\Omega_p(t) \equiv \bar{\Omega}_p \lb[ 1 + \delta \Omega_p(t) \rb]$ where $\langle \delta \Omega_p(t) \rangle \equiv 0$, we have $\bar{\Omega}_p \approx 9\times 10^{17}\,\text{m}^{-1}\,\text{s}^{-2}$~\cite{celias-mtofARTICLE,celias-mtofDATASET,celias-mtofURL}.
Therefore, for \Alice\ we have the net acceleration fluctuation of
%%%%%%
\begin{align}
    \delta a_{\alice}(t) \sim \frac{ 3 m_p }{ 4 \rho_{\alice} R_{\alice} } \bar{\Omega}_p \lb( \frac{r_{\oplus}}{r_{\alice}} \rb)^{2} \delta \Omega_p(t).
    \label{eq:solarWindAccn}
\end{align}
%%%%%%
This is identical to \eqref{eq:accnTimeSeriesAlice} under the replacements $\bar{I}_{\odot} \rightarrow m_p \bar{\Omega}_p$ and $\delta I_{\odot}(t) \rightarrow \delta \Omega_p(t)$.

The acceleration ASD is thus obtained from \eqref{eq:accnASDAlice} under the replacement on the RHS of $\bar{I}_{\odot} \rightarrow m_p \bar{\Omega}_p$ and $\sqrt{S[\delta I_{\odot}](f)}\rightarrow \sqrt{S[\delta \Omega_p](f)}$, where the solar wind fluctuation ASD $\sqrt{S[\delta \Omega_p](f)}$ is directly computed from the CELIAS/MTOF PM SOHO time series data by Fast Fourier Transform (FFT).%
%%%%%%%%%%%%
\footnote{\label{ftnt:ASDsolarComp}%
    The data stream from the CELIAS/MTOF PM is not `complete', in the sense that there are durations over which data sampled at a uniform temporal spacing are not available.
    This presents issues for the FFT.
    One na\"ive way to deal with this is to linearly interpolate available data for $\Omega_p(t)$ to a regular grid before performing the FFT, and the results we present in this paper are based on that approach.
    Alternative, more sophisticated, approaches such as the non-uniform fast Fourier transform (NUFFT) are available (see, e.g., \citeR[s]{2018arXiv180806736B,2020arXiv200109405B,FINUFFTwebsite} for one implementation); however, in the relevant frequency band, we have explicitly checked that the NUFFT gives results for $h_c(f)$ (smoothed in log-frequency space by a sliding Gaussian kernel with a standard deviation parameter of $0.05\,\log_{10}[\text{Hz}]$) that differ from the na\"ive approach by only an $\order{1}$ factor at worst, and are typically in much better agreement than even that.
} %
%%%%%%%%%%%%

Under the same assumptions discussed in the text between \eqref[s]{eq:accnASDAlice} and (\ref{eq:hCalice}), we then obtain the characteristic strain from the solar wind fluctuations from \eqref{eq:hCaliceSimplified} with the replacements ${\bar{I}_{\odot} \rightarrow m_p \bar{\Omega}_p}$ and $\sqrt{S[\delta I_{\odot}](f)}\rightarrow \sqrt{S[\delta \Omega_p](f)}$.

This noise curve is shown by the green band in \subfigref{fig:noise_contributions}{b}, with the solid green line being the sliding average noise curve smoothed over a Gaussian kernel in\linebreak (log-)frequency space.
Similar comments as in \sectref{sect:TSI} regarding real-asteroid modifications to these results apply here as they do for the TSI noise source.

%%%%%%%%%%%%%%%%%%%%%%%%%%%%%%%%%%%%%%%%%%%%%%%%%%%%%%
\subsubsection{Asteroid gravity gradient noise}
\label{sect:GGN}
%%%%%%%%%%%%%%%%%%%%%%%%%%%%%%%%%%%%%%%%%%%%%%%%%%%%%%
Because this detection proposal makes use of local test masses located in the inner Solar System, it is subject to the asteroid gravity gradient noise (GGN) that we previously identified and estimated in \citeR{Fedderke:2020yfy}. 
Here, we adopt the noise curve from \citeR{Fedderke:2020yfy} for circular $1\,$AU detector orbits with a 1\,AU baseline (middle panel of Fig.~2 of \citeR{Fedderke:2020yfy}); despite the mismatch with the assumed $1.5\,$AU orbits here for \Alice\ and \Bob, we expect that this is an appropriate estimate for this noise contribution within some $\mathcal{O}(1)$ factor, since this orbital radius is still far outside the main belt.
In a future detailed technical design study for this mission concept, this noise source should be recomputed assuming the real (elliptical, inclined) orbits of the asteroids selected.
This noise curve is shown by the dark blue line in \subfigref{fig:noise_contributions}{d}, with the lighter blue shaded band giving the close-pass noise estimate discussed in Sec.~V\,E of \citeR{Fedderke:2020yfy} and shown in Fig.~5 of that reference.
In both cases, we have only shown these curves for $h_c \gtrsim 10^{-20}$ [horizontal black dotted line in \subfigref{fig:noise_contributions}{d}], where the asteroid GGN is already a highly subdominant noise contribution for this proposal; this is done in order to avoid clear numerical artefacts that occurred in our simulations from \citeR{Fedderke:2020yfy} at smaller values of $h_c$.%
%%%%%%%%%%%%
\footnote{\label{ftnt:differentCut}%
    In \citeR{Fedderke:2020yfy}, we imposed a cutoff $\sqrt{S_n} \gtrsim 10^{-17}/\sqrt{\text{Hz}}$; this is a roughly equivalent criterion to the one used in this work, since the cutoff occurs for $f\sim 10^{-6}\,$Hz and $h_c = \sqrt{fS_n}$.
    } %
%%%%%%%%%%%%

%%%%%%%%%%%%%%%%%%%%%%%%%%%%%%%%%%%%%%%%%%%%%%%%%%%%%%
\subsubsection{Collisions}
\label{sect:collisions}
%%%%%%%%%%%%%%%%%%%%%%%%%%%%%%%%%%%%%%%%%%%%%%%%%%%%%%

%%%%%%%%%%%%%%%%%%%%%%%%%%%%%%%%%%%%%%%%%%%%%%%%%%%%%%%%%%%%%%%%%%%%%%%%%%%%%%%%
\begin{figure}[!p]
    \centering
    \includegraphics[width=\columnwidth]{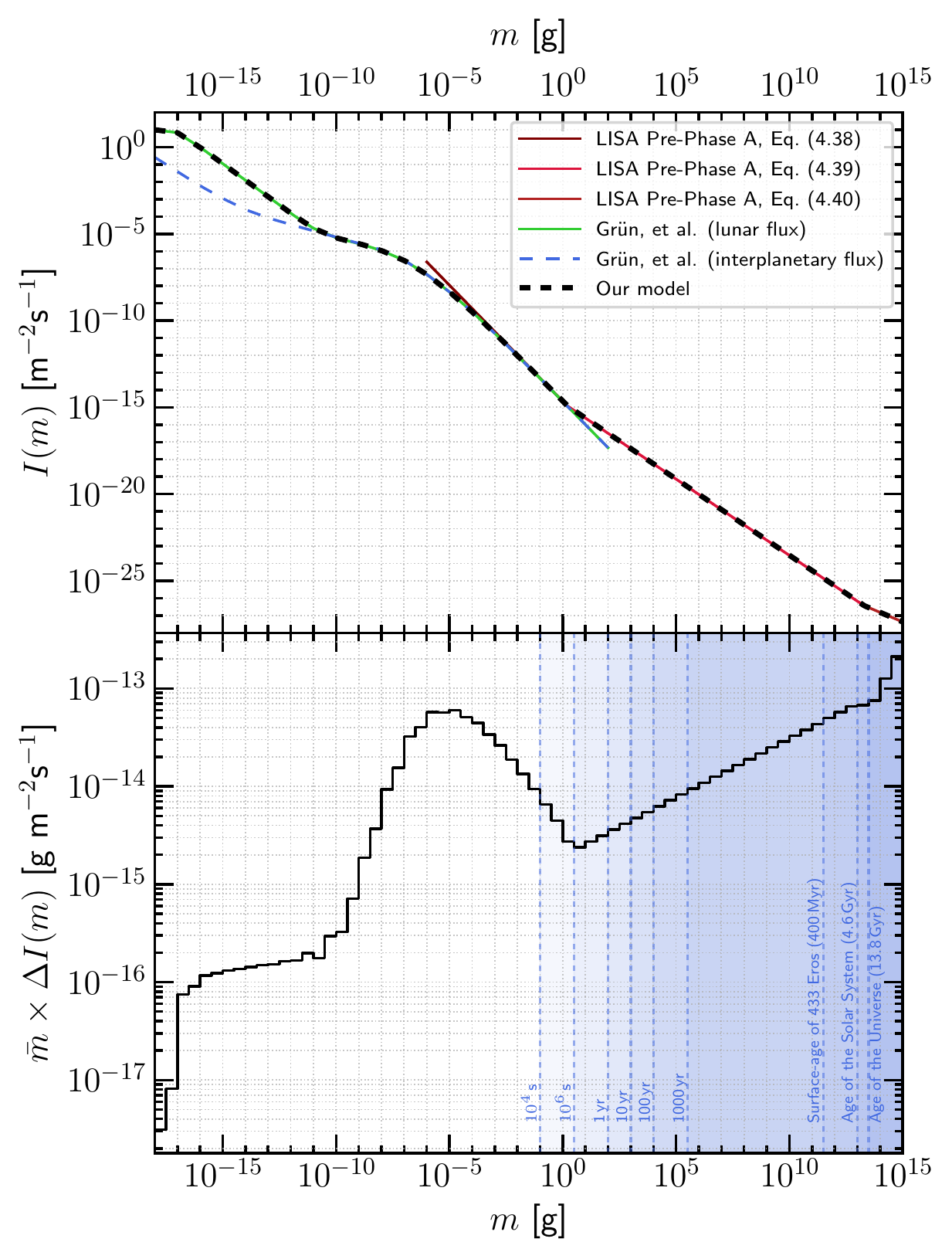}
    \caption{\label{fig:IM}%
    \textsc{Upper panel:} Integral number-flux density $I(m)$, defined as at \eqref{eq:dustNumberFluxDensityDefn}, of dust in the interplanetary medium in the vicinity of Earth's orbit.
    The various colored lines are: the dust model adopted in \citeR{LISA-Pre-Phase-A} (various shades of red, as annotated), and both the lunar (green) and interplanetary (dashed blue) flux models from \citeR{GRUN1985244} (identified in the legend as `Gr{\"u}n, et al.').
    We adopt the conservative estimate shown in thick, dashed black; the exact construction of this curve is discussed in the text.
    \textsc{Lower panel:} The mass-weighted difference number-flux density (black). 
    We take $\bar{m} = \sqrt{m_1 m_2}$ and $\Delta I(m) \equiv I(m_1) - I(m_2)$, where $m_{2}>m_{1}$ are the bin edges (we present these results using two bins per decade of mass as measured in grams).
    Note also that $I(m)$ is a decreasing function of $m$: $I(m_1)>I(m_2)$.
    Each vertical dotted blue line shows the upper edge of the largest mass bin for which one collision with \Alice\ can be expected in the amount of time annotated on the line; fewer than one collision in the indicated time is expected to occur for objects with masses in each mass bin in the blue shaded regions to the right of each of these lines.
    We draw lines at $10^4\,\text{s}$ (approx.~the highest frequency of interest for our proposal), $10^6\,\text{s}$, 1 year, 10 years (a typical mission duration), 100\,yrs, 1000\,yrs, the surface age of 433 Eros (approx.~400\,Myr)~\cite{RICHARDSON2005325},%
    %%%%%%%%%%%%
    \footnote{\label{ftnt:ErosAgeWandering}%
        Eros itself has however likely spent a significant fraction of this period in the Main Belt, where its surface cratering/collisional rates would be much higher than in its current near-Earth environment~\cite{1998AJ....116.2023M}.
        } %
    %%%%%%%%%%%%
    the age of the Solar System (4.6\,Gyr)~\cite{Bouvier:2010dgq}, and the age of the Universe (13.8\,Gyr)~\cite{Planck:2018vyg}.
    Excluding rare events that would not be expected to occur within a mission duration, the dominant mass-weighted differential number flux is in the region $m\sim 10^{-6}$--$10^{-5}\,\text{g}$.
    }
\end{figure}
%%%%%%%%%%%%%%%%%%%%%%%%%%%%%%%%%%%%%%%%%%%%%%%%%%%%%%%%%%%%%%%%%%%%%%%%%%%%%%%%

Asteroids are subject to external perturbations from collisions with dust particles and meteoroids in the interplanetary medium (IPM), as well as their much rarer, but more catastrophic, collisions with other asteroids (see, e.g., \citeR{RICHARDSON2005325}).
The dust and meteoroid density and flux are measured using a variety of techniques appropriate to different mass ranges, including measurements of meteor impacts with Earth's atmosphere, measurements of the zodiacal light, direct measurements of high-velocity impacts on experiments deployed on deep-space missions, and counts of the number and size of micro and macro impact craters on the Lunar surface~\cite{GRUN1985244,KRUGER2014657,Shoemaker:1983qae}.
The total dust density in the vicinity of Earth's orbit is $\rho_{\text{dust}}(m_{\text{dust}}\lesssim 10^{2}\,\text{g}) \sim 10^{-16}\text{g/m}{}^3$~\cite{GRUN1985244,KRUGER2014657}, with around half of that dust-mass being particles in the mass range  $10^{-6}\,\text{g}\lesssim m_{\text{dust}}\lesssim 10^{-4}\,\text{g}$~\cite{GRUN1985244}; tens of tonnes of dust enter Earth's atmosphere within a typical 24-hour period.

We adopt a conservative dust model in order to estimate the impact of these collisions on the asteroid CoM position.
The assumed integral number-flux density, 
%%%%%%
\begin{align}
    I(m) &\equiv \int_{m}^{\infty} v \frac{dn}{dm'} dm',
    \label{eq:dustNumberFluxDensityDefn}
\end{align}
%%%%%%
where $v$ is the dust speed, $m$ is the dust-particle mass, and $n$ is the dust number density, is shown in the upper panel of \figref{fig:IM}.
This model for $I(m)$ is constructed as follows: 
for $10^{-18}\,\text{g}\lesssim m \lesssim 1\,\text{g}$, we adopt the `Lunar flux model' from Fig.~1 and Table 1 of \citeR{GRUN1985244}; this is known~\cite{GRUN1985244} to be an overestimate by a factor of $\mathcal{O}(10^2)$ of the IPM dust density for $m \lesssim 10^{-10}\,\text{g}$ owing to secondary impacts of ejecta generated by primary Lunar cratering increasing the micro-crater count on the Moon, but it is conservative to adopt this curve instead of the `Interplanetary flux model' from the same reference, and it has little impact on our results to do so.
For $m \gtrsim 1\,\text{g}$, we adopt the procedure of \citeR{LISA-Pre-Phase-A} and adopt the broken power law given by Eqs.~(4.38)--(4.40) in \citeR{LISA-Pre-Phase-A}.
Eq.~(4.38) in that reference is based on the same dust results as in  \citeR{GRUN1985244}, whereas Eq.~(4.39) and (4.40) are based on (or extrapolated from) lunar impact-crater data~\cite{Shoemaker:1983qae}. 
The matching between Eqs.~(4.38) and (4.39) occurs [continuously in $I(m)$] at $m \approx 1.48\,\text{g}$, and the matching between Eqs.~(4.39) and (4.40) occurs [again, continuously in $I(m)$] at $m \sim 1.91\times 10^{13}\,\text{g}$.
Per \citeR{LISA-Pre-Phase-A}, this is expected to be a conservative overestimate for $m \lesssim 2\times 10^{7}\,\text{g}$.

Using this number-flux density, we make two estimates the collisional influence on the asteroid CoM.
The first is the more realistic estimate, and the second a conservative overestimate.

%%%%%%%%%%%%%%%%%%%%%%%%%%%%%%%%%%%%%%%%
\bparagraph{Realistic estimate}
%%%%%%%%%%%%%%%%%%%%%%%%%%%%%%%%%%%%%%%%
Consider the mass range $i$ defined by $m_{1i} \leq m \leq m_{2i}$.
The number of objects in mass range $i$ that collide with \Alice\ in a GW period $T_{\textsc{gw}} = 1/f_{\textsc{gw}}$ is given by
%%%%%%
\begin{align}
    N_i \sim 4 \pi R_{\alice}^2 T_{\textsc{gw}} \big[ I(m_{1i}) - I(m_{2i}) \big].
    \label{eq:nAstInBin}
\end{align}
%%%%%%
Note that it is appropriate (and also conservative) to use the full surface area of \Alice\ here, $4\pi R_{\alice}^2$, instead of the cross-sectional area.
The flux numbers in \citeR{GRUN1985244} are, assuming an isotropic flux, stated for a spinning flat plate with an effective solid angle acceptance of $\pi$\,sr: every area element on the asteroid surface of size $dA = R_{\alice}^2d\Omega_{\alice}$ therefore sees an impact-angle-averaged incoming rate of objects larger than mass $m$ of $d^2N/(dtdA) = I(m)$.
Integrating over the asteroid surface area, GW period, and mass bin then gives \eqref{eq:nAstInBin}.

\begin{figure}[!t]
    \centering
    \includegraphics[width=\columnwidth]{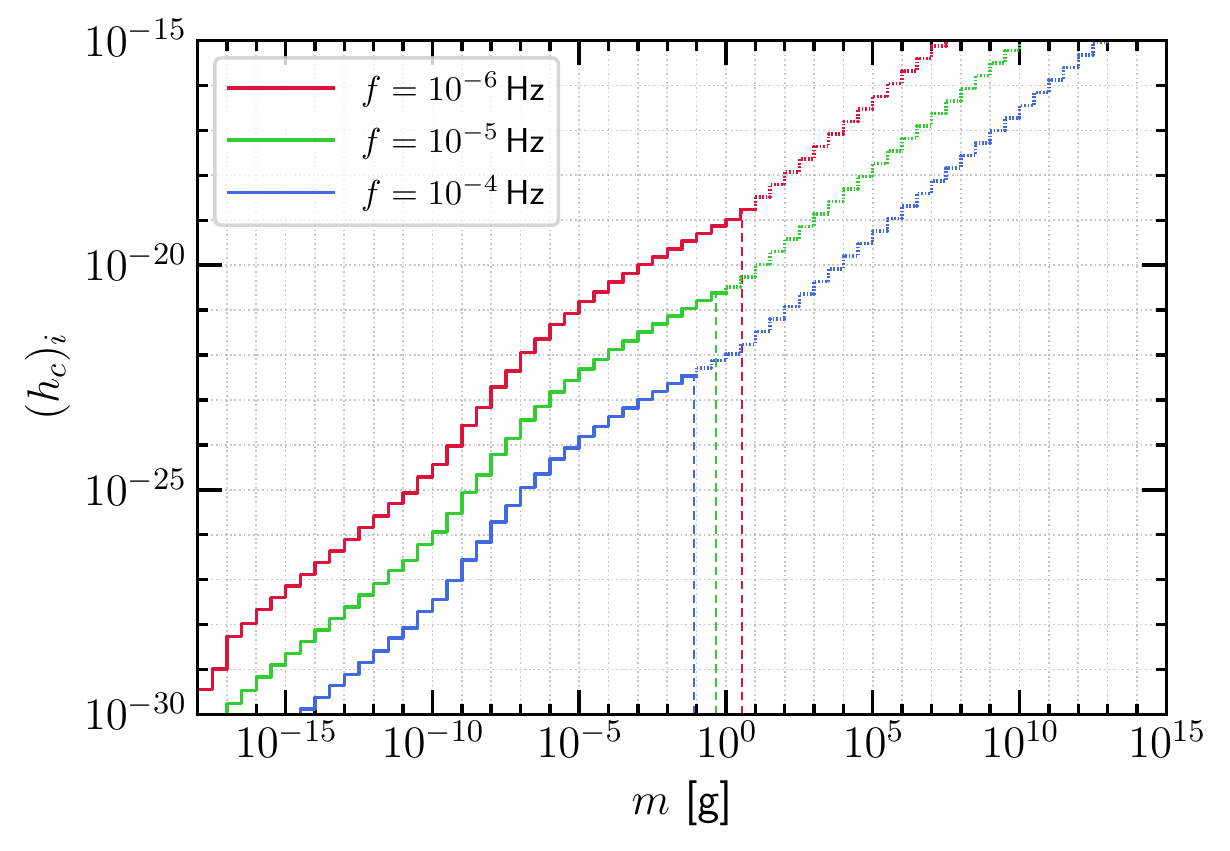}
    \caption{\label{fig:collisionalContributions}%
        Contributions to the strain noise $(h_c)_i$ given in \eqref{eq:oneMassBinStrain} from each mass-bin $i$, presented using two mass bins per decade of mass as measured in grams.
        The red, green, and blue lines are, respectively, results for GW frequencies $f_{\textsc{gw}} = 10^{-6}$, $10^{-5}$, and $10^{-4}\,\text{Hz}$.
        The lines are drawn solid up to and including the bin containing the maximum mass ($m_{\text{max}}$) object as defined at \eqref{eq:maxMass} in the text; the maximum mass for each frequency is denoted by the thin vertical dashed line of like color. 
        Above the bin containing the maximum mass, the per-bin strain results $(h_c)_i$ are shown by dotted lines; this estimate is actually not correct in that mass range, and we do not use it.
        The overall strain estimate at \eqref{eq:collisionalStrain} only includes contributions up to and including the maximum-mass bin (i.e., we use only the solid parts of the various colored lines).
        Because $(h_c)_i$ is an increasing function of $m$, the rarest and largest collisions dominate this noise estimate.
    }
\end{figure}

Each of these $N_i$ collisions will impart an impulsive velocity kick to the asteroid of order $\delta v \sim \bar{m}_i v_{\text{coll}} / m_{\alice}$, where $\bar{m}_i \equiv \sqrt{m_{1i} m_{2i}}$ and $v_{\text{coll}} \sim 30\,\text{km/s}$ is a conservatively high typical collision speed in the inner Solar System (we ignore here orientation and finite-size effects).
However, these impulsive velocity kicks will be directed in random directions, and so will cancel out up to a residual, randomly directed overall velocity kick of order $\Delta v_i \sim \delta v \times \Delta N_i$ where $\Delta N_i = \sqrt{N_i}$ is the Poisson fluctuation in the number of collisions from this bin. We then assume that this velocity kick acts for a time $T_{\textsc{gw}}$ to give a displacement of order $\Delta x_i \sim T_{\textsc{gw}} \cdot \Delta v_i$; we multiply this by $\sqrt{2}$ to account for perturbations on both asteroids, yielding a strain noise estimate from mass `bin' $i$ of order%
%%%%%%%%%%%%
\interfootnotelinepenalty=-100000
\footnote{\label{ftnt:scalingWithAsteroidSize}%
        It is useful to understand the full scaling of this result with~$R_{\alice}$.
        Per \figref{fig:collisionalContributions}, $(h_c)_i$ is dominated by the largest logarithmic mass bin, so we can replace the $\sqrt{\,\cdots}$ factor in \eqref{eq:oneMassBinStrain} with $\sqrt{dI(m_{\text{max}})/d\log m\times \Delta\log m}$.
        Since $I(m_{\text{max}})$ is a power law in the relevant mass range, both $dI(m_{\text{max}})/d\log m$ and $I(m_{\text{max}})$ have the same scaling with $m_{\text{max}}$.
        Moreover, from \eqref{eq:maxMass} we have $I(m_{\text{max}}) \propto R_{\alice}^{-2}$; that is, smaller $R_{\alice}$ will be accompanied by larger $I(m_{\text{max}})$.
        In the relevant range of masses $m$ applicable for impacts on \Alice\ if $R_{\alice}$ is in the vicinity of our 8\,km fidicial value, we have $I(m) \propto m^{-1.34}$ per Eq.~(4.38) of \citeR{LISA-Pre-Phase-A} for $R\lesssim R_{\alice}$.
        Since $1.34 \sim 4/3$, we therefore roughly have $m_{\text{max}}\propto R_{\alice}^{3/2}$.
        This scaling holds until $R_{\alice}$ is small enough that Eq.~(4.38) ceases to be self-consistently valid in this estimate (see \figref{fig:IM}).
        Putting this all together, we find that $(h_c)_i \propto R_{\alice}^{3/2} \sqrt{ R_{\alice}^{-2} } / R_{\alice}^2 \propto R_{\alice}^{-3/2}$. 
        We have also verified this scaling numerically for an 800\,m radius asteroid; find that it has a noise $\approx 30$ times larger than that of \Alice.
    } %
\interfootnotelinepenalty=100
%%%%%%%%%%%%
%%%%%%
\begin{align}
    (h_c)_i \sim \frac{\sqrt{2}\Delta x_i}{L} = \frac{3\bar{m}_i v_{\text{coll}}\sqrt{ \big[ I(m_{1i}) - I(m_{2i}) \big] }  }{\sqrt{2\pi} \rho_{\alice} R_{\alice}^2 Lf^{3/2}_{\textsc{gw}}}.
    \label{eq:oneMassBinStrain}
\end{align}
%%%%%%
We plot curves showing $(h_c)_i$ for selected values of $f_{\textsc{gw}}$ in \figref{fig:collisionalContributions}, again with two mass bins per decade of mass as measured in grams.
It is clear that $(h_c)_i$ is an increasing function of $m_i$, so larger collisions will dominate the noise estimate; see discussion below.

Of course, each mass range $i$ contributes a randomly directed motion of this type, so the correct estimate accounting for all mass ranges of interest would sum the contributions from \eqref{eq:oneMassBinStrain} in quadrature over all bins $i$:
%%%%%%
\begin{align}
    (h_c)_{\text{coll}} 
    &\sim \sqrt{ \sum_{i=i_{\text{min}}}^{i_{\text{max}}} (h_c)_i^2 };
    \label{eq:collisionalStrain}
\end{align}
%%%%%%
where $i_{\text{min}}$ is the index such that $m_{1,i_\text{min}} = 10^{-18}\,\text{g}$ is a fixed lower-mass cutoff to the available flux model (see \figref{fig:IM}), and $i_{\text{max}}$ is the index such that $m_{2,i_{\text{max}}} = m_{\text{max}}(f_{\textsc{gw}})$, where $m_{\text{max}}(f_{\textsc{gw}})$ is a frequency-dependent high-mass cutoff to this estimate which is fixed by requiring that the number of collisions with objects $m\geq m_{\text{max}}(f_{\textsc{gw}})$ in a period $T_{\textsc{gw}}$ is less than 0.5 (a similar, by not identical, criterion is shown by the blue shaded bands in \figref{fig:IM}):
%%%%%%
\begin{align}
    4\pi  R_{\alice}^2 T_{\textsc{gw}} I\big(m_{\text{max}}(f_{\textsc{gw}})\big)\equiv 0.5.
    \label{eq:maxMass}
\end{align}
%%%%%%
This high-mass cutoff is of course relevant to the estimate of the overall noise level since the results of \figref{fig:collisionalContributions} indicate that the sum at \eqref{eq:collisionalStrain} is dominated by the largest few mass bins; we note that by estimating the noise to include all objects for which there is a probability of more than 0.5 collisions to occur during the GW period is thus likely conservative.
The results of this collisional noise estimation procedure, obtained using 100 bins per decade of mass in grams, are shown by the solid turquoise line in \subfigref{fig:noise_contributions}{g}; note that the scaling is faster than the $f_{\textsc{gw}}^{-3/2}$ scaling that might be expected from \eqref{eq:collisionalStrain}, because $I(m_{\text{max}}(f_{\textsc{gw}}))$ also depends on $f_{\textsc{gw}}$ non-trivially, as is clear from \figref{fig:collisionalContributions}.

%%%%%%%%%%%%%%%%%%%%%%%%%%%%%%%%%%%%%%%%%%%%%%%%%%%%%%%%%%%%%%%%%%%%%%%%%%%%%%%%
%%%%%%%%%%%%%%%%%%%%%%%%%%%%%%%%%%%%%%%%%%%%%%%%%%%%%%%%%%%%%%%%%%%%%%%%%%%%%%%%

%%%%%%%%%%%%%%%%%%%%%%%%%%%%%%%%%%%%%%%%
\bparagraph{Conservative estimate}
%%%%%%%%%%%%%%%%%%%%%%%%%%%%%%%%%%%%%%%%
Because the realistic estimate above is dominated by the largest objects we include, it is sensitive to the high-mass cutoff.
Moreover, the dust might have structure, and this could lead to fluctuations larger than $\sqrt{N}$.
We thus also construct a very conservative estimate of the largest possible collisional noise that might be reasonable to assume.

\begin{figure}[!t]
    \centering
    \includegraphics[width=\columnwidth]{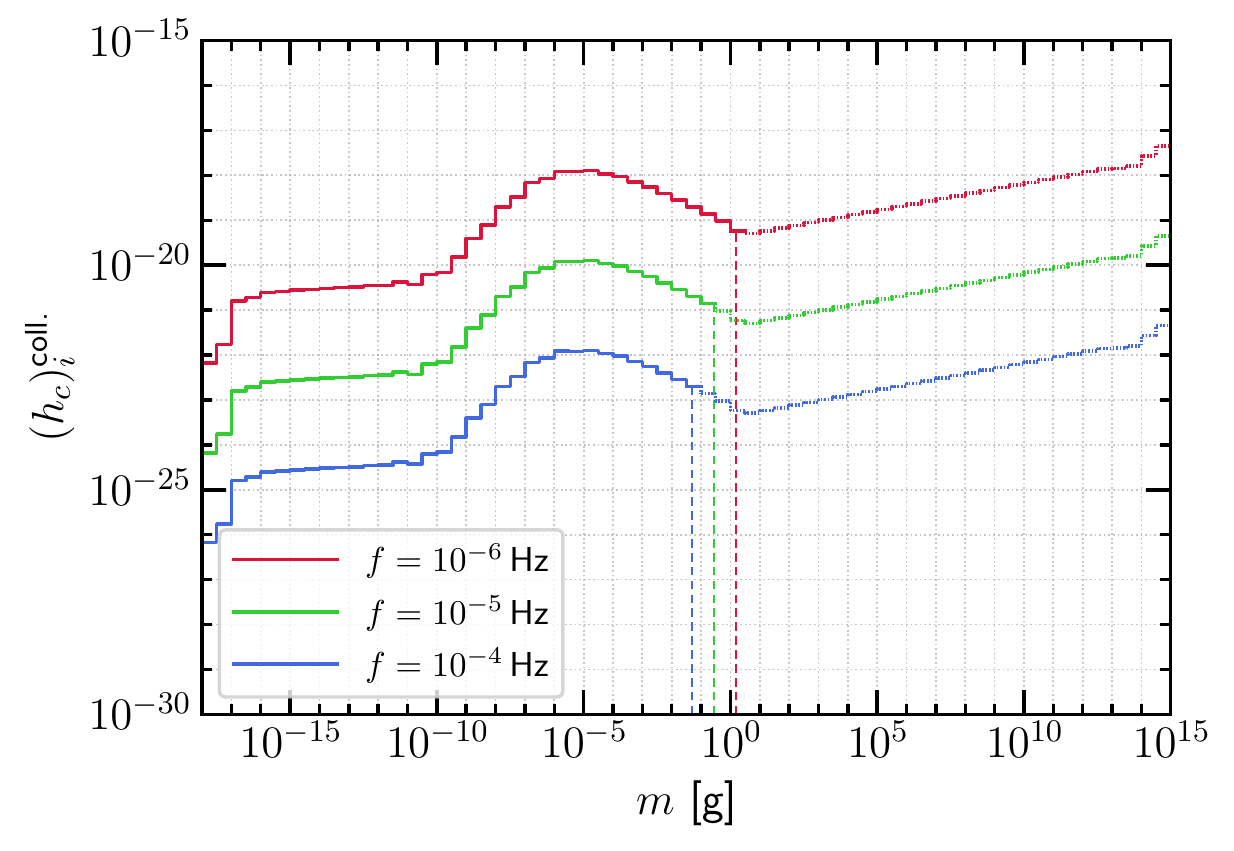}
    \caption{\label{fig:collisionalContributionsCons}%
        As for \figref{fig:collisionalContributions}, but using instead the conservative noise estimate given at \eqref{eq:oneMassBinStrainCons} and discussed in the text.
        For this estimate, collisions with objects with $m\sim 10^{-5}\,\text{g}$ always dominate the estimate for the GW frequencies in our band.
    }
\end{figure}

Suppose that the dust in the interplanetary medium exhibited $\mathcal{O}(1)$ density fluctuations on exactly the right length scale, $\ell \sim v_{\text{ast}} T_{\textsc{gw}}$, to supply a fluctuation in the force applied to the asteroid on a period of exactly $T_{\textsc{gw}}$. 
Suppose also that, instead of the stochastic fluctuation we assumed in our realistic estimate, this dust density fluctuation results in a coherently directed force on the asteroid: i.e., all dust particles impact on the asteroid from the same direction (as might be expected if, e.g., the asteroid were moving into a dust overdensity).
In this case, the velocity kick suffered by the asteroid due to impacts with objects in mass-bin $i$ is given by $\Delta v_i \sim (N_i/2) \delta v_i$, with $N_i$ still given by \eqref{eq:nAstInBin}, where the $1/2$ roughly accounts for the flux impinging on the asteroid from only one direction [this is likely incorrect by an $\mathcal{O}(1)$ factor, as we are converting isotropic flux numbers from \citeR{GRUN1985244} to a directional flux here; this error is however within the uncertainty on this estimate].
Given the assumption that the dust density varies by $\mathcal{O}(1)$ on the GW period, the magnitude of this velocity kick thus also varies by $\mathcal{O}(1)$ on the GW period, so it becomes the relevant kick to use to estimate the noise for GW detection at that period.
Then, following the same logic as for the realistic estimate, we would estimate the strain noise contribution to be
%%%%%%
\begin{align}
    (h_c)_i^{\text{cons.}} \sim \frac{3 \bar{m}_i v_{\text{coll}}}{\sqrt{2} L \rho_{\alice}R_{\alice}Lf_{\textsc{gw}}^2} \big[ I(m_{1,i}) - I(m_{2,i}) \big].
    \label{eq:oneMassBinStrainCons}
\end{align}
%%%%%%
For the GW frequencies of interest, this estimate is now dominated by objects with $m_{\text{dust}}\sim10^{-5}\,\text{g}$; see \figref{fig:collisionalContributionsCons}.

%%%%%%%%%%%%%%%%%%%%%%%%%%%%%%%%%%%%%%%%%%%%%%%%%%%%%%%%%%%%%%%%%%%%%%%%%%%%%%%%
%%%%%%%%%%%%%%%%%%%%%%%%%%%%%%%%%%%%%%%%%%%%%%%%%%%%%%%%%%%%%%%%%%%%%%%%%%%%%%%%

Moreover, because we assume the collisions are all coming from the same direction, the net strain contribution is the linear sum over all bins: 
%%%%%%
\begin{align}
    (h_c)_{\text{coll}}^{\text{cons.}}
    &\sim \sum_{i=i_{\text{min}}}^{i_{\text{max}}} (h_c)_i^{\text{cons.}}.
    \label{eq:collisionalStrainCons}
\end{align}
%%%%%%
We again estimate $m_{\text{max}}=m_{2,i_{\text{max}}}$ as at \eqref{eq:maxMass}, but with a numerical factor of 2 replacing the numerical factor of 4 on the LHS (again to account for the smaller asteroid surface area exposed to this assumed directional flux); the result is however somewhat insensitive to this cutoff now, owing to being dominated by collisions with objects with $m_{\text{dust}} \sim10^{-5}\,\text{g}$ (see \figref{fig:collisionalContributionsCons}).
The result of this conservative collisional estimate are shown by the dashed turquoise line in \subfigref{fig:noise_contributions}{g}.
Because we constructed this estimate only to provide an extremely conservative upper bound on this noise source, we do not include it in our results further.
It is clear moreover that this upper bound is only slightly worse, by an $\mathcal{O}(1)$ factor, than the enveloped noise curves (see \sectref{sect:sensitivity} and \figref{fig:results}) that were constructed without including it, in the frequency range around $f_{\textsc{gw}}\sim 3\,\mu\text{Hz}$.

We conclude that collisions are most probably a subdominant noise source, and are at worst a noise at approximately the same level as other sources we estimate in this paper in the relevant frequency range.

%%%%%%%%%%%%%%%%%%%%%%%%%%%%%%%%%%%%%%%%%%%%%%%%%%%%%%
\subsubsection{Electromagnetic forces}
\label{sect:EMforces}
%%%%%%%%%%%%%%%%%%%%%%%%%%%%%%%%%%%%%%%%%%%%%%%%%%%%%%
Asteroids are also potentially subject to electromagnetic forces that perturb their CoM.

Interplanetary space is permeated by the interplanetary medium, one component of which is the hot plasma of protons (and other ions) and electrons in the solar wind (see \sectref{sect:solarWind}).
This plasma is quasi-neutral~\cite{Verscharen:2019vsd}: neutral on macroscopic scales, with violations of neutrality at Debye-length scales, $\lambda_{\text{Debye}} = \sqrt{T/(4\pi\alpha \bar{n}_p)}$ assuming $q=1$; see, e.g., \citeR[s]{Verscharen:2019vsd,LandauLifshitzVol5,Brydges_1999}.
Taking a typical temperature $10^5\,\text{K}\lesssim T\lesssim 10^6\,\text{K}$~\cite{DSCOVRwebsite}, and an average proton number density $n_p \sim 5\,\text{cm}^{-3}$~\cite{celias-mtofDATASET}, this length scale is $10\,\text{m}\lesssim \lambda_{\text{Debye}} \lesssim 30\,\text{m}$.%
%%%%%%%%%%%%
\footnote{\label{ftnt:debyeLength}%
    Note that the proton temperature is about a factor of 10 lower than the translational kinetic energy associated with the bulk wind outflow, but even if the Debye length is estimated using that energy in place of the temperature, the length scale is still $\mathcal{O}(100\,\text{m})$.
    } %
%%%%%%%%%%%%
As such, large-scale electric fields are screened in the Solar System, also on Debye-length scales~\cite{Verscharen:2019vsd}; a large-scale heliospheric magnetic field (HMF)%
%%%%%%%%%%%%
\footnote{\label{ftnt:altName}%
    Historically, also called the interplanetary magnetic field (IMF)~\cite{Owens:2013edh}.
    } %
%%%%%%%%%%%%
$B_{\text{HMF}}$ is however maintained in the plasma~\cite{Zurbuchen:2007onw,Balogh:2013onq,Owens:2013edh}.

In this subsection, we consider possible CoM motion effects arising from electromagnetic fields:
\hyperlink{para:charging}{(a)} if the asteroid were to become charged, it would experience a magnetic Lorentz force as it moves through the HMF at speeds $v_{\alice}\sim 30\,\text{km/s}$; \hyperlink{para:gradBforce}{(b)} if the asteroid itself is permanently magnetized, it will be subject to a force due to magnetic field gradients; and \hyperlink{para:EfieldFluctuations}{(c)} if the asteroid is charged by the solar wind and there are fluctuations in the electric field in the IPM on Debye-length scales, this could also give a force on the asteroid.

\begin{figure}[!t]
    \centering
    \includegraphics[width=\columnwidth]{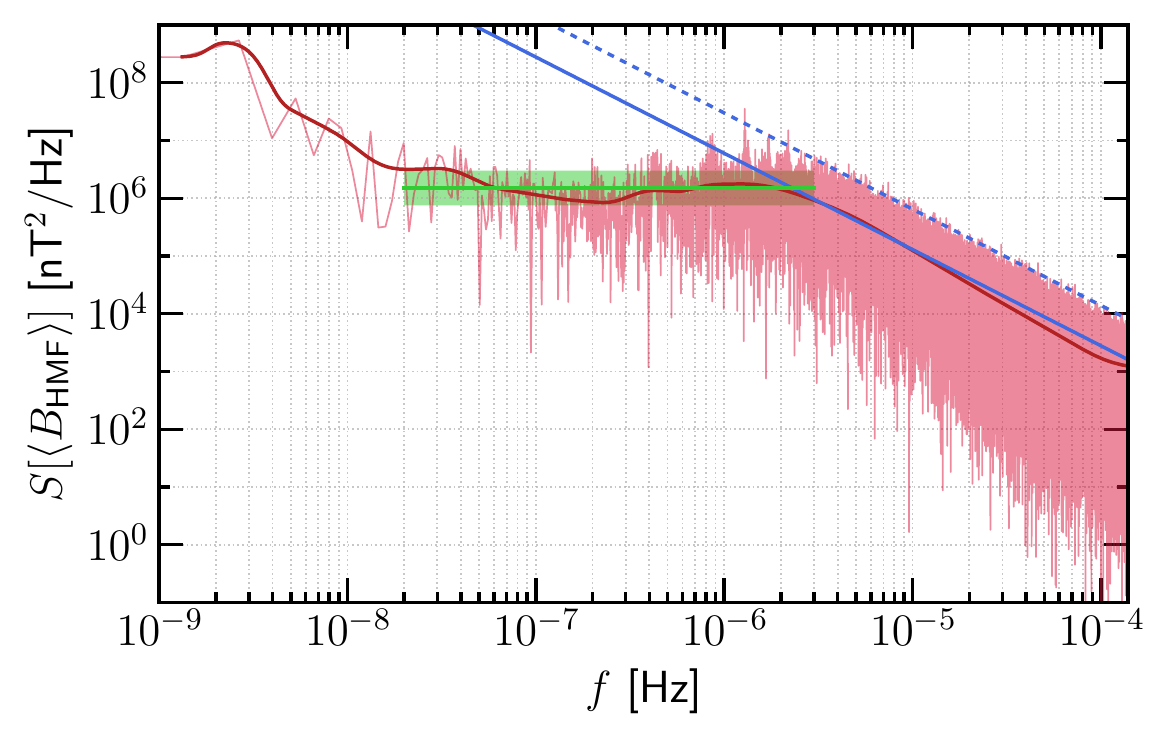}
    \caption{\label{fig:ACE_B_PSD}%
        One-sided power spectral density (PSD) of the one-hour-averaged heliospheric magnetic field (HMF) as measured by the ACE satellite~\cite{Stone:1998dga,ACEwebsite} at its orbital location around the Earth--Sun L1 Lagrange point over the time period 1997--2021 (light-red line merging into the light-red band).
        The sliding average of the PSD taken over a Gaussian kernel in log-frequency space with a width of $0.1 \log_{10}[\text{Hz}]$ is shown by the thick red line.
        The solid blue line shows $S[\langle B_{\text{HMF}}\rangle] = 60\,\text{nT}^2/\text{Hz} \times (f/\text{mHz})^{-5/3}$, while the dotted blue line is $5\times$~that same analytical expression; see discussion in text.
        The horizontal green line shows the value $S[\langle B_{\text{HMF}}\rangle] = 1.5\times 10^{6}\,\text{nT}^2/\text{Hz}$, with the green band covering a range from half to twice that value (i.e., a factor-of-2 variation).
    }
\end{figure}

%%%%%%%%%%%%%%%%%%%%%%%%%%%%
\paragraph{Electrical charging and Lorentz force}
\label{sect:charging}\hypertarget{para:charging}{}
%%%%%%%%%%%%%%%%%%%%%%%%%%%%
Suppose that \Alice\ and \Bob\ were each subject to white-noise charge fluctuations with an rms charge $Q e$ over a frequency band of order $f_{\textsc{gw}}$; here, $e\equiv\sqrt{4\pi\alpha}$ is the fundamental unit of charge.
Then the single-asteroid acceleration ASD from the magnetic Lorentz force would be of order
%%%%%%
\begin{align}
    \sqrt{ f_{\textsc{gw}} S_a } \sim \frac{ Q e v_{\alice} B_{\textsc{HMF}}}{ M_{\alice} },
    \label{eq:accnCharge}
\end{align}
%%%%%%
leading to an in-band strain noise estimate (assuming equal-magnitude noises on each asteroid) of
%%%%%%
\begin{align}
    h_c \sim \frac{2\sqrt{2} Q e v_{\alice} B_{\textsc{HMF}}}{ M_{\alice} (2\pi f_{\textsc{gw}})^2 L }.
    \label{eq:hcCharge1}
\end{align}
%%%%%%
Here, we have assumed that the HMF itself does not display fluctuations much larger than its average value in the band of interest; although the HMF can display $\order{1}$ fluctuations in amplitude and large directional changes~\cite{DSCOVRwebsite}, this is generally a reasonably well-motivated approximation (see also \figref{fig:ACE_B_PSD}).

It remains to estimate the size of the charge fluctuation~$Q$. 
The asteroid can only become charged on large scales by the ionized solar wind that impinges on its surface.
Let us take a na\"ive model of the solar wind as comprised as packets of volume $\sim \lambda_{\text{Debye}}^3$ that alternate in the sign of the charge; this is of course not realistic, but it is a conservative model as far as solar-wind-induced charge fluctuations are concerned.
These packets of charges are constantly blowing past the asteroid, randomly charging up various parts of the surface.
We can estimate the fluctuation of the asteroid charge by asking for the Poisson fluctuations in the total charge of the asteroid arising by counting the number of packets of charge $eQ_{\text{Debye}}$, where $Q_{\text{Debye}} = (4\pi/3)\lambda_{\text{Debye}}^3 \bar{n}_p$, whose cross-sectional area would blanket the asteroid surface, $N_{\text{patches}} \sim 4\pi R_{\alice}^2 /(\pi \lambda_{\text{Debye}}^2)$, and estimating the rms charge fluctuation as
%%%%%%
\begin{align}
    Q &\sim Q_{\text{Debye}} \times \sqrt{ N_{\text{patches}} } \\
    &\sim \frac{8\pi}{3} R_{\alice} \lambda_{\text{Debye}}^2 \bar{n}_p\\
    &\sim \frac{2}{3} R_{\alice} \frac{T}{\alpha}\\
    &\sim 3\times 10^{14} \times \lb( \frac{T}{10^6\,\text{K}} \rb) \times \lb( \frac{R_{\alice}}{8\,\text{km}} \rb).\label{eq:Qrms}
\end{align}
%%%%%%
While we have maintained $\order{1}$ numerical factors here, this estimate is only accurate at the order-of-magnitude level.
Note also that this estimate, up to $\order{1}$ factors, is the same as that which would be obtained by equating the thermal kinetic energy of a solar wind particle with its electrostatic potential energy computed assuming a $1/r$ Coulomb potential for the charged asteroid (i.e., ignoring the plasma screening).
If one replaces the thermal kinetic energy $T$ with the translational bulk outflow kinetic energy $\bar{K}_p \sim m_p\bar{v}_p^2/2$ of the wind (up to an $\order{1}$ factor arising from the average of the square vs.~the square of the average), this estimate increases by only a factor of $\sim 10$.

An alternative estimate for the total asteroid charge would be to take $Q\sim Q_{\text{local}}\sqrt{N_{\text{patches}}}$ with $N_{\text{patches}}$ estimated as before, but with $Q_{\text{local}}$ being the maximum surface charge within each such `packet' area that can be built up given the incoming solar wind speed.
This can be obtained by an energetics argument: because the electric field generated by this patch of charge is screened in the radial direction within the length scale $\lambda_{\text{Debye}}$, it takes a potential energy of $\sim \alpha Q_{\text{local}}/\lambda_{\text{Debye}}$ to introduce an additional proton to the asteroid surface if the patch is charged to $+|Q_{\text{local}}|$; but each incoming proton has roughly $\bar{K}_p \sim m_p\bar{v}_p^2/2$ of kinetic energy, so we can estimate $Q_{\text{local}} \sim \bar{K}_p\lambda_{\text{Debye}}/\alpha = 3(\bar{K}_p/T)Q_{\text{Debye}}$; the rms charge fluctuation $Q$ obtained from this estimate is just \eqref{eq:Qrms} under the replacement $T\rightarrow 3\bar{K}_p \sim 30 T$, which is similar to the \emph{ad hoc} estimate based on the replacement of $T$ by the bulk outflow kinetic energy $\bar{K}_p$ that was outlined at the end of the previous paragraph.

Taking $Q$ from \eqref{eq:Qrms}, the strain noise estimate is thus
%%%%%%
\begin{align}
    h_c &\sim \frac{\sqrt{2} ev_{\alice} B_{\textsc{HMF}} T }{ \pi R_{\alice}^2\rho_{\alice} (2\pi f_{\textsc{gw}})^2 L \alpha}  \\
    &\sim 1.4 \times 10^{-24} \times \lb( \frac{\mu\text{Hz}}{f_{\textsc{gw}}} \rb)^2 \times \lb( \frac{ B_{\text{HMF}}}{10\,\text{nT}} \rb) \times \lb( \frac{T}{10^6\,\text{K}} \rb),
    \label{eq:hcCharge2}
\end{align}
%%%%%%
where we conservatively took $B_{\text{HMF}} \sim 10\,\text{nT}$~\cite{Behannon:1978fhs,DSCOVRwebsite,Owens:2013edh}.
By comparison to the results in \figref{fig:noise_contributions}, one can see that this is a negligible noise source by some $\sim 7$ orders of magnitude; even were the estimate repeated with $T \rightarrow c \bar{K}_p$ with $c$ an $\mathcal{O}(1$--$3)$ numerical factor, this would still be a highly subdominant noise source.

%%%%%%%%%%%%%%%%%%%%%%%%%%%%
\paragraph{Magnetic field gradient}
\label{sect:gradBforce}\hypertarget{para:gradBforce}{}
%%%%%%%%%%%%%%%%%%%%%%%%%%%%
If \Alice\ has a permanent dipolar magnetic moment $\bm{\mu}_{\alice}$, it is subject to a force~\cite{Jackson1999} $\bm{F} = \bm{\nabla}(\bm{\mu}_{\alice}\cdot\bm{B}_{\text{HMF}}) \approx (\bm{\mu}_{\alice}\cdot\bm{\nabla})\bm{B}_{\text{HMF}}$ in the gradient of the HMF, or an acceleration of $\bm{a} \approx (\bm{\hat{\mu}}_{\alice}\cdot\bm{\nabla})\bm{B}_{\text{HMF}}$, where $\bm{\hat{\mu}}_{\alice} \equiv \bm{\mu}_{\alice}/M_{\alice}$ is the specific (per mass) magnetic moment.

The near-surface magnetic-field environment of 433~Eros was characterized by the NEAR--Shoemaker mission while orbiting and during final descent to the asteroid surface~\cite{ACUNA2002220}.%
%%%%%%%%%%%%
\footnote{\label{ftnt:SIinThisParagraph}%
    We quote values in SI units in this paragraph.
    The conversion from SI to natural Heaviside--Lorentz units is $1\,\text{A\,m}^2 \approx 3.2 \times 10^{16}\text{eV}^{-1}$.
    Recall also that $1\,\text{T} \approx 195.4\,\text{eV}^2$.
    } %
%%%%%%%%%%%%
These data place an upper limit of $B(35\,\text{km}) \lesssim 10^{-10}\,\text{T}$ on the magnetic field measured by the satellite while in a 35\,km-radius orbit around the CoM of 433~Eros, which would place a limit on the magnetic moment of $\mu_{\text{Eros}} \lesssim 4.3\times 10^{10}\,\text{A\,m}^2$~\cite{ACUNA2002220}; further data taken during the final descent to landing on the asteroid surface improve this limit by a factor of $\sim 3$ to be $\mu_{\text{Eros}} < 1.3\times 10^{10}\,\text{A\,m}^2$~\cite{ACUNA2002220}, which corresponds to a specific magnetic moment limit $\hat{\mu}_{\text{Eros}} \equiv \mu_{\text{Eros}}/M_{\text{Eros}} \lesssim 1.9 \times 10^{-6} \,\text{A\,m}^2\,\text{kg}^{-1}$~\cite{ACUNA2002220}.
On the other hand, some other large asteroids such as 951~Gaspra (S class~\cite{JPL-SBD}) and 9969~Braille (Q class~\cite{JPL-SBD}) are known to have significantly higher specific magnetic moments, as high as $\hat{\mu} \sim 3\times 10^{-2}$~\cite{Kivelson:1993rsw,Richter:2001ehe,Hercik:2020svs}; other large asteroids such as 162173~Ryugu (Cg class~\cite{JPL-SBD}) and 21~Lutetia (M class~\cite{JPL-SBD}) are however known to have global moments lower than that of Eros~\cite{Hercik:2020svs}.
Although 433~Eros is an ideal example target for one end of the baseline for this mission, and one can select asteroid targets based on their magnetization properties, we will nevertheless give noise estimates assuming that the specific magnetic moment of \Alice\ lies between a conservatively high value of $\hat{\mu}_{\alice}^{\text{high}} \sim 3\times 10^{-2}\,\text{A\,m}^2\,\text{kg}^{-1}$, and an 433~Eros-like value of $\hat{\mu}_{\alice}^{\text{low}} \sim 2\times 10^{-6}\,\text{A\,m}^2\,\text{kg}^{-1}$.

We ignore the vectorial orientation of the asteroid moment and the HMF, and take the parametric estimate $a_{\alice} \sim \hat{\mu}_{\alice} \Delta B_{\text{HMF}} / \lambda_{\text{HMF}}$ where we have assumed that the HMF has fluctuations of order $\Delta B_{\text{HMF}}$ on length-scales $\lambda_{\text{HMF}}$.
As an initial, order-of-magnitude estimate, let us assume that there are broadband, approximately white-noise fluctuations in the HMF with an rms size of $B_{\text{HMF}}\sim 10\,\text{nT}$ over a bandwidth of $f_{\textsc{gw}}$.
We will take $\lambda_{\text{HMF}} \sim \bar{v}_p / f_{\textsc{gw}}$ to be a typical gradient scale associated the HMF field lines, which are entrained in the solar wind; we take $\bar{v}_p \sim 400\,\text{km/s}$, giving $\lambda_{\textsc{HMF}} \sim 2.7\,\text{AU} \times (\mu{\text{Hz}}/f_{\textsc{gw}})$, which is also roughly the same AU length scale on which the static HMF itself falls off by an $\mathcal{O}(1)$ factor in the vicinity of Earth's orbit~\cite{1958ApJ...128..664P,Zurbuchen:2007onw,Balogh:2013onq,Owens:2013edh}.
Then, $\sqrt{f_{\textsc{gw}} S[a_{\alice}]} \sim \hat{\mu}_{\alice} B_{\text{HMF}} f_{\textsc{gw}} / \bar{v}_p$, and so
%%%%%%
\begin{align}
    h_c &\sim  \frac{2\sqrt{2} f_{\textsc{gw}} \hat{\mu}_{\alice} B_{\text{HMF}} }{ (2\pi f_{\textsc{gw}})^2 L \bar{v}_p } \\
    &\sim 4\times 10^{-22} \times \lb( \frac{ \mu\text{Hz} }{ f_{\textsc{gw}} } \rb) \times \lb( \frac{ \hat{\mu}_{\alice} }{ \hat{\mu}_{\alice}^{\text{high}} } \rb)\\
    &\sim 2\times 10^{-26} \times \lb( \frac{ \mu\text{Hz} }{ f_{\textsc{gw}} } \rb) \times \lb( \frac{ \hat{\mu}_{\alice} }{ \hat{\mu}_{\alice}^{\text{low}} } \rb).
    \label{eq:magForce1}
\end{align}
%%%%%%
Comparison to \figref{fig:noise_contributions} indicates that the estimate using the high (respectively, low) specific magnetic moment is safe: it is sub-dominant to existing noise sources by some 4--5 (respectively, 9) orders of magnitude.
We note that this large margin of safety supplies an \emph{a posteriori} justification for some of the vaguer approximations used in this estimate: they would need to be violated by many orders of magnitude to invalidate the estimate.

Nevertheless, however safe the above estimate is, it is na\"ive, and we can improve it: the HMF, as measured on a heliocentric orbital trajectory, more typically exhibits a PSD following a power law $S[B_{\text{HMF}}](f;r) \sim S[B_{\text{HMF}}](f_*;r) (f/f_*)^{-5/3}$ for $f\gtrsim 10^{-5}\,\text{Hz}$, with $S[B_{\text{HMF}}](f_*\sim\text{mHz};r=1.75\,\text{AU}) \sim 10\,\text{nT}^2/\text{Hz}$~\cite{ACUNA2002220}.
Data spanning 1997--2021 (i.e., over slightly more than two full solar cycles) from the ACE mission~\cite{Stone:1998dga,ACEwebsite}%
%%%%%%%%%%%%
\footnote{\label{ftnt:DSCOVRalso}%
    Similar data are also in principle available for the DSCOVR mission~\cite{DSCOVRwebsite,DSCOVRdata}.
    } %
%%%%%%%%%%%%
located on a Lissajous orbit near the Earth--Sun L1 Lagrange point ($r\sim r_{\oplus} \sim 1\,\text{AU}$ within 1\%) indicate a very similar $-5/3$ power-law spectral index for $f\gtrsim 3\,\mu\text{Hz}$; see \figref{fig:ACE_B_PSD}.
However, the smoothed normalization is $S[B_{\text{HMF}}](f_*=\text{mHz};r\approx r_{\oplus}) \sim 60\,\text{nT}^2\text{/Hz}$; frequency-to-frequency fluctuations are however large, although the largest upward fluctuations are still within a factor of $\sim 5$ of this value.
The spectral index of this PSD however flattens for lower frequencies and it is almost flat for $2\times 10^{-8}\,\text{Hz} \lesssim f \lesssim 3\times 10^{-6}\,\text{Hz}$, taking the value $S[B_{\text{HMF}}](f;r\approx 1\,\text{AU}) \sim 1.5\times 10^{6}\,\text{nT}^2\text{/Hz}$ in this frequency range [the smoothed PSD varies by a factor of $\order{2}$ around this value in this range].
The higher-frequency spectral normalizations at the different heliocentric radii are broadly consistent, within $\order{3}$ factors, with the expected drop-off in the HMF with radius: for the \emph{static} HMF, we would have $B_{\phi}(r_{\oplus}) \sim B_r(r_{\oplus})$, while $B_r(r) \propto r^{-2}$ while $B_{\phi}(r) \propto r^{-1}$~\cite{1958ApJ...128..664P,Zurbuchen:2007onw,Balogh:2013onq,Owens:2013edh}.

%%%%%%%%%%%%%%%%%%%%%%%%%%%%%%%%%%%%%%%%%%%%%%%%%%%%%%%%%%%%%%%%%%%%%%%%%%%%%%%%
%%%%%%%%%%%%%%%%%%%%%%%%%%%%%%%%%%%%%%%%%%%%%%%%%%%%%%%%%%%%%%%%%%%%%%%%%%%%%%%%

Overall, we conservatively take $S[B_{\text{HMF}}](f) \sim 60 \,\text{nT}^2/\text{Hz} \times ( f / \text{mHz} )^{-5/3}$ over the entire frequency range of interest, as this is a good average value for Earth-radius orbits at higher frequencies, and an overestimate at lower frequencies ($f\lesssim 3\times 10^{-6}\,\text{Hz}$).
We use the Earth-radius value rather than a value at $r\sim r_{\alice} \sim 1.5\,\text{AU}$ to be conservative.
We also update our approach to estimating the gradient of $B_{\text{HMF}}$: because the HMF field lines are entrained with the solar wind, we replace $\nabla B_{\text{HMF}} \sim \bar{v}_p^{-1} \dot{B}_{\text{HMF}}$.
And we can use%
%%%%%%%%%%%%
\footnote{\label{ftnt:S2pif}%
    In the Fourier domain, temporal differentiation brings down a factor of $(2\pi f)$ on the Fourier transform $\tilde{B}$, and $S[B] \propto |\tilde{B}|^2$.
    } %
%%%%%%%%%%%%
$S[\dot{B}] \sim (2\pi f)^2 S[B]$; we will continue to take $\bar{v}_p \sim 400\,\text{km/s}$.
The noise estimate is then
%%%%%%
\begin{align}
    h_c &\sim  \frac{2\sqrt{2} \hat{\mu}_{\alice} \sqrt{f_{\textsc{gw}} S[B_{\text{HMF}}](r_{\oplus},f_{\textsc{gw}}) } }{ 2\pi f_{\textsc{gw}} L \bar{v}_p } \\
    &\sim 6\times 10^{-22} \times \lb( \frac{ \mu\text{Hz} }{ f_{\textsc{gw}} } \rb)^{4/3} \times \lb( \frac{ \hat{\mu}_{\alice} }{ \hat{\mu}_{\alice}^{\text{high}} } \rb) \\
    &\sim 4\times 10^{-26} \times \lb( \frac{ \mu\text{Hz} }{ f_{\textsc{gw}} } \rb)^{4/3} \times \lb( \frac{ \hat{\mu}_{\alice} }{ \hat{\mu}_{\alice}^{\text{low}} } \rb).
    \label{eq:magForce2}
\end{align}
%%%%%%
These estimates are similar in magnitude to the previous ones.
This is thus not a relevant noise source by some 4--9 orders of magnitude, depending on the assumed specific magnetic moment.

Alternatively, the relevant speed in the HMF gradient estimation may be the Alfv\'en speed, which is $v_{A} = B_{\text{HMF}}/\sqrt{m_p\bar{n}_p} \sim 50\,\text{km/s} \times (B_{\text{HMF}}/5\,\text{nT})\times(\bar{n}_p/5\,\text{cm}^{-3})^{-1/2}$, or roughly  $v_{A} \sim \bar{v}_p /  10$.
We note that even if, in our length-scale estimate, we replaced $\bar{v}_p \rightarrow v_{\alice} \sim \bar{v}_p / 10$, or took the Alfv\'en speed $v_A \sim \bar{v}_p/10$ instead, and also took the absolute largest normalizations of $S[B_{\text{HMF}}]$ (i.e., a factor of 5 larger than we used here), the same qualitative conclusion that this is not a relevant noise source would still result, and still by at least 2 orders of magnitude.

%%%%%%%%%%%%%%%%%%%%%%%%%%%%
\paragraph{Electric field fluctuations}
\label{sect:EfieldFluctuations}\hypertarget{para:EfieldFluctuations}{}
%%%%%%%%%%%%%%%%%%%%%%%%%%%%
Although large-scale electric fields are screened, such fields can exist on length scales of order the Debye length.
Let us return to our mock model of the solar wind as comprised of alternating quasi-spherical packets of charge with radius $\lambda_{\text{Debye}}$ and charge $eQ_{\text{Debye}}$ with $Q_{\text{Debye}} = (4\pi/3)\lambda_{\text{Debye}}^3\bar{n}_p$ impinging on the asteroid.
Suppose one such packet has just transferred all its charge to the surface, and consider the electromagnetic force that is then exerted on that patch of the surface of radius $\lambda_{\text{Debye}}$ by the next incoming charge packet (of opposite charge); parametrically, this will be $F_{\text{patch}} \sim \alpha Q_{\text{Debye}}^2 / \lambda_{\text{Debye}}^2$.
At any given instant, there are $N_{\text{patches}}$ such randomly oriented forces acting on the asteroid, leading to a net instantaneous force of order $F_{\alice}\sim F_{\text{patch}}\sqrt{N_{\text{patches}}}$.
This force will vary in both magnitude and direction by an $\mathcal{O}(1)$ factor on a timescale $\tau \sim \lambda_{\text{Debye}}/v_p$, since the incoming solar wind will randomly alter the asteroid surface charge on the timescale it takes the solar wind to cross the Debye length.
The displacement of the asteroid in this time will be of order $\delta x \sim (F_{\alice}/M_{\alice}) \tau^2$, and this displacement will random walk over timescales $T_{\textsc{gw}}= 1/f_{\textsc{gw}}$ to give a net displacement of order $\Delta x \sim \delta x \sqrt{T_{\textsc{gw}}/\tau} $, leading to a strain noise of order
%%%%%%
\begin{align}
    h_c &\sim \frac{ \alpha Q_{\text{Debye}}^2 }{ \lambda_{\text{Debye}}^2 M_{\alice} L } \sqrt{N_{\text{patches}}} \lb( \frac{\lambda_{\text{Debye}}}{\bar{v}_p} \rb)^2 \sqrt{ \frac{\bar{v}_p }{ f_{\textsc{gw}} \lambda_{\text{Debye}} }}\\
     &\sim \frac{ T^{9/4}  }{6\sqrt{2}  \lb( \pi\alpha \rb)^{5/4} \bar{n}_p^{1/4} R_{\alice}^2 \rho_{\alice} L \bar{v}_p^{3/2} f_{\textsc{gw}}^{1/2} }\\
     &\sim 4\times 10^{-35} \times \lb( \frac{f_{\textsc{gw}}}{\mu\text{Hz}} \rb)^{-1/2} \times \lb( \frac{T}{10^6\,\text{K}} \rb)^{9/4}, 
\end{align}
%%%%%%
which is an extremely small noise source.
Had we assumed that the force $F_{\alice}$ acted coherently for a time $T_{\textsc{gw}}$ instead, the estimate would increase by a factor of $\sim 10^{15}$, but that would still not make it a relevant noise source, and would itself be a dramatic overestimate.

%%%%%%%%%%%%%%%%%%%%%%%%%%%%%%%%%%%%%%%%%%%%%%%%%%%%%%
%%%%%%%%%%%%%%%%%%%%%%%%%%%%%%%%%%%%%%%%%%%%%%%%%%%%%%
\subsection{Fluctuating torques: rotational motion}
\label{sect:externaltorques}
%%%%%%%%%%%%%%%%%%%%%%%%%%%%%%%%%%%%%%%%%%%%%%%%%%%%%%
%%%%%%%%%%%%%%%%%%%%%%%%%%%%%%%%%%%%%%%%%%%%%%%%%%%%%%
External perturbations also apply fluctuating torques to an asteroid owing to asteroids' non-spherical surface geometries, non-uniform surface albedos, and non-uniform mass distributions. 
This can alter the rotational state of the asteroid which gives rise to additional noise sources on the baseline measurement because each asteroid CoM must be located indirectly by referencing it to the location of one or more points on the surface of the asteroid.

In this subsection, we estimate the impact of torques arising from fluctuating external sources: \hyperref[sect:TSIandWindTorque]{(1)} solar radiation pressure and the solar wind, \hyperref[sect:EMforcesTorque]{(2)} electromagnetic forces, \hyperref[sect:collisionsTorque]{(3)} collisions with dust, and \hyperref[sect:flybysTorque]{(4)} close fly-bys with larger objects.

While we show that the fluctuating torque-noise sources are no more problematic for our purposes than the direct CoM motions induced by external forces, we also discuss various mitigation possibilities where appropriate.

%%%%%%%%%%%%%%%%%%%%%%%%%%%%%%%%%%%%%%%%%%%%%%%%%%%%%%
\subsubsection{Solar radiation and wind torques}
\label{sect:TSIandWindTorque}
%%%%%%%%%%%%%%%%%%%%%%%%%%%%%%%%%%%%%%%%%%%%%%%%%%%%%%
The fluctuations in the solar radiation and solar wind can be characterized as a fluctuating pressure acting on the surface of \Alice. 
Let us suppose that the pressure fluctuation has an in-band amplitude of $\delta P \sim \sqrt{f S[P]}$, where $S[P]$ is the relevant pressure PSD; we will return below to what the relevant frequency band is to consider.
To be conservative, we will consider an $\mathcal{O}(1)$ asymmetry in how this pressure is applied to two halves of \Alice\ that lie on either end of some chosen axis $\hat{n}$: for instance, this could occur for the solar radiation pressure if one half of \Alice\ is much lighter (respectively, darker) than the other and therefore has a higher (lower) albedo---note that this an effect that would be absent if we imposed the \Alice\ simplifying assumption that the surface of the asteroid were uniform, so we must relax that assumption here.
Up to an $\mathcal{O}(1)$ geometrical factor $c_1$ that we do not compute as it depends in detail on the asteroid surface geometry, a conservatively large estimate for the torque that this fluctuating pressure asymmetry would induce is $\delta \tau \sim c_1 R_{\alice} A_{\alice} \delta P \sim c_1 \pi R_{\alice}^3 \sqrt{f S[P]}$.

The effect of this torque depends on the axis about which it is applied; we consider in turn the cases where the torque is applied \hyperlink{case:TSIwindTorquePara}{(a)} along the existing angular momentum axis, or \hyperlink{case:TSIwindTorquePerp}{(b)} perpendicular to that axis.

%%%%%%%%%%%%%%
\paragraph{Torque along angular momentum axis}
\hypertarget{case:TSIwindTorquePara}{}
%%%%%%%%%%%%%%
In the case where this torque is aligned along the rotational axis of \Alice, the relevant frequency range of pressure fluctuations that given rise to in-band noise around $f_{\textsc{gw}} \sim \mu$Hz will actually be $f\sim f_{\alice}$, where $f_{\alice}$ is the \Alice\ rotational frequency.
This is because we have $ f_{\textsc{gw}} \ll f_{\alice}$ except at the highest frequencies of interest in our band, and because the origin of any surface asymmetry of \Alice\ that is giving rise to the differential torque along the rotational axis must necessarily be co-rotating with \Alice. 
Therefore, low-frequency angular motion perturbations will occur as a beat note between a pressure fluctuation near the rotational period and the rotational period itself (i.e., at $f \sim f_{\alice} \pm f_{\textsc{gw}}$).
Therefore, we will take $\delta P \sim \sqrt{ f_{\textsc{gw}} S[P](f_{\alice})}$ in estimating the torque $\delta \tau$; note that the relevant bandwidth around $f_{\alice}$ is still only $\sim f_{\textsc{gw}}$ wide, which is why $ f_{\textsc{gw}}$ and not $f_{\alice}$ appears in the square root. 
Because for both the solar radiation pressure and the solar wind pressure, the PSD $S[P]$ is a falling function of frequency between $f_{\textsc{gw}}$ and $f_{\alice}$, this would only aid to suppress noise.

Because this torque acts along the rotational axis, it gives rise to a straightforward angular acceleration $\delta \alpha$ that gives rise to a fluctuation in the rotational rate: $\delta\alpha \sim \delta\tau / I_{\alice}$ where $I_{\alice} \sim c_2 M_{\alice}R_{\alice}^2$ is the moment of inertia of \Alice\ about the rotational axis, with $c_2$ being $\mathcal{O}(1)$ numerical factor that depends on the exact asteroid geometry; therefore, $\delta \alpha \sim 3 c_2 \delta \tau / ( 4\pi R_{\alice}^5 \rho_{\alice})$.
This gives rise to a fluctuation in the location of a point on the surface of the asteroid with an in-band rms amplitude of $\delta x \sim r_{\parallel} \delta \alpha /(2\pi f_{\textsc{gw}})^{2}$, where $0\leq r_{\parallel} \leq R_{\alice}$ is the shortest distance from the rotational axis of the asteroid to the relevant point on the surface.
In the worst case, the noise is similar in magnitude at both ends of the baseline, yielding a $\sqrt{2}$ larger noise than if the larger of the two noise contributions is assumed.
Putting this all together, the approximate two-asteroid contribution to the characteristic strain noise is of order%
%%%%%%%%%%%%
\footnote{\label{ftnt:Leading2}%
    The additional factor of 2 has the origin discussed in footnote \ref{ftnt:leadingFactorOfTwo}; despite these estimates being rough at the level of $\mathcal{O}(1)$ factors, we consistently account for that factor here in order to make the comparison to results in \sectref[s]{sect:TSI} and \ref{sect:solarWind} fair.
    } %
%%%%%%%%%%%%
%%%%%%
\begin{align}
    h_c^{\text{rot}, \parallel} \sim \frac{2\sqrt{2}\delta x}{L}     &\sim \frac{3\sqrt{2} c_3}{8 \pi^2} \frac{r_{\parallel}}{R_{\alice}}\frac{ \sqrt{ f_{\textsc{gw}} S[P](f_{\alice})} }{ \rho_{\alice}  R_{\alice} f_{\textsc{gw}}^{2} L} \\
    &\sim c_3 \frac{r_{\parallel}}{R_{\alice}} \frac{\sqrt{ S[P](f_{\alice})}}{\sqrt{ S[P](f_{\textsc{gw}})} }  \times h_c^{\text{CoM}}
    \label{eq:rotStrainNoise},
\end{align}
%%%%%%
where $c_3$ is another $\mathcal{O}(1)$ geometric factor that folds in both $c_1$ and $c_2$ and baseline-projection effects, and $ h_c^{\text{CoM}}$ is the cognate two-asteroid CoM noise estimate from either the solar radiation pressure [cf.~\eqref{eq:hCaliceSimplified}, recalling that $\delta P_{\odot} \sim \delta I_\odot \times (r_{\oplus}/r_{\alice})^2$ for solar radiation pressure], or the solar wind [see comments below \eqref{eq:solarWindAccn}].

We have written the final form of \eqref{eq:rotStrainNoise} in that way  
because, for both of these noise sources, we have $S[P](f) \gtrsim S[P](f_{\alice})$ for $f<f_{\alice}$; because we also have $r_{\parallel} \leq R_{\alice}$, the factor multiplying $h_c^{\text{CoM}}$ in \eqref{eq:rotStrainNoise} is thus at worst an $\mathcal{O}(1)$ factor, and is likely actually a suppression, especially if the base station is intentionally located near a rotational pole, so that $r\ll R_{\alice}$.

The qualitative conclusion is that the strain noise arising from fluctuating torques along the angular momentum direction from the solar radiation pressure and the solar wind is in the worst case no larger than the cognate CoM strain estimate arising from the same sources.

%%%%%%%%%%%%%%
\paragraph{Torque perpendicular to angular momentum axis}
\hypertarget{case:TSIwindTorquePerp}{}
%%%%%%%%%%%%%%
A torque applied perpendicular to the axis of rotation can in principle arise from a non-rotationally modulating difference in response of the asteroid to the solar radiation field or the solar wind: e.g., for the solar radiation, the `northern' hemisphere of the asteroid could be permanently lighter [higher albedo] than the `southern' one.
The pressure fluctuation to consider in this case in estimating the torque $\delta \tau$ should be taken to be $\delta P \sim \sqrt{ f_{\textsc{gw}} S[P](f_{\textsc{gw}})}$; it is possibly smaller than this if, for instance, the origin of the asymmetry is a rotating light or dark spot on one hemisphere, but we will proceed under this conservative assumption.
Other than this change, the conservative torque fluctuation would be estimated in the same way as for the parallel case: $\delta \tau \sim c'_1 \pi R_{\alice}^3 \sqrt{ f_{\textsc{gw}} S[P](f_{\textsc{gw}})}$ where $c'_1$ is again an $\mathcal{O}(1)$ factor.

However, the response of the asteroid to this torque is of course different: it will cause the asteroid's angular momentum vector to precess.
In response to a sinusoidal torque perturbation at frequency $f_{\textsc{gw}}$ applied perpendicular to the existing angular momentum vector $\bm{L}_\alice$, the asteroid will wobble by an angle of order $\delta \theta \sim \delta L / L_{\alice}  \sim c_3'  \delta \tau / ( 2\pi f_{\textsc{gw}} L_\alice )$ over a GW period, where $c_3'$ is an $\mathcal{O}(1)$ factor, and $L_{\alice} \sim \omega_{\alice}I_{\alice}$ is the magnitude of the angular momentum of \Alice\ around the rotational axis, with $I_{\alice} \sim c_2 M_{\alice}R_{\alice}^2$ being the moment of inertia of \Alice\ around that same axis.
This leads to an in-band motion of a point on the asteroid surface of order $\delta x \sim c_3' r_{\perp} \delta \tau / ( 4\pi^2 f_{\textsc{gw}} f_{\alice} I_{\alice} )$ and $0 \leq r_\perp \leq R_{\alice}$ is the shortest distance from the relevant point on the asteroid surface to the axis about which the torque is applied.

Again, in the case where the noise arising from each asteroid is similar in size, the combined noise is at worst a factor of $\sqrt{2}$ larger than the larger of the two single-asteroid contributions, so we have a two-asteroid strain-noise contribution of order%
%%%%%%%%%%%%
\footnote{\label{ftnt:Leading2again}%
    We again consistently account for the factor of 2 arising from footnote \ref{ftnt:leadingFactorOfTwo}, as at \eqref{eq:rotStrainNoise}.
    } %
%%%%%%%%%%%%
%%%%%%
\begin{align}
    h_c^{\text{rot}, \perp} & \sim\frac{3\sqrt{2}c_4'}{8\pi^2}  \frac{r_\perp}{R_{\alice}}  \frac{ \sqrt{ f_{\textsc{gw}} S[P](f_{\textsc{gw}})} }{ \rho_{\alice} R_{\alice} f_{\textsc{gw}}f_{\alice}  L }\\
    &\sim c_4' \frac{r_\perp}{R_{\alice}} \frac{f_{\textsc{gw}}}{f_{\alice}} \times h_c^{\text{CoM}},
    \label{eq:hcRotPerp}
\end{align}
%%%%%%
where $c_4'$ is an $\mathcal{O}(1)$ factor that subsumes $c_1'$, $c_2$, and $c_3'$ and accounts for baseline-projection effects, and $h_c^{\text{CoM}}$ is again the cognate two-asteroid CoM strain noise estimate [see discussion below \eqref{eq:rotStrainNoise}].
We have again written \eqref{eq:hcRotPerp} in this form to demonstrate that the result is the CoM noise estimate multiplied by a suppression factor:
we have $f_{\textsc{gw}}\lesssim f_{\alice}$ and $r_\perp \leq R_{\alice}$.
Note however that although $r_{\perp} \leq R_{\alice}$, we do have $r_{\perp}^2 + r_{\parallel}^2 \sim R_{\alice}^2$, so that one cannot simultaneously suppress both $\parallel$ and $\perp$ responses using these radius-ratio factors.

The fluctuating torques from the solar radiation pressure and the solar wind that act perpendicular to the angular-momentum vector thus give rise to a strain noise that is again even in the worst case no worse than the cognate CoM noise estimate arising from the same external perturbation.

%%%%%%%%%%%%%%%%%%%%%%%%%%%%%%%%%%%%%%%%%%%%%%%%%%%%%%
\subsubsection{Electromagnetic torques}
\label{sect:EMforcesTorque}
%%%%%%%%%%%%%%%%%%%%%%%%%%%%%%%%%%%%%%%%%%%%%%%%%%%%%%
The heliospheric magnetic field will also give rise to fluctuating torquing of any permanent magnetic moment of \Alice: $\delta\bm{\tau}_{\alice} = \bm{\mu}_{\alice} \times \bm{B}_{\textsc{hmf}}$.
As in \sectref{sect:gradBforce}, for the purposes of presenting \emph{analytical} estimates in this section, we again conservatively assume that the HMF has a power spectrum $S[B_{\text{HMF}}](f) \sim 60 \,\text{nT}^2/\text{Hz} \times ( f / \text{mHz} )^{-5/3}$; we will however use the actual HMF PSD~\cite{Stone:1998dga,ACEwebsite} shown in \figref{fig:ACE_B_PSD} for graphical presentation of these noise estimates in \subfigref{fig:noise_contributions}{g}.
As in \sectref{sect:TSIandWindTorque}, will consider two cases: assuming the torque has magnitude $\delta\tau_{\alice} \sim \mu_{\alice} B_{\textsc{hmf}}$ either \hyperlink{case:HMFtorquePara}{(a)} along the angular momentum axis or \hyperlink{case:HMFtorquePerp}{(b)} perpendicular to it.

%%%%%%%%%%%%%%
\paragraph{Torque along angular momentum axis}
\hypertarget{case:HMFtorquePara}{}
%%%%%%%%%%%%%%
The torque gives rise to a fluctuating angular acceleration $\delta \alpha_{\alice} \sim \delta \tau_{\alice} / I_{\alice} \sim \hat{\mu}_{\alice} B_{\textsc{hmf}} R_{\alice}^{-2}$, where we have used $I_{\alice} \sim M_{\alice} R_{\alice}^2$ ignoring $\mathcal{O}(1)$ geometrical factors, and $\hat{\mu}_{\alice}$ is again the specific (per-mass) magnetic moment; see \sectref{sect:gradBforce}.
Taking into account the frequency modulation effects discussed for this case in \sectref{sect:TSIandWindTorque}, and estimating similarly sized noises on both asteroids, we obtain a strain noise contribution of order
%%%%%%
\begin{align}
    &h_c^{\textsc{hmf}, \parallel}\nonumber\\ 
    &\sim \frac{r_{\parallel}}{R_{\alice}} \frac{\sqrt{2} \hat{\mu}_{\alice}} {(2\pi f_{\textsc{gw}})^2 LR_{\alice}} \sqrt{ f_{\textsc{gw}} S[B_{\textsc{hmf}}](f_{\alice}) } \label{eq:HMFparallelTorque}\\
    &\sim 8\times 10^{-17} \times \lb( \frac{r_{\parallel}}{R_{\alice}} \rb) \times \lb( \frac{ \hat{\mu}_{\alice} }{ \hat{\mu}_{\alice}^{\text{high}} } \rb) \times \lb( \frac{f_{\textsc{gw}}}{\mu\text{Hz}} \rb)^{-3/2}\\
    &\sim 5\times 10^{-21} \times \lb( \frac{r_{\parallel}}{R_{\alice}} \rb) \times \lb( \frac{ \hat{\mu}_{\alice} }{ \hat{\mu}_{\alice}^{\text{low}} } \rb) \times \lb( \frac{f_{\textsc{gw}}}{\mu\text{Hz}} \rb)^{-3/2},
\end{align}
%%%%%%
where $0\leq r_{\parallel}\leq R_{\alice}$ is again the distance from the station location on the asteroid surface to the rotational axis, and we have taken $f_{\alice} \sim (5\,\text{hrs})^{-1}$.
Importantly, note that the HMF PSD in \eqref{eq:HMFparallelTorque} is evaluated at $f_{\alice}$, and not at $f_{\textsc{gw}}$, for the reasons discussed above in \sectref{sect:TSIandWindTorque}.

Once again, as in \sectref{sect:gradBforce}, we have given two point-estimates assuming either a 433~Eros-like specific magnetic moment $\hat{\mu}_{\alice}^{\text{low}} \sim 2\times 10^{-6}\,\text{A\,m}^2/\text{kg}$, or a much higher generic asteroid magnetic moment $\hat{\mu}_{\alice}^{\text{low}} \sim 3\times 10^{-2}\,\text{A\,m}^2/\text{kg}$; see discussion in \sectref{sect:gradBforce}.
For the lower, 433~Eros-like specific magnetic moment ($\hat{\mu}_{\alice} = \hat{\mu}_{\alice}^{\text{low}}$), this noise is sub-dominant to other noise sources we have already estimated, as indicated by the solid purple line in \subfigref{fig:noise_contributions}{g}, which assumes $r_{\parallel} = R_{\alice}$ and is drawn using in \eqref{eq:HMFparallelTorque} the (smoothed) HMF PSD value for $S[B_{\textsc{hmf}}](f_{\alice})$ that is shown in \figref{fig:ACE_B_PSD}.
However, for the higher magnetic moment ($\hat{\mu}_{\alice} = \hat{\mu}_{\alice}^{\text{high}}$), it could end up being a dominant noise source by a factor of up to $\sim 30$ (at the worst-case frequencies), as indicated by the dotted purple line in \subfigref{fig:noise_contributions}{g} which again assumes $r_{\parallel} = R_{\alice}$ and is again drawn using in \eqref{eq:HMFparallelTorque} the (smoothed) HMF PSD value for $S[B_{\textsc{hmf}}](f_{\alice})$ that is shown in \figref{fig:ACE_B_PSD}.
This motivates finding at least one other asteroid TM candidate with specific magnetic moment properties similar to those of 433~Eros, in order to avoid this potential noise problem; it could also be mitigated somewhat by locating the station near (e.g., within $\sim 3$\% of) the rotational pole of \Alice.

%%%%%%%%%%%%%%
\paragraph{Torque perpendicular to angular momentum axis}
\hypertarget{case:HMFtorquePerp}{}
%%%%%%%%%%%%%%
Consistent with our discussion in \sectref{sect:TSIandWindTorque}, the strain-noise result for the case of a torque perpendicular to the angular momentum axis is obtained from the estimate parallel to the axis---at least up to $\mathcal{O}(1)$ numerical factors that we are ignoring here---by replacing $r_{\parallel} \rightarrow r_{\perp}$, $\sqrt{ S[B_{\textsc{hmf}}](f_{\alice})} \rightarrow \sqrt{ S[B_{\textsc{hmf}}](f_{\textsc{gw}})}$, and $(2\pi f_{\textsc{gw}})^2 \rightarrow (2\pi f_{\textsc{gw}})(2\pi f_{\alice})$ in \eqref{eq:HMFparallelTorque}.
Therefore,
%%%%%%
\begin{align}
    &h_c^{\textsc{hmf}, \perp}\nonumber \\ 
    &\sim \frac{r_{\perp}}{R_{\alice}} \frac{\sqrt{2} \hat{\mu}_{\alice}} {4\pi^2 f_{\textsc{gw}} f_{\alice} LR_{\alice}} \sqrt{ f_{\textsc{gw}} S[B_{\textsc{hmf}}](f_{\textsc{gw}}) } \label{eq:HMFperpTorque}\\
    &\sim 4\times 10^{-17} \times \lb( \frac{r_{\perp}}{R_{\alice}} \rb)  \times \lb( \frac{ \hat{\mu}_{\alice} }{ \hat{\mu}_{\alice}^{\text{high}} } \rb) \times \lb( \frac{f_{\textsc{gw}}}{\mu\text{Hz}} \rb)^{-4/3} \label{eq:hcHMFperpHigh}\\
    &\sim 3\times 10^{-21} \times \lb( \frac{r_{\perp}}{R_{\alice}} \rb) \times \lb( \frac{ \hat{\mu}_{\alice} }{ \hat{\mu}_{\alice}^{\text{low}} } \rb) \times \lb( \frac{f_{\textsc{gw}}}{\mu\text{Hz}} \rb)^{-4/3}.\label{eq:hcHMFperpLow}
\end{align}
%%%%%%
Again, for the lower (433~Eros-like) specific magnetic moment ($\hat{\mu}_{\alice} = \hat{\mu}_{\alice}^{\text{low}}$), this noise is sub-dominant to other noise sources we have already estimated; see the lower orange band (with the solid orange line being the log-frequency-space Gaussian-kernel smoothing of the band) in \subfigref{fig:noise_contributions}{g}, which is drawn assuming $r_{\perp} = R_{\alice}$ and using in \eqref{eq:HMFperpTorque} the full HMF PSD $S[B_{\textsc{hmf}}]$ that is shown in \figref{fig:ACE_B_PSD}.%
%%%%%%%%%%%%
\footnote{\label{ftnt:disagreementExpected}%
    Note that this means that the numerical values quoted at \eqref[s]{eq:hcHMFperpHigh} and (\ref{eq:hcHMFperpLow}) will, when evaluated at $f_{\textsc{gw}}\lesssim 3\times 10^{-6}\,\text{Hz}$, disagree numerically with the results plotted in \subfigref{fig:noise_contributions}{g}, because the analytical model used for $S[B_{\text{HMF}}](f)$ in arriving at \eqref[s]{eq:hcHMFperpHigh} and (\ref{eq:hcHMFperpLow}) over-estimates the HMF PSD in that frequency range, as shown in \figref{fig:ACE_B_PSD}. 
    } %
%%%%%%%%%%%%
However, for the higher specific magnetic moment ($\hat{\mu}_{\alice} = \hat{\mu}_{\alice}^{\text{high}}$), it again ends up being about a factor of up to $\sim 30$ (at the worst-case frequencies) larger than the other, dominant noise sources we have already estimated; see the upper orange band (with the dotted orange line being the log-frequency-space Gaussian-kernel smoothing of the band) in \subfigref{fig:noise_contributions}{g}, which is again drawn assuming $r_{\perp} = R_{\alice}$ and again using in \eqref{eq:HMFperpTorque} the full HMF PSD $S[B_{\textsc{hmf}}]$ that is shown in \figref{fig:ACE_B_PSD}.
Again, this motivates looking for 433~Eros-like TM candidates to avoid this noise contribution.

%%%%%%%%%%%%%%
\paragraph{Comment}
%%%%%%%%%%%%%%
While the estimates above obtained using the 433~Eros-like specific magnetic moment $\mu_{\alice}^{\text{low}} \sim 2\times 10^{-6}\,\text{A\,m}^2/\text{kg}$ are easily sub-dominant to other noise sources, the estimates obtained from the higher specific magnetic moments $\mu_{\alice}^{\text{high}} \sim 3\times 10^{-2}\,\text{A\,m}^2/\text{kg}$ are larger than other dominant noise sources we have estimated for $ 10^{-6}\,\text{Hz} \lesssim f_{\textsc{gw}} \lesssim 3 \times 10^{-5}\,\text{Hz}$; see \subfigref{fig:noise_contributions}{g}.
Because specific asteroids with such large specific magnetic do exist~\cite{Kivelson:1993rsw,Richter:2001ehe,ACUNA2002220,Hercik:2020svs}, care must be taken when selecting asteroid targets for this mission to identify low-magnetization asteroids, with specific magnetic moments closer to that of 433~Eros.
Alternatively, because the noise source we have identified here is only a factor of at worst $\sim 30$ larger than other other sources at the worst-case frequencies, \emph{in situ} measurements of the local HMF field fluctuations to 2--3 significant figures by a magnetometer on board the base-station package would be sufficient to allow modeling of this noise source, allowing it to be mitigated to levels no worse than the other noise sources, even assuming that an asteroid with a large specific moment must be selected for other operational reasons (e.g., size, location, rotational characteristics, etc.).

%%%%%%%%%%%%%%%%%%%%%%%%%%%%%%%%%%%%%%%%%%%%%%%%%%%%%%
\subsubsection{Collisions}
\label{sect:collisionsTorque}
%%%%%%%%%%%%%%%%%%%%%%%%%%%%%%%%%%%%%%%%%%%%%%%%%%%%%%
Similar to the case of the solar radiation and wind torques, torques from collisions can be reduced to estimates similar to the cognate CoM motion estimate.
Consider objects with mass $m_{i}$ in the range $m_{1i}\leq m_{i} \leq m_{2i}$ colliding with \Alice\ at a distance $\sim R_{\alice}$ from the rotational axis of the asteroid at a relative impact speed of $v_{\text{coll}}$.
Up to $\mathcal{O}(1)$ geometrical factors, immediately prior to the collision, any such object carries an angular momentum $L_{\text{obj},i} \sim R_{\alice} m_{i} v_{\text{coll}}$ relative to an axis passing through CoM of \Alice.
Neglecting spallation of particles from the asteroid surface upon collision, this angular momentum is transferred to the asteroid: $(\delta L_{\alice})_{1,i} \sim R_{\alice} m_{i} v_{\text{coll}}$. 
Suppose that during a time $T_{\textsc{gw}}$, $N_i(T_{\textsc{gw}})$ collisions of objects in this mass-range occur in a randomly directed fashion, with $N_i(T_{\textsc{gw}})$ still given by \eqref{eq:nAstInBin}. 
This will lead to a net change in the  angular momentum of \Alice\ over a GW period of order $(\delta L_{\alice})_i \sim R_{\alice} \bar{m}_i v_{\text{coll}} \sqrt{N(T_{\textsc{gw}})}$ where $\bar{m}_i \equiv \sqrt{m_{1i}m_{2i}}$, with similar magnitude changes occurring along all three inertial axes.
Note that we make this estimate under the `realistic' case for the collision noise discussed in \sectref{sect:collisions}; we will not discuss the conservative case here, as that was an overestimate.
We again treat the cases \hyperlink{case:collTorquePara}{(a)} along and \hyperlink{case:collTorquePerp}{(b)} perpendicular to the angular momentum axis separately.

%%%%%%%%%%%%%%
\paragraph{Torque along angular momentum axis}
\hypertarget{case:collTorquePara}{}
%%%%%%%%%%%%%%
For torques along the angular momentum axis, the change in the angular momentum of \Alice\ leads to a net change in its angular velocity of order $(\delta \omega_{\alice})_i \sim (\delta L_{\alice})_i / I_{\alice}$.
Over a period $T_{\textsc{gw}}$, this causes a position error on the location of a station a distance $r_{\parallel}$ from the rotational axis of order $(\delta x)_i \sim r_{\parallel} (\delta \omega_{\alice})_i T_{\textsc{gw}}$ [note: $(\delta\omega_{\alice})_iT_{\textsc{gw}}\ll 1$], leading to a two-asteroid strain noise contribution from this mass-bin (assuming roughly equal-magnitude noise at both ends of the baseline) of
%%%%%%
\begin{align}
    (h_c)_i^{\text{rot coll},\parallel} &\sim 
    \frac{r_{\parallel}}{R_{\alice}} \frac{3 \bar{m}_i v_{\text{coll}}  \sqrt{  \lb[ I(m_{1i}) - I(m_{2i}) \rb] } }{\sqrt{2\pi} \rho_{\alice} L R_{\alice}^2 f^{3/2}_{\textsc{gw}} }  \\
    &\sim \frac{r_{\parallel}}{R_{\alice}} (h_c)_i^{\text{coll, CoM}} \\
    &\lesssim (h_c)_i^{\text{coll, CoM}},
\end{align}
%%%%%%
where $(h_c)_i^{\text{coll, CoM}}$ is the collisional CoM strain noise estimate from the same mass-bin given at \eqref{eq:oneMassBinStrain}.
The same discussion that follows \eqref{eq:oneMassBinStrain} in \sectref{sect:collisions} regarding the dominant mass bin thus applies here too.

Similar to the torque results for solar radiation and wind noise sources, we see a suppression of the rotational part of the collisional strain noise contribution as compared to the CoM noise contribution, by the ratio of the typical distance between the relevant point on the asteroid surface and the rotational axis, to the typical radius of \Alice.
The rotational noise contribution from collisions is thus no worse than the cognate CoM noise contribution.
As shown in \subfigref{fig:noise_contributions}{g}, the realistic collisional CoM noise estimate is already quite safe, so its rotational cognate is not a source of additional problematic noise.

Were we to re-run this argument with the conservative noise estimate, we would find a similar parametric scaling would arise relating the rotational and CoM cases, and so the conservative estimate of the rotational part of the collisional strain noise contribution is again no worse than the CoM contribution.
This is in the same ballpark as other dominant noises we estimated, although it can be larger by up to a factor of $\sim 30$ for the worst-case frequencies [see \sectref{sect:collisions} and \subfigref{fig:noise_contributions}{g}]; we however know that estimate to be a vast overestimate / absolute upper bound.

%%%%%%%%%%%%%%
\paragraph{Torque perpendicular to angular momentum axis}
\hypertarget{case:collTorquePerp}{}
%%%%%%%%%%%%%%
Consider now a similar-magnitude torque noise applied to either of the two axes perpendicular to the angular momentum direction, which gives rise to precessional motion of the angular momentum axis, causing an angular movement of the asteroid axis by an amount $\delta \theta_{\alice} \sim \delta L_{\alice} / L_{\alice}$.
This leads to a two-asteroid strain noise estimate (assuming similarly sized noises arise from each asteroid, and along each of the 2 perpendicular axes) of order
%%%%%%
\begin{align}
    (h_c)_i^{\text{rot coll},\perp} &\sim \sqrt{2}
    \frac{r_{\perp}}{R_{\alice}} \frac{f_{\textsc{gw}}}{ 2 \pi f_{\alice}} \frac{ 3\bar{m}_i v_{\text{coll}} \sqrt{  \lb[I(m_{1i}) - I(m_{2i})\rb] } }{ \sqrt{2\pi} \rho_{\alice}L R_{\alice}^2  f^{3/2}_{\textsc{gw}} } \\
    &\sim \sqrt{2} \frac{r_{\perp}}{R_{\alice}} \frac{f_{\textsc{gw}}}{ 2 \pi f_{\alice}} (h_c)_i^{\text{coll, CoM}}\\
    &\lesssim (h_c)_i^{\text{coll, CoM}},
\end{align}
%%%%%%
where $r_\perp$ is the distance of the station location from the relevant rotational axis, which is of order $r_\perp \sim \sqrt{ R_{\alice}^2 - r_{\parallel}^2 }$.
This is again suppressed as compared to the CoM estimate from the same bin, \eqref{eq:oneMassBinStrain}, since $r_{\perp} \leq R_{\alice}$ and $f_{\textsc{gw}} < f_{\alice}$ in our GW frequency band.

Again, running the same argument on the conservative collisional noise estimate from \sectref{sect:collisions} would yield a similar parametric suppression of  the rotational result as compared to the cognate conservative CoM result.

This is thus again no more problematic a noise source than the direct CoM strain noise estimate.

%%%%%%%%%%%%%%%%%%%%%%%%%%%%%%%%%%%%%%%%%%%%%%%%%%%%%%
\subsubsection{Flybys}
\label{sect:flybysTorque}
%%%%%%%%%%%%%%%%%%%%%%%%%%%%%%%%%%%%%%%%%%%%%%%%%%%%%%
The estimate for torquing of \Alice\ from flybys of small objects can be based on the GGN noise estimate from \citeR{Fedderke:2020yfy}: this is because, for a close flyby, the GGN noise is dominated by the force applied to a single asteroid of the pair forming the baseline.
The differential force applied to \Alice\ during a flyby gives rise to a torque on the asteroid.
The parametrics of the estimate will go through in much the same way as the parametrics for the estimates for the solar radiation pressure, solar wind, and collisions, leading to a torque-induced strain noise contribution from flybys that can be parametrically estimated based on the cognate CoM contribution from flybys. 

There is however one exception here: the relevant fluctuating differential force applied to \Alice\ during a single-object flyby that gives rise to a torque is the \emph{tidal} gravitational force acting across the asteroid, not the full gravitational force: the torque-induced strain noise estimate for a single flyby is thus parametrically suppressed by an additional factor of $\sim R_{\alice} / b$ as compared to the cognate CoM estimate, where $b$ is the impact parameter for the flyby. 
Because $R_{\alice} / b \ll 1$ for all flybys except for those of the most minute dust grains which do not dominate the estimate (see Sec.~V of \citeR{Fedderke:2020yfy}), the flyby torque noise contribution can thus be estimated to be tidally suppressed as compared to the GGN noise level shown in \subfigref{fig:noise_contributions}{d}.
This noise is thus not relevant.

%%%%%%%%%%%%%%%%%%%%%%%%%%%%%%%%%%%%%%%%%%%%%%%%%%%%%%
\subsection{Asteroid orbital and rigid-body kinematics}
\label{sect:asteroidKinematics}
%%%%%%%%%%%%%%%%%%%%%%%%%%%%%%%%%%%%%%%%%%%%%%%%%%%%%%
In additional to their response to fluctuating torques, asteroids are also subject to \hyperref[sect:orbitalKinematics]{(1)} intrinsic orbital and \hyperref[sect:rotationalMotion]{(2)}~torque-free rotational motions that will limit their utility as test masses in certain frequency bands.
We discuss these motions in this subsection.

%%%%%%%%%%%%%%%%%%%%%%%
\subsubsection{Orbital motion}
\label{sect:orbitalKinematics}
%%%%%%%%%%%%%%%%%%%%%%%
For inner Solar System asteroids of the type that we consider as TM candidates in this work, typical orbital periods are $\mathcal{O}(\text{years})$; see \tabref{tab:asteroids}.
This places their orbital frequencies in the band $f_{\text{orb}}\lesssim 3\times 10^{-8}\,\text{Hz}$, well below our band of interest.
When computing the baseline distance between two such asteroids, both asteroid periods, as well as all of their higher harmonics, and the sum and difference frequencies of those frequencies, will all also generally enter in the variation of the baseline distance.
Heuristically, because the eccentricity enters in the heliocentric radius expression as $r[\theta(t)] = \mathfrak{a}\lb(1-\mathfrak{e}^2\rb)\lb(1+\mathfrak{e}\cos[\theta_\mathfrak{e}(t)]\rb)^{-1}$ where $\theta_\mathfrak{e}(t)$ is the solution to the eccentric orbit equation (see Appendix A.4 of \citeR{Fedderke:2020yfy}), it is generically the case that higher harmonics of the orbital period enter in the baseline distance expression accompanied by higher powers of the eccentricity.%
%%%%%%%%%%%%
\footnote{\label{eq:roughArgument}%
    To flesh out this heuristic argument: the lowest power of a term $\sim \cos(\omega t)$ containing a term with frequency content at $\omega_n = n\omega$ is $\cos^n(\omega t)$, and each higher power of $\cos[\theta_\mathfrak{e}(t)]$ in an $\mathfrak{e}\ll 1$ expansion of the radial distance $r[\theta(t)]$ enters with one higher power of $\mathfrak{e}$, so the $n$-th harmonic is suppressed by $\sim \mathfrak{e}^n$.
    This argument is of course an heuristic explanatory tool and does not capture the full dependence of the baseline distance on $\mathfrak{e}$; in particular, it does not address the $\mathfrak{e}$-dependence residing in $\theta_{\mathfrak{e}}(t)$ itself.
    The point of the numerical results presented in this section is to verify explicitly that this exponential suppression does occur in the full computation.
    } %
%%%%%%%%%%%%
Asteroid eccentricities are typically non-trivially large, but generically not so large as to be $\mathcal{O}(1)$; see \tabref{tab:asteroids}.
There is therefore an exponential suppression of the power in higher harmonics (see also Sec.~II of \citeR{Fedderke:2020yfy} for a much more detailed discussion of this effect).
We thus expect little of the orbital motion to have direct frequency overlap with our band of interest.
Some overlap is however in principle possible, and we wish to quantify this.

\begin{figure*}[!p]
    \centering
    \includegraphics[width=\textwidth]{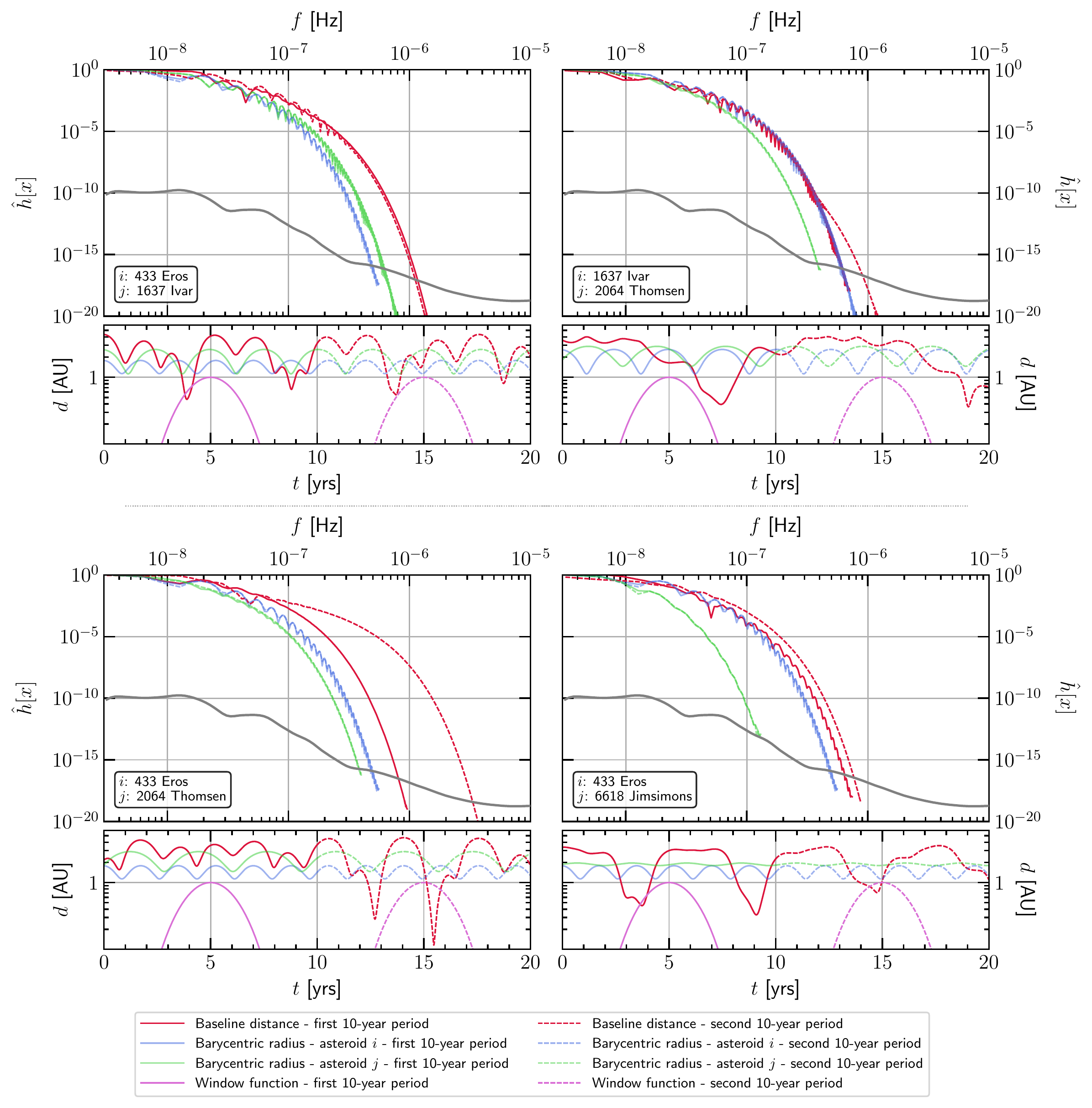}
    \caption{\label{fig:baselineASDs}%
        For each of four pairs $i,j$ of asteroids selected from \tabref{tab:asteroids}, and annotated in each quadrant of this figure, we present two sets of results.\quad
        \textsc{Upper panel for each asteroid pair:} We show $\hat{h}[x]$ as defined in the text, for the quantities $x$ taken to be the baseline separation distance $L_{i,j}(t)$ between the asteroids (red lines), and the Solar-System-barycentric radius of each asteroid orbit $r_k(t)$ for $k=i,j$ (blue and green lines, respectively).
        These results are presented for two separate, contiguous simulated 10-year mission durations: one set of results as solid lines and the other as dashed lines.
        Note that these results are obtained after processing the time-series data through a window function $w(t) = \sin^8\lb[\pi (t-t_0)/T \rb]$ to reduce spectral leakage and increase the dynamic range of the PSD~\cite{Fedderke:2020yfy,Harris:1978wdg}, but are corrected upward by a factor $\zeta \equiv 128/35$ to partially account for the amplitude-suppression effects of the window (see discussion in text) and are thus directly comparable to results for~$h_c$.
        Also shown in gray is the smoothed and enveloped $h_c$ noise curve arising from all other relevant noise sources, as shown in \figref{fig:results}.         
        We note that a number of the results curves are truncated at values of $\hat{h}[x]$ either at, or below, the noise curve; this is to avoid showing unphysical results that are impacted by residual spectral leakage effects that our windowing procedure reduced but did not completely eliminate (see discussion in the text). 
        No conclusions are modified by this truncation.\quad
        \textsc{Lower panel for each asteroid pair:} We show the time-series data for $L_{i,j}(t)$ (red lines) and $r_k(t)$ for $k=i,j$ (blue and green lines, respectively), with the two contiguous 10-year mission durations shown again as solid and dashed lines, respectively, and matching the same line styles as used in the upper panels.
        Also shown in purple (with the same solid/dashed definition) is the envelope of the window function $w(t)$.
        These results are discussed at length in the text.
    }
\end{figure*}

We have explicitly computed the baseline distance $L_{i,j}(t)$ between some of the pairs of asteroids $i,j$ listed in \tabref{tab:asteroids}, and have examined their frequency content; see \figref{fig:baselineASDs}.
We perform this computation twice, over two disjoint, contiguous simulated missions each of $T=10$\,yr duration, in order to illustrate some interesting variation of the results.
We have assumed in doing this that the asteroids in question follow exactly elliptical orbits specified by their instantaneous osculating elliptical orbital parameters specified in the NASA JPL Small-Body Database~\cite{JPL-SBD}; although we know this to be imprecise for the timescales of interest here, quantifying even this idealized case is useful.
In order to address issues of spectral leakage and enlarge the dynamic range of our results (see discussion in Appendix D of \citeR{Fedderke:2020yfy}, and also \citeR{Harris:1978wdg}), we apply a window function $w(t) = \sin^8(\pi (t-t_0) /T)$ to $L_{i,j}(t)$: $L^w_{i,j}(T) \equiv w(t) L_{i,j}(t)$.
We then compute the PSD of the windowed timeseries $S[L^w_{i,j}]$, and construct the quantity 
%%%%%%
\begin{align}
    \hat{h}[L_{i,j}] \equiv \zeta \langle L_{i,j} \rangle^{-1} \sqrt{f S[L^w_{i,j}](f)},
    \label{eq:hathDefn}
\end{align}
%%%%%%
where $\langle\,\cdots\rangle$ is the temporal average over the simulated mission duration and $\zeta$ is a factor designed to account for the PSD-amplitude suppression effect of the window function, so that the resulting quantity $\hat{h}_{i,j}$ can roughly be compared to the characteristic strain $h_c$ that would be detectable.
Given our window function, we take $\zeta = 128/35$, which is correct for a narrowband ($\Delta f \lesssim 1/T$) source; see discussion in Appendix D of \citeR{Fedderke:2020yfy}.
These results are shown in the upper panel in each quadrant of \figref{fig:baselineASDs}.

Generalizing, we define $\hat{h}[x]$ as at \eqref{eq:hathDefn}, but with $L_{i,j}(t) \rightarrow x(t)$ and $L^w_{i,j}(t) \rightarrow x^w(t) = w(t) x(t)$ for any function $x(t)$.
For comparative purposes, we also show in the upper panel of each quadrant in \figref{fig:baselineASDs} the results for $\hat{h}[r_k]; \; k=1,2$, where $r_k(t)$ is the Solar System barycentric radius of the orbit of each asteroid in the relevant pair.
We also show our final noise curve result from \figref{fig:results} for comparison.

Additionally, in the lower panels of each quadrant of \figref{fig:baselineASDs}, we show the (unwindowed) time series data for all the relevant functions for each simulated mission duration, along with the window function envelope.

We have truncated a number of our $\hat{h}[x]$ results curves (for relevant $x$) in the upper panel of each quadrant of \figref{fig:baselineASDs} at thresholds either at or below where they cross our noise curve from \figref{fig:results}. 
This is because the window function we have used, while excellent at suppressing spectral leakage and thereby widening the dynamic range over which we can present results, does not completely eliminate that leakage; it only mitigates it.
Had we continued to present the curves to lower values of $\hat{h}[x]$ than shown without making any changes to the procedures discussed above that we used to compute them, they would \emph{unphysically} transition from falling exponentials to smooth, falling power laws; this is a feature that (in this context, at least) is known to be diagnostic of spectral leakage effects.
We note that this is purely a signal-processing issue, and could in principle be mitigated by resorting to a different window function that more severely trades off for dynamic range at the expense of frequency resolution.
However, because the existing results suffice to track $\hat{h}[x]$ for relevant $x$ at least to the level of our noise curve (and in many cases a few orders of magnitude below it), the case for resorting to a different window function is weak.
The results as presented in \figref{fig:baselineASDs} are valid where shown, physical, and track the curves to or beyond the point where they are needed; no conclusions are altered by our truncation, nor would they be had we implemented a different windowing optimization to explore the results to smaller values of $\hat{h}[x]$.

%%%%%%%%%%%%%%%%%%%%%%%%%%%%%%%%%%%%%%%%%%%%%%%%%%%%%%%%%%%%%%%%%%%%%%%%%%%%%%%%
%%%%%%%%%%%%%%%%%%%%%%%%%%%%%%%%%%%%%%%%%%%%%%%%%%%%%%%%%%%%%%%%%%%%%%%%%%%%%%%%

These results illustrate a number of interesting features: 
(1) for most cases, the orbital motion gives rise to exponentially suppressed frequency content in the baseline distance variation above $f\sim\mu$Hz; 
(2) there is however a large frequency content of the baseline distance variation in the 0.1--1$\mu$Hz band; 
(3) there can be, but there is not necessarily, some large variation in the frequency content of the baseline distance for a fixed asteroid pair from one 10-year mission duration to the next, including in some cases still quite large power at frequencies above $\mu$Hz; and
(4) there is a reasonably large pair-to-pair variation in the quantitative results, albeit with clear common qualitative features, such as exponential drop-off in the noise as a function of frequency.

Observations (1) and (2) can be explained reasonably directly from our comments above regarding the eccentricity of the asteroid orbits generically leading to $\mathfrak{e}^n \ll 1$ suppression for the $n$-th harmonics in the motion.

Observation (3) can also be easily explained: there is a clear correlation between having higher-amplitude high-frequency components in $\hat{h}[L_{i,j}]$ in a given mission duration, and the existence within that same mission duration of a closer encounter between the two asteroids (especially a close encounter near the peak of the window function).
Such a close encounter obviously induces higher-frequency changes in the baseline distance.
Conversely, the mission durations with lower-amplitude high-frequency components in $\hat{h}[L_{i,j}]$ exhibit an absence of such close passes (or their appearance is far from the peak of the window function).
What is clear from \figref{fig:baselineASDs}, however, is that there is no corresponding large change to the way in which the radii of the asteroid orbits themselves, as measured from the Solar System barycenter, vary from one mission duration to the next: they never exhibit large amounts of power above $\sim \mu$Hz, even in cases where $\hat{h}[L_{i,j}]$ shows higher power above $\mu$Hz.
The most potentially troublesome high-frequency content (i.e., that above $\sim \mu$Hz) of the baseline distance variation thus arises only from the effective `projection' of the orbits onto each other that is inherent in making a distance measurement between two TMs on those orbits.
Most na\"ively, one could simply window out those durations of time in the data stream when the asteroids are known to be executing more rapid close encounters, and focus on analyzing the durations of time when the asteroids are further apart.
Alternatively, mission planning could select durations of time during which such close passes are not expected to occur.

Note that this effect does not occur for all asteroid pairs.
With reference to the lower panels in each quadrant of \figref{fig:baselineASDs}, it is clear that, depending on the orbital parameters, close passes can (a) occur many times during each simulated 10-year mission and thus always be present, leading to higher high-frequency components in any simulated mission of that length; (b) not occur at all and thus always be absent, leading to lower high-frequency components; or (c) can occur infrequently, leading to a fluctuation between in the amplitude of the high-frequency components depending on whether or not such a pass happened to occur in any given relevant mission duration.
This also easily explains observation (4): particular pairs of asteroid orbits and orbital orientations will just give rise to larger numbers of close encounters, leading to larger high-frequency components.
Again, mission planning could select asteroids for which such close passes are not expected to occur, or are expected to occur only rarely so that they can be windowed out without causing a loss of large amounts of useful mission time.

We should emphasize however that the overlap in frequency space of the upper end of the frequency content of the baseline distance variation with the lower end of our GW band of interest, for some asteroids during some chosen mission durations, occurs here because we have made no attempt in this discussion to model and remove the orbital motion.
The orbital motion is \emph{deterministic}, so this frequency content overlap is only a noise to the extent to which the modeling of that motion is inaccurate or imprecise: in a full analysis, one would of course try to fit the full time series of the measured baseline distance by a model that includes orbital motions and a GW signal (plus other kinematic motions such as rotations [see next subsection]), and not just take a PSD of the raw (windowed) baseline separation measurements and look for excess narrowband power.
For instance, suppose the asteroids actually followed exactly elliptical orbits around the Solar System barycenter: one could then fit out the \emph{entirety} of the orbital motion with a simple 12-parameter fit (six orbital elements per asteroid: semimajor axis, eccentricity, time of perihelion passage, and three Euler angles that define the orbital orientation).
That is, the entirety of the results shown for $\hat{h}[L_{i,j}]$ in \figref{fig:baselineASDs} could be modeled and removed by a 12-parameter fit for each asteroid pair.
In practice, of course, asteroids do not follow exactly elliptical orbits due to $N$-body gravitational interactions and other perturbations, but these effects can be reasonably well modeled (see, e.g., \citeR[s]{Greenberg:2020wrg,Evans:2018wgy}).

Overall, these results make clear that the frequency separation between the orbital frequency band and the $1$--$10\,\mu$Hz band is in many cases already sufficient to allow for GW detection in our band \emph{without} the need for any modeling of the orbital motion.
Orbital modeling, or more sophisticated analysis strategies, may however be needed for some chosen sets of asteroids when analyzed during some time periods if full access to the 1--10\,$\mu$Hz band is desired.
Below $f \sim \mu$Hz, the overlap of the high-frequency tail of the orbital motion with a GW signal might present a greater difficulty; to the extent that orbital modeling cannot mitigate this, this mission concept may lose some coverage in the $0.1$--$1\,\mu$Hz band.

%%%%%%%%%%%%%%%%%%%%%%%
\subsubsection{Rotational motion}
\label{sect:rotationalMotion}
%%%%%%%%%%%%%%%%%%%%%%%
\newcommand{\Trot}{\ensuremath{T_{\text{rot}}}}
%%%%%
Asteroids are naturally found to be in a state of rotational motion, with periods \Trot\ that can vary widely, depending also on the asteroid population under consideration.

To select just some long-period examples in the main belt: 288~Glauke has a $\sim 29\,$km diameter~\cite{2011ApJ...741...68M} and $\Trot\sim 1.2\times 10^3\,$hrs~\cite{JPL-SBD}; and 5644~Hyakutake, which is of $\sim 16\,$km diameter~\cite{2011ApJ...741...68M}, has $\Trot\sim 2\times 10^2\,$hr~\cite{JPL-SBD}. 
On the other hand, various similarly sized main-belt asteroids spin much faster: for instance, 10263~Vadimsimona has a $\sim 16\,$km diameter~\cite{2011ApJ...741...68M} and $\Trot \sim 0.55\,$hr.

Of course, we do not wish to select main-belt asteroids owing to the noise environment there, but there is some variation even for asteroids outside the main belt: we have listed a variety of such asteroid rotational periods in \tabref{tab:asteroids} that are in the range of $2.4\,\text{hrs}\lesssim \Trot\lesssim 12.1\,\text{hrs}$.
The rotational state of 433~Eros in particular has been extremely well characterized using data from the NEAR--Shoemaker mission~\cite{MILLER20023,Yeomans:2000eth,FRENCH2020113537,KONOPLIV2002289}.

The reason this is relevant is that the ranging reference point on the asteroid will necessarily lie at (or near) the surface of the asteroid, and will thus be executing periodic motion at the asteroid rotational period, which may hamper the ability of the mission to extract a GW signal, absent accurate modeling or measurement of the rotational state.
As we have characterized the relevant torque noises and have shown that they are all sufficiently small, these rotational motions are all highly stable, but given that the amplitude of the rotational motion of the asteroid `equator' for a \Alice-class asteroid leads to a na\"ive strain variation of order $h_c \sim 8\,\text{km}/1\,\text{AU} \sim 5\times 10^{-8}$ at the rotational period, any such rotational characterization would need to be accurate at many significant figures to completely remove this additional rotational contribution to the baseline distance variation if one were searching for a GW at the rotational period.

As such, we consider GW measurements at the rotational period of the asteroid to be unlikely to be robust.
Moreover, any number of mechanisms could also give rise to additional baseline distance variation at higher harmonics of the rotational period, and so we would also assume that extraction of a GW signal at an harmonic of the rotational period is also unlikely to be robust, although the motion at higher harmonics may be suppressed.
As a result, we assume that GW detection is severely inhibited/blinded at all frequencies $f^i_n = n f^i_{\text{rot}}$ for $n=1,2,\ldots$, where $f^i_{\text{rot}} = 1/\Trot \approx 6.9\times 10^{-5}\,\text{Hz} \times ( 4\,\text{hrs} / \Trot )$, as well as in some frequency bands around these frequencies.
The width of these inhibited bands in frequency space will be set by the rotational stability of the asteroids.
For the purposes of this paper, we assume that the inhibited bands around each $f_n^i$ are fractionally $\pm 5\,\%$ wide.
Future detailed mission planning for specific asteroids would refine that estimate.
Of course, with two such asteroids with non-commensurate periods, the overlap between the inhibited bands will eventually fill an $\order{1}$ fraction of the frequency range as one moves to frequencies above the first few harmonics, supplying a natural high-frequency cutoff to the sensitivity of this proposal that starts at or around the rotational period or its first few harmonics, $f \sim 10^{-4}\,\text{Hz}$.
We plot vertical shaded bands in \subfigref{fig:noise_contributions}{h} showing the frequency ranges $f^i_n(1 -\delta_f) \leq f_{\textsc{gw}} \leq f^i_n (1+\delta_f)$ for $n=1,\ldots, 11$ (i.e., fundamental and 10 harmonics), assuming $\delta_f = 0.05$, both for $\Trot \sim T_{\alice} \sim 5$\,hrs for \Alice\ (darker, blue-gray bands) and $\Trot \sim 4$\,hrs for \Bob\ (lighter, gray bands); see also \figref[s]{fig:results_combo} and \ref{fig:results}.

Moreover, because asteroids are not in general spherically symmetric bodies, they are not necessarily simply in stable rotational motion around one of their principal axes (i.e., the orthogonal eigenvectors of the moment of inertia tensor); as such, additional rotational motions are possible.
This motion has in particular been very carefully considered and characterized for 433~Eros~\cite{MILLER20023,Yeomans:2000eth,FRENCH2020113537,KONOPLIV2002289}.

First, there is the torque-free Eulerian wobbling of the body-fixed frame around the fixed angular momentum vector that can occur for such a non-symmetrical body~\cite{Goldstein:2002eas}.
433~Eros is found to be in a state of rotational motion very nearly aligned with the third principal axis (largest principal moment of inertia), up to a wobble angle in the rotational axis that has been constrained, using two-way radio Doppler data from the Deep Space Network (DSN) and the NEAR--Shoemaker spacecraft after it landed on the surface of 433~Eros, to be $\Delta\theta \lesssim 10^{-3}\,\text{deg}$~\cite{FRENCH2020113537}, and which would have a period $T_{\text{wobble}}\sim 14.8\,$hrs ($f_{\text{wobble}}\sim 1.88\times 10^{-5}$\,Hz).
Such a wobble would however still possibly constitute a gigantic signal of order $h_c \lesssim 10^{-12}$, and would fall into our band of interest. 
This would undoubtedly blind the GW search at some additional, slightly lower than rotational, frequencies.
However, this motion will once again be stable, and will only blind certain narrow frequency bands; moreover, this motion does not reach down into the band below $10\,\mu$Hz.

433~Eros also exhibits longer-period precession and nutation of its rotational pole under the action of the solar gravity gradient torque~\cite{MILLER20023}.%
%%%%%%%%%%%%
\footnote{\label{ftnt:gradientForTorque}%
    This is similar to the precession and nutation of a spinning top, with the difference that a constant gravitational field is sufficient to cause precession of the top owing to the surface support point of the latter not being the CoM of the top~\cite{Goldstein:2002eas}.
    For an asteroid unsupported in free space, a constant gravitational force does not supply a torque through the CoM (this is by definition of the CoM); it is only the gradient of the Sun's gravity across the asteroid that can do this.
    } %
%%%%%%%%%%%%
This motion is estimated to be at the $10^{-2}\,\text{deg}$ level with a 9-month period~\cite{KONOPLIV2002289} ($f \sim 4.3\times 10^{-8}\,$Hz); this motion has not been unambiguously detected, but is approximately at the limit of detectability given surface-landmark tracking performed using NEAR--Shoemaker data taken during the 433~Eros-orbital phase of that mission~\cite{KONOPLIV2002289}.
While the amplitude is thus large, it is out of band on the low-frequency side; again, it will also largely be stable, and so confined to narrow bands, with higher harmonics suppressed.

The situation is thus that the relevant rotational motions for an ideal candidate asteroid such as 433~Eros appear to blind certain frequency bands to GW detection above $10\,\mu$Hz, with the problem likely becoming quite severe by $f\sim 10^{-4}\,$Hz owing to blinded-band overlap.
Other rotational motion of the asteroid may also blind low-frequency bands below $\sim 0.1\mu$Hz, but their overlap with our band of interest will be small.
We do not expect strong rotational blinding within the $1$--$10\,\mu$Hz band.

Moreover, there are potentially mitigations can could in principle be deployed were rotational motion to be more of a severe issue.

First, the rotational motion does not depend at all on the instantaneous baseline distance $L(t)$, which a GW signal of course does: rotational motion and a GW signal are thus not completely degenerate.
This could in principle be used to separate out a GW signal in a fit of a model of orbital and rotational motion plus a GW signal against the time series of measured baseline distances.
Additionally, rotational motion is asteroid-specific: if the mission concept were extended to include more than two asteroids, additional isolation of asteroid-specific (both rotational and orbital) noise sources could be possible, while the GW signal would be common (up to GW polarization and orientation effects).

Second, we again reiterate that these rotational motions, as with orbital motion, are deterministic and can be modeled to some degree: because we have separately estimated in \sectref{sect:externaltorques} the relevant torque noises to be small enough, the deterministic rotational motion is only a potential inhibitor to this mission concept (and even then, only at the relevant frequencies outside our main band of interest) to the extent that the deterministic rotational modeling is inaccurate. 
For instance, it is possible that the rotational state of the asteroid could itself be measured independently of the GW strain measurement, and this data included to constrain the rotational model in a global fit of asteroid baseline separation data.
Indeed, some of the existing 433~Eros rotational modeling relied on such approaches, using natural surface landmarks (craters, etc.) as tracking points to constrain the rotational state~\cite{KONOPLIV2002289}.
It is mentioned in \citeR{MILLER20023} that one idea would be to land a number of human-made transponders on the surface of the asteroid, and use them as precise reference points for, e.g., an asteroid-orbiting satellite to track.
This would be an idea that would fit in naturally with one of the possible strategies we propose in \sectref[s]{sect:concept} and \ref{sect:missionDesign} for a design for this mission concept: landing on the surface of the asteroid a large number of transponders or retroreflectors, keeping the main base-station systems (clock, ranging) in an asteroid-orbiting satellite, and ranging the satellite to the landed transponders/retroreflectors on the asteroid surface via a secondary ranging system; see further discussion in \sectref{sect:missionDesign} below.
Of course, performing the necessary angular measurement would require rotational stabilization (or real-time monitoring) of the orbiting spacecraft relative to an inertial reference frame (e.g., a distant set of stars and/or quasars; the approach used, e.g., on Gravity Probe B [GP-B]~\cite{Everitt:2011hp,GPBscope1,GWO20031401,Wang_2015}).

To independently measure out any possible rotational motion down to a level that would completely remove it as a potential source of concern would however require the combined asteroid angular tracking relative to the spacecraft and the spacecraft tracking relative to the inertial frame to be performed at the level of $\delta \theta \sim h_c \times L/R_{\alice} \sim 2\times 10^{-12} \times (h_c / 10^{-19})$ at $f_{\textsc{gw}} \sim 10^{-5}\,$Hz; this corresponds to an angular drift measurement of order%
%%%%%%%%%%%%
\footnote{\label{ftnt:masDefn}%
    Recall: $1 \,\text{mas} = 10^{-3}\text{arcsec} \approx 2.8\times10^{-7}\,\text{deg} \approx 4.8\times 10^{-9}\,\text{rad}$.
    } %
%%%%%%%%%%%%
$\partial_t(\delta\theta) \sim (2\pi f_{\textsc{gw}}) \delta \theta \sim 7\times 10^{-15}\,\text{deg/s} \approx 0.8\,\text{mas/yr}$.
This is at or around the level of stellar tracking accuracy that was achieved by GP-B~\cite{Everitt:2011hp}, and so would in principle be achievable here (potentially by using alternative technologies to those employed in GP-B), albeit at the cost of some additional engineering.
Given our considerations above regarding the frequency content of the rotational motion, such tracking may not be independently necessary.

Finally, because some small number of (typically, smaller) asteroids can have active surfaces where boulders and other objects could move on the surface (see, e.g., \citeR{Lauretta:2019erb}), we also estimate how large an object would need to move on an asteroid in order to perturb the rotational rate of the asteroid sufficiently so as to be a noise source for our mission concept.
We note that this estimate is provided with the understanding that (1) the asteroids chosen for this mission may be devoid of this sort of effect (indeed, it might be one criterion to use in selection), and (2) this may not even be a relevant noise source, because for an event of this nature to actually be a noise (as opposed to a rare event that could be vetoed on), it would need to occur at least as often as once per GW period.
Nevertheless, to be conservative, we estimate the maximum size of the relevant effects if this were to be a noise.

Let us assume that \Alice\ has located on its pole a spherical boulder of mass $m_{\text{b}}\ll M_{\alice}$ and radius $R_{\text{b}}$, so that $m_{\text{b}} \sim (4\pi/3)\rho_{\alice}R_{\text{b}}^3$, assuming it is comprised of the same material as \Alice.
Assume further that, owing to some unknown perturbation, this boulder is perturbed from that location and ends up a distance $r_{\text{b}}$ from the rotational axis.
The initial moment of inertia of \Alice\ plus the boulder about the rotational axis is $I_i \sim \frac{2}{5} M_{\alice}R_{\alice}^2 + \frac{2}{5} m_{\text{b}}R_{\text{b}}^2 \sim \frac{8\pi}{15} \rho_{\alice} \lb( R_{\alice}^5 + R_\text{b}^5 \rb)$.
Invoking the displaced-axis theorem, the final momenta of inertia  about the original rotational axis of \Alice\ plus the boulder will be approximately $I_f \sim \frac{2}{5} M_{\alice}R_{\alice}^2 + \frac{2}{5} m_{\text{b}}R_{\text{b}}^2 + m_{\text{b}} r_{\text{b}}^2 \sim \frac{8\pi}{15} \rho_{\alice} \lb( R_{\alice}^5 + R_{\text{b}}^5 + \frac{5}{2}  R_\text{b}^3 r_{\text{b}}^2 \rb)$.
This change in momenta of inertia  will be occasioned by a change in the rotational rate ($\omega_i \rightarrow \omega_f$) about the original axis.
Using the conservation of angular momentum about the original rotational axis, we have
%%%%%%
\begin{align}
    I_f \omega_f &= I_i \omega_i \\
    \Rightarrow \Delta \omega = \omega_f-\omega_i &\sim - \frac{5}{2} \frac{m_{\text{b}}}{M_{\alice}} \lb(\frac{r_{\text{b}}}{R_{\alice}}\rb)^2 \omega_{\alice},
\end{align}
%%%%%%
where at the last $\sim$ sign we dropped corrections suppressed by at least one more power of $m_\text{b}/M_{\alice}$.
Over a GW period, this will cause a change in the location of a reference point on the equator of the asteroid of order
%%%%%%
\begin{align}
    \Delta x \sim R_{\alice} T_{\textsc{gw}} \Delta \omega  \sim - \frac{5}{2} \frac{m_{\text{b}}}{M_{\alice}} R_{\alice}  \lb(\frac{r_{\text{b}}}{R_{\alice}}\rb)^2 \frac{2\pi T_{\textsc{gw}}}{T_{\alice}},
\end{align}
%%%%%%
leading to a strain contribution of order
%%%%%%
\begin{align}
    h_c \sim \frac{5}{2} \frac{m_{\text{b}}}{M_{\alice}} \frac{ R_{\alice} }{ L }  \lb(\frac{r_{\text{b}}}{R_{\alice}}\rb)^2 \frac{2\pi}{T_{\alice}f_{\textsc{gw}}},
\end{align}
%%%%%%
or
%%%%%%
\begin{align}
   & m_{\text{b}}\nonumber \\
   &\sim \frac{M_{\alice}}{5\pi}  \frac{L}{R_{\alice}} \lb(\frac{R_{\alice}}{r_{\text{b}}}\rb)^2 (T_{\alice}f_{\textsc{gw}})\, h_c \label{eq:mb1}\\
   &\sim 10^2\,\text{kg}\times \lb( \frac{h_c}{10^{-19}} \rb)\times\lb( \frac{f_{\textsc{gw}}}{10\,\mu\text{Hz}} \rb) \times \lb(\frac{ r_{\text{b}} }{R_{\alice}}\rb)^{-2}\\
    &\sim 10^4\,\text{kg}\times \lb( \frac{h_c}{10^{-19}} \rb)\times\lb( \frac{f_{\textsc{gw}}}{10\,\mu\text{Hz}} \rb) \times \lb(\frac{ r_{\text{b}} }{800\,\text{m}}\rb)^{-2}\\
   &\sim 3\times 10^7\,\text{kg}\times \lb( \frac{h_c}{10^{-19}} \rb)\times\lb( \frac{f_{\textsc{gw}}}{10\,\mu\text{Hz}} \rb) \times \lb(\frac{ r_{\text{b}} }{15\,\text{m}}\rb)^{-2},\label{eq:mb3}
\end{align}
%%%%%%
where the numerical estimates are given at the order of magnitude level.
A spherical boulder of density $\rho_{\alice}$ and that mass would have a radius of 
%%%%%%
\begin{align}
    &R_{\text{b}} \nonumber \\
    &\sim 20\,\text{cm} \times \lb( \frac{h_c}{10^{-19}} \rb)^{1/3} \times\lb( \frac{f_{\textsc{gw}}}{10\,\mu\text{Hz}} \rb)^{1/3} \times \lb(\frac{ r_{\text{b}} }{R_{\alice}}\rb)^{-2/3} \\
    &\sim 1\,\text{m} \times \lb( \frac{h_c}{10^{-19}} \rb)^{1/3} \times\lb( \frac{f_{\textsc{gw}}}{10\,\mu\text{Hz}} \rb)^{1/3} \times \lb(\frac{ r_{\text{b}} }{800\,\text{m}}\rb)^{-2/3}\\
    &\sim 15\,\text{m} \times \lb( \frac{h_c}{10^{-19}} \rb)^{1/3} \times\lb( \frac{f_{\textsc{gw}}}{10\,\mu\text{Hz}} \rb)^{1/3} \times \lb(\frac{ r_{\text{b}} }{15\,\text{m}}\rb)^{-2/3}.
\end{align}
%%%%%%

Even in the worst case here (smallest boulder moving the full radius of the asteroid, which is unlikely to occur), such a boulder is likely large enough that it could be visually imaged by an asteroid-orbiting spacecraft: see, e.g., \citeR{rs13071315}, which reports characterization of surface features of 101955~Bennu globally with a pixel resolution of 42\,cm using images taken during asteroid approach, and locally in targeted regions with a pixel resolution of 1\,cm using images taken during the orbital phase (earlier work reports complete surface feature characterization of the same asteroid down to objects at the 8\,m scale, with imaging pixel resolution as small as 33\,cm~\cite{10.1038/s41550-019-0731-1}).
Consider also that in order to obtain $\delta \sim 20\,$cm imaging resolution at near-UV optical wavelengths ($\lambda \sim 400\,\text{nm}$) from a distance of $\ell \sim 40\,\text{km}$ (rough surface elevation of the NEAR--Shoemaker satellite during the period for which it was in a 50\,km radius orbit around 433~Eros, as measured from the CoM of the latter~\cite{NearShoemakerWebsite,Yeomans:2000eth,MILLER20023}), an imaging system with a telescope of diameter $D \sim 1.22 \lambda  \ell / \delta \sim 10\,\text{cm}$ suffices to exceed the diffraction limit.
It therefore seems likely that it would be possible to know whether, within any GW period, such a large surface disturbance had occurred (assuming that the surface can be imaged completely in this time); one could thus veto such a period, or otherwise attempt to account for or model the effects of the change.
Additionally, motion of an object of this type on an asteroid surface is likely to give rise to seismic disturbances, which could also be actively be monitored for; see comments in \sectref{sect:exploratory}.

Additionally, this movement of the boulder will induce a small change in the orientation of the rotational axis.
Under the assumptions here, it is a relatively straightforward exercise in rigid-body kinematics to demonstrate that, as long as $T_{\textsc{gw}}\ll T_{\alice}(M_{\alice}/m_\text{b})$ and $R_\text{b} \ll R_{\alice}$, the angle by which the axis will shift in a time $T_{\textsc{gw}}$ is of order%
%%%%%%%%%%%%
\footnote{\label{ftnt:periodicMotionActually}%
    The angular motion is of course actually periodic over longer timescales, with angular frequency $\Omega \sim \omega_{\alice} (5/4) (m_{\text{b}}/M_{\alice})[ 1 - (r_{\text{b}}/R_{\alice})^2 ]$.
    Of course, the frequency content of that motion is far out of band if $m_{\text{b}} \ll (f_{\textsc{gw}}T_{\alice}) M_{\alice}$.
    } %
%%%%%%%%%%%%
$\delta \theta \sim (5/2)(m_\text{b}/M_{\alice}) (r_{\text{b}}/R_{\alice})^2[ 1 - (r_{\text{b}}/R_{\alice})^2 ] \omega_{\alice} T_{\textsc{gw}}$.
Such a motion, if repeated due to stochastically occurring boulder-movement events occurring roughly once per GW period, would give a strain noise of order
%%%%%%
\begin{align}
    h_c &\sim \frac{R_{\alice} \delta \theta }{L}\\
    &\sim \frac{ m_\text{b}}{M_{\alice}} \frac{r_\text{b}}{R_{\alice}} \frac{ r_{\text{b}}}{ L} \lb[ 1 - \lb(\frac{r_{\text{b}}}{R_{\alice}}\rb)^2 \rb] \frac{5\pi}{T_{\alice}f_{\textsc{gw}}} \\
    &\sim  \begin{cases}
            \dfrac{ m_\text{b}}{M_{\alice}} \dfrac{r_\text{b}}{R_{\alice}} \dfrac{ r_{\text{b}}}{ L} \dfrac{5\pi}{T_{\alice}f_{\textsc{gw}}}; & r_{\text{b}} \ll R_{\alice} \\[4ex]
            \dfrac{1}{4} \dfrac{ m_\text{b}}{M_{\alice}} \dfrac{ R_{\alice} }{ L} \dfrac{5\pi}{T_{\alice}f_{\textsc{gw}}}; & r_{\text{b}} \sim R_{\alice}/\sqrt{2}
            \end{cases},
\end{align}
%%%%%%
where in the latter estimate we took the worst-case result.
Perhaps unsurprisingly, in either case, this is parametrically a shift of the same size of those given previously for the change in the rotational rate at \eqrefRange{eq:mb1}{eq:mb3}, and similar comments apply.

Moreover, because the motion of a boulder on the surface by a distance $\sim r_{\text{b}}$ is an internal change to the asteroid plus boulder system, the overall CoM position does not change.
However, there will be a shift to the position of the CoM of the rigid (i.e., non-boulder) part of the asteroid to which the base station or retroreflectors are rigidly attached, relative to the overall CoM of the asteroid plus boulder system.
This effect is of order $\Delta x \sim (m_{\text{b}}/M_{\alice}) r_{\text{b}}$, which would give a strain contribution
%%%%%%
\begin{align}
   h_c \sim \frac{m_{\text{b}}}{M_{\alice}} \frac{ r_{\text{b} } }{ L },
\end{align}
%%%%%%
leading to 
%%%%%%
\begin{align}
  m_{\text{b}}
  &\sim h_c \frac{L}{r_{\text{b}}} {M_{\alice}}\\
    &\sim 10^4\,\text{kg} \times \lb( \frac{ h_c }{10^{-19}} \rb) \times \lb( \frac{ r_{\text{b}} }{ R_{\alice} } \rb)^{-1}\\
    &\sim 10^5\,\text{kg} \times \lb( \frac{ h_c }{10^{-19}} \rb) \times \lb( \frac{ r_{\text{b}} }{ 800\,\text{m} } \rb)^{-1}\\
    &\sim 5\times 10^6\,\text{kg} \times \lb( \frac{ h_c }{10^{-19}} \rb) \times \lb( \frac{ r_{\text{b}} }{ 15\,\text{m} } \rb)^{-1},
\end{align}
%%%%%%
with corresponding boulder radii
%%%%%%
\begin{align}
  R_{\text{b}}
    &\sim 1\,\text{m} \times \lb( \frac{ h_c }{10^{-19}} \rb)^{1/3} \times \lb( \frac{ r_{\text{b}} }{ R_{\alice} } \rb)^{-1/3}\\
    &\sim 2\,\text{m} \times \lb( \frac{ h_c }{10^{-19}} \rb)^{1/3} \times \lb( \frac{ r_{\text{b}} }{ 800\,\text{m} } \rb)^{-1/3}\\
    &\sim 8\,\text{m} \times \lb( \frac{ h_c }{10^{-19}} \rb)^{1/3} \times \lb( \frac{ r_{\text{b}} }{ 15\,\text{m} } \rb)^{-1/3}.
\end{align}
%%%%%%
Imaging of the surface to $\sim 1\,$m resolution would thus suffice to notice this change to the asteroid surface.

None of these potential surface-boulder movement effects would appear to be a seriously problematic noise source.

%%%%%%%%%%%%%%%%%%%%%%%%%%%%%%%%%%%%%%%%%%%%%%%%%%%%%%
\subsection{Excitation of internal degrees of freedom}
\label{sect:internalDOF}
%%%%%%%%%%%%%%%%%%%%%%%%%%%%%%%%%%%%%%%%%%%%%%%%%%%%%%
Asteroids, being extended bodies, have internal degrees of freedom which can contribute as noise sources for the strain measurement by virtue of the fact that our mission concept uses reference points on the asteroid surface in order to infer the CoM position of the asteroid TM.
We estimate and/or discuss a number of these noise sources in this section: \hyperref[sect:gravGradientStretching]{(1)} tidal stretching of the asteroid TMs, \hyperref[sect:thermal]{(2)} thermal expansion and cycling, and \hyperref[sect:seismics]{(3)} seismic noise.

%%%%%%%%%%%%%%%%%%%%%%%%%%%%%%%%%%%%%%%%%%%%%%%%%%%%%%
\subsubsection{Stretching by external gravity gradients}
\label{sect:gravGradientStretching}
%%%%%%%%%%%%%%%%%%%%%%%%%%%%%%%%%%%%%%%%%%%%%%%%%%%%%%
Asteroids will be stretched by gravitational tidal forces acting across their diameter, which arise from other massive bodies in the Solar System.
This could in principle constitute a noise source, but it is relatively easy to show that the amplitude of any such stretching is bounded to be small enough at the strain levels of interest.

Consider the tidal gravitational force between the CoM of \Alice\ and a point on its equator: $\Delta F \sim 2G_{\text{N}} M_{\alice} M_{\text{source}} R_{\alice} / d^3$, where $d$ is the distance from the source of the gravitational force to \Alice\ and $M_{\text{source}}$ is the mass of the source.
This applies a bulk stress on the asteroid material of order $\sigma \sim \Delta F / ( \pi R_{\alice} ^2)$, which will give rise to a stretching of the distance from the CoM to the point on the equator of order $\Delta x \sim \sigma R_{\alice}/E_{\alice}$, where $E_{\alice}$ is Young's modulus for the asteroid material (of course, in general, the stress and strain are related by a tensor modulus, but we are interested here only in the magnitude of the stretching, so we ignore such subtleties).
While there is some uncertainty as to what one should assume for $E_{\alice}$ given that not all asteroids are solid monolithic rock objects, we will give an estimate assuming that $10\,\text{GPa} \lesssim E_{\alice} \lesssim 100\,\text{GPa}$, which brackets the typical range of solid rock in Earth samples~\cite{YoungsModulus},
and is consistent with the (reduced%
%%%%%%%%%%%%
\footnote{\label{ftnt:Er}%
    The reduced Young's modulus $E_r$ is obtained by measuring the penetration depth under a known applied load of a probe tip into the surface of the sample of the material~\cite{Tanbakouei:2019efg}. 
    It is related~\cite{Tanbakouei:2019efg} to the Young's moduli $E_{(i)}$ of the sample and probe by the relationship $E_r^{-1} = \sum_i \lb[ E_{(i)} / (1-\nu_{(i)}^2) \rb]^{-1} $ where $(i)=\text{probe},\ \text{sample}$ and $\nu_{(i)}$ is the Poisson ratio of the relevant material (this measures the tendency of that material to expand or contract in the dimensions perpendicular to an applied stress).
    Since probe tips are typically taken to a material such as diamond (e.g., ~\citeR{Tanbakouei:2019efg}) which has $E_{\text{probe}} \gg E_{\text{sample}}$, we can approximate $E_r = E_{\text{sample}} / (1-\nu_{\text{sample}}^2)$.
    Moreover, at the level of accuracy to which we are working here, we then employ $E$ and $E_r$ interchangeably, since $\nu \sim 0.2$--$0.3$ is typical for many types of rock~\cite{GERCEK20071}, leading to only modest ($\lesssim 10\%$) differences between $E$ and $E_r$.
    }%
%%%%%%%%%%%%
) Young's moduli measured in small samples of material returned from 25143~Itokawa by the Hayabusa mission ($82\,\text{GPa}\lesssim E_r \lesssim 111\,\text{GPa}$)~\cite{Tanbakouei:2019efg} (although we note of course that it does not necessarily follow that the modulus for the asteroid as a whole is similar to that of small samples), and with the calculated Young's modulus of the Chelyabinsk meteorite ($71\,\text{GPa}\lesssim E_r \lesssim 136\,$GPa)~\cite{2017ApJ...835..157M} and measured Young's moduli of chrondrite meteorites that have impacted Earth ($9.5\,\text{GPa}\lesssim E \lesssim 138\,$GPa)~\cite{Kiyoshi:1983saa}.

This stretching of the asteroid gives rise to a GW strain noise contribution (taken across the baseline $L$) of order
%%%%%%
\begin{align}
    h_c \sim \frac{8}{3} \frac{G_{\text{N}} \rho_{\alice} R_{\alice}^3 M_{\text{source}} }{  d^3 E_{\alice} L  }.
\end{align}
%%%%%%
For the Sun as the source we take $d\sim 1.5\,\text{AU}$ as the typical \Alice\ heliocentric radius, and obtain $3\times 10^{-21}\lesssim h_c\lesssim 3\times 10^{-20}$ for $100\,\text{GPa} \gtrsim E_{\alice} \gtrsim 10\,\text{GPa}$, which is small enough.
At perihelion, 433~Eros has $d \sim 1.2\,\text{AU}$~\cite{JPL-SBD}, which would increase these estimates by a factor of $\sim 2$, but they are still safe.

Jupiter is too far from the inner Solar System ($\mathfrak{a}_{\text{Jupiter}}\sim 5.2\,\text{AU}$ and  $\mathfrak{e}_{\text{Jupiter}} \sim 0.048$~\cite{JPL-SBD}) to ever dominate over the solar tidal force for inner Solar System asteroids, but close encounters with other major bodies could in principle dominate; this would however need to checked on an asteroid-by-asteroid basis.
For instance, 433~Eros has a closest possible Earth-approach distance of $d\sim0.15\,$AU~\cite{JPL-SBD}, but $M_{\odot}/(1.5\,\text{AU})^3 \sim 1.8\times 10^{-4}\,\text{kg/m}^3$, while $M_{\oplus}/(0.15\,\text{AU})^3 \sim 5.3\times 10^{-7}\,\text{kg/m}^3$, so this will cause a smaller degree of stretching than the solar tidal force.

Note also that this strain contribution is driven by an external force that changes on orbital timescales, but this low-frequency driving force could also be modulated up to the asteroid rotational frequency and appear as a side-band around the latter; for the relevant asteroids, it is thus both small enough and likely out of band for GW searches in the $\mu$Hz band.

%%%%%%%%%%%%%%%%%%%%%%%%%%%%%%%%%%%%%%%%%%%%%%%%%%%%%%
\subsubsection{Thermal expansion and cycling}
\label{sect:thermal}
%%%%%%%%%%%%%%%%%%%%%%%%%%%%%%%%%%%%%%%%%%%%%%%%%%%%%%
Another large effect that the solar radiation flux has on an asteroid is of course to heat the surface.
For instance, 2867~Steins has day-side asteroid surface temperatures as high as $T\sim 250\,$K, and night-side temperatures that fall below $T\lesssim 100\,$K~\cite{2011A&A...531A.168L}.
As another example, around perihelion, 101955~Bennu can have a surface temperature as high as $T\sim390\,$K at certain points on its surface, with a day-night temperature variation $\Delta T \sim 140\,$K~\cite{Rozitis:2020wga}.

This heating and cooling of course causes reversible thermal expansion of the asteroid surface layers, as well as possible irreversible noise sources owing to cracking or surface material slippage.
This will exhibit strong frequency content at and above the rotational frequencies of the asteroid, which is another good reason to select asteroids whose rotational frequencies lie above our band of interest.

There is however also a more subtle long-term effect that can occur.
Because the maximum day-side surface temperature of the asteroid is set by the balance of incoming radiation flux against Stephan--Boltzmann re-radiation from the surface,
%%%%%%
\begin{align}
    I_{\odot}(r_\alice)\sim T_{\alice}^4,
\end{align}
%%%%%%
a periodic fluctuation in the solar radiation flux will be occasioned by a fluctuation in the largest possible day-side surface temperature of the asteroid at the same period. 
At linear order, this is estimated as
%%%%%%
\begin{align}
    \frac{ \Delta I_{\odot}(r_\alice,f) }{ \bar{I}_{\odot}(r_\alice) } \sim 4 \frac{\Delta T_{\alice}(f) } { \bar{T}_{\alice}},
    \label{eq:PowerEqualsTempFluctuation}
\end{align}
%%%%%%
where $\bar{T}_{\alice} \sim 300\,$K is a conservatively high dayside surface temperature estimate for the asteroid, and $\Delta I_{\odot}(r_\alice,f) = \bar{I}_{\odot}(r_\alice) \delta I_{\odot}(f)$ is the fluctuation in the solar TSI at the location of \Alice.

This extra heating and cooling of the asteroid surface will cause the surface layers of the asteroid to expand or contract thermally, and because the solar power fluctuates stochastically, this fluctuation will be stochastic.
Because the base station (or reference point) is located on the asteroid surface, this will constitute an additional noise source for the strain measurement.

We can however bound the possible magnitude of this additional noise.
Let us assume that the largest possible fluctuation in the day-side surface temperature of the asteroid given at \eqref{eq:PowerEqualsTempFluctuation} is actually realized.
However, this is a \emph{boundary condition} fluctuation: the temperature of the entire bulk of the asteroid is of course not fluctuating by this amount.
Heat penetration into the asteroid is governed by the heat equation, and the characteristic depth of penetration $d_{\text{therm}}$ can be estimated parametrically from the heat equation as 
%%%%%%
\begin{align}
    d_{\text{therm}} \sim \sqrt{\kappa_{\text{therm}} \tau},
\end{align}
%%%%%%
where $\tau$ is the characteristic timescale of interest and $\kappa_{\text{therm}}$ is the thermal diffusivity of the asteroid material.
We will take $\tau \sim 1/f_{\textsc{gw}}$ to allow the maximum depth of heat penetration that can occur within a GW period; in this case, we have
%%%%%%
\begin{align}
    d_{\text{therm}} &\sim \sqrt{\frac{\kappa_{\text{therm}}}{ f_{\textsc{gw}}}} \\
    &\sim 1\,\text{m} \times \lb( \frac{ \mu\text{Hz} }{ f_{\textsc{gw}} } \rb)^{1/2} \times \lb( \frac{\kappa_{\text{therm}} }{ 10^{-2}\,\text{cm}^2/\text{s} } \rb)^{1/2},
    \label{eq:dTherm}
\end{align}
%%%%%%
where we have taken $\kappa_{\text{therm}} \sim 10^{-2}\,\text{cm}^2/\text{s}$ as a typical order of magnitude of solid-rock thermal diffusivity (see Table 13 of \citeR{PropertiesOfRocks}), which is a conservative estimate given that the surface layers of bodies such as asteroids are typically not bare solid rock, or are brecciated by historic surface impacts (e.g., \citeR[s]{LunarSeismicReport,rs13071315,10.1038/s41550-019-0731-1,TSUDA202042}).
Note that the estimate at \eqref{eq:dTherm} is validated by \emph{in situ} thermal-flux measurements made on the Moon during the Apollo 17 mission~\cite{HeatFlowExperiment}: a borehole drilled into the Moon and instrumented with temperature sensors indicated negligible day-night periodic temperature variation below $\sim 0.5\,$m from the surface~\cite{HeatFlowExperiment,Dunnebier:1974srg} [in fact, the estimate at \eqref{eq:dTherm} is conservatively large on the basis of those data, but of course this depends on the details of the surface layers].

Let us then take a model in which the upper layer of thickness $d_{\text{therm}}$ undergoes a temperature excursion $\Delta T_{\alice}(f)$ given by \eqref{eq:PowerEqualsTempFluctuation}.
We assume that the absolute vertical expansion of the surface rock layer $\Delta y$ can be obtained as $\Delta y / d_{\text{therm}}  \sim (1/3) (\Delta V / V)$, where $\Delta V = k_{\text{therm}} V \Delta T$ is the thermal volumetric expansion of a volume of rock $V$ with $k_{\text{therm}}$ the volumetric thermal expansion coefficient of the rock in question.
Typical solid-rock volumetric thermal expansion coefficients for material on Earth are of the order of $k_{\text{therm}} \sim 3\times 10^{-5}\,\text{K}^{-1}$ (see Tables 9 and 10 of \citeR{PropertiesOfRocks}); this number is again likely an overestimate if the surface layers of an asteroid are not solid bare rock.
This leads to the estimate for the amplitude of this vertical motion:
%%%%%%
\begin{align}
    \Delta y(f_{\textsc{gw}}) \sim \frac{1}{12} k_{\text{therm}} T_{\alice} d_{\text{therm}} \delta I_{\odot}(f_{\textsc{gw}}) .
\end{align}
%%%%%%

Up to an $\order{1}$ numerical factor accounting for geometrical baseline-projection effects and possible equal-sized noise from each end of the baseline that we ignore here, the typical size of the characteristic strain noise across a baseline $L$ will thus be
%%%%%%
\begin{align}
    h_c &\sim \frac{k_{\text{therm}} \bar{T}_{\alice} d_{\text{therm}}}{12L}  \sqrt{ f_{\textsc{gw}} S[\delta I_{\odot}](f_{\textsc{gw}})} \\[1ex]
    &\sim \frac{k_{\text{therm}} \bar{T}_{\alice} }{12L} \sqrt{ \kappa_{\text{therm}} S[\delta I_{\odot}](f_{\textsc{gw}})},
\end{align}
%%%%%%
where $S[\delta I_{\odot}](f)$ is again the fractional TSI fluctuation PSD; see \sectref{sect:TSI}.

This estimate is shown in \subfigref{fig:noise_contributions}{c}, for the same variety of fractional TSI fluctuation PSDs~\cite{Froehlich:2004wed} discussed in \sectref{sect:TSI};
using the same numbering as in that section to identify the TSI PSDs, the lines are shown in (1) magenta, (2) darker pink, and (3) lighter pink.
As discussed above, we have assumed in drawing these curves that $\bar{T}_{\alice} \sim 300\,$K, $\kappa_{\text{therm}} \sim 10^{-2}\,\text{cm}^2/\text{s}$ and $k_{\text{therm}} \sim 3\times 10^{-5}\,\text{K}^{-1}$, all of which should be conservatively large values; we have taken the other parameter values to be those of \Alice, where necessary.
Even this conservatively large estimate for this noise source is subdominant, except right around $f_{\textsc{gw}}\sim 10\,\mu$Hz, where it is of the same magnitude as other noise sources we have estimated.

Note that we have assumed in making these estimates that only the expansion in the vertical direction is relevant: expansion in the horizontal directions is of course constrained by mechanical forces applied by the neighboring rock, and will typically lead to stressing of the rock without it necessarily expanding.
This may lead to seismic events; see the next section.

%%%%%%%%%%%%%%%%%%%%%%%%%%%%%%%%%%%%%%%%%%%%%%%%%%%%%%
%%%%%%%%%%%%%%%%%%%%%%%%%%%%%%%%%%%%%%%%%%%%%%%%%%%%%%
\subsubsection{Seismic noise}
\label{sect:seismics}
%%%%%%%%%%%%%%%%%%%%%%%%%%%%%%%%%%%%%%%%%%%%%%%%%%%%%%
%%%%%%%%%%%%%%%%%%%%%%%%%%%%%%%%%%%%%%%%%%%%%%%%%%%%%%
Direct knowledge of seismic activity on asteroids is presently non-existent since seismometers were not present on any of the missions that have hitherto landed on asteroids~\cite{Yeomans2085,NearShoemakerWebsite,Baker1327,HayabusaWebsite,TSUDA202042,Hayabusa2Website,Lauretta:2017erq,Smith672,OsirisRex,OririsRexWebsite1,OririsRexWebsite2,Bibring493,Taylor387,RosettaWebsite}. 
To estimate the level of seismic activity on an asteroid, we use measurements that were made on the Moon~\cite{1975LPSC....6.2863C,Dunnebier:1974nox,Dunnebier:1974srg,1981JGR....86.5061G,Apollo15PSR,Lammlein:1974dbf,Nunn:2020bf,LunarSeismicReport,10.2307/74874,Nakamura:2005dfa,DIMECH20174457,NAKAMURA2003197,LAMMLEIN1977224,1982LPSC...13..117N,2006AGUFM.U44B..06N} and Mars~\cite{INSIGHT:2020sdg} to infer the sources of such activity on ancient rock such as asteroids that have lost their heat of formation. 
These estimates are likely most applicable to stony asteroids that are rigid, as opposed to loosely bound rubble piles that are likely to be far more susceptible to seismic activity. 

Analysis of lunar seismology shows that the primary sources of seismic activity on the Moon are 
(a) `deep moonquakes'~\cite{Lammlein:1974dbf,LAMMLEIN1977224,NAKAMURA2003197} originating from well-identified hypocenters that appear to lie on global lunar fractures at depths of 800--1000\,km, that are triggered by tidal strains associated with the $\sim 27$--$28$\,day librational and orbital periods of the Earth--Moon system (with longer-period effect arising from solar gravitational perturbations); 
(b) `shallow moonquakes'~\cite{LAMMLEIN1977224,1979LPSC...10.2299N,1980LPSC...11.1847N,1982LPSC...13..117N,Watters:2019erh}%
%%%%%%%%%%%%
\footnote{\label{ftnt:akaHFT}%
    Also originally known as `high-frequency teleseismic' (HFT) events~\cite{LAMMLEIN1977224}.
    } %
%%%%%%%%%%%%
that are much rarer than the deep moonquakes but appear to originate at depths of 0--300\,km in locations correlated with the fractures identified by the deep quakes, and which may be tectonic%
%%%%%%%%%%%%
\footnote{\label{ftnt:tectonic}%
    `Tectonic' here means that they would occur as a result of the secular accumulation of tidal strain over time in fault structures in the lunar material, with the accumulated strain being released by some (possibly tidal) trigger.
    } %
%%%%%%%%%%%%
in origin as they are too large to be explained by tidal strains (although they are potentially triggered by tidal compressional stress~\cite{Watters:2019erh}); (c) `thermal moonquakes'~\cite{Dunnebier:1974srg,1975LPSC....6.2863C} induced by the extreme differential heating of the lunar surface associated with the 29.5\,day synodic period (i.e., lunar day) that are likely sourced by thermal-stress-induced soil slumping on lunar slopes (although other explanations might be possible~\cite{1974LPI.....5..151C}); and (d) impacts (both local and distant) on the lunar surface~\cite{Dunnebier:1974nox}.%
%%%%%%%%%%%%
\footnote{\label{ftnt:LMEvents}%
    So-called `LM events' (where LM stands for `Lunar Module') were also identified as a clearly distinct class of seismic events in Apollo data: these are also thermally induced, but arise from the rapid heating and cooling of the LM at lunar sunrise/set~\cite{Dunnebier:1974srg}.
    These are obviously specific to the human-engineered hardware deployed during Apollo.
    } %
%%%%%%%%%%%%

In the following, where possible, we estimate the size of these phenomena on an asteroid and find that the expected amplitude of such quakes is less than the required $\sim 0.1 \, \mu \text{m}$ required for this mission. 

The deep moonquakes are caused by tidal strain (and also depend in some detail on the lunar internal structure).
Tidal energy is however a strong function of the size of the asteroid and its distance to the object causing the tide. 
While significant for the Moon, for a $\sim 10$\,km asteroid that is $\sim 1$\,AU away from perturbing bodies, we find that the tidal energy on the asteroid is orders of magnitude smaller than the differential thermal energy available for thermal quakes on the asteroid. 
Thus, we expect thermal quakes to dominate over quakes caused by tidal strain on asteroids.
The estimate is as follows (see also \sectref{sect:gravGradientStretching}): the tidal force $F_T$ on an asteroid of density $\rho_{\alice}$ and radius $R_{\alice}$ at a distance $d_p$ from a planet of mass $M_p$ is, parametrically, $F_T \sim (G_{\text{N}} M_p \rho_{\alice}R_{\alice}^3/d_p^2) \times  \left(R_{\alice}/d_p\right)$. 
This tidal force causes a strain on the asteroid causing its overall size to change, parametrically, by $\delta r \sim  ( G_{\text{N}} M_p R_{\alice}^3 \rho_{\alice})/(d_p^3 E_{\alice})$ where $E_{\alice}$ is the Young's modulus of the asteroid. 
This direct stretching effect was considered in \sectref{sect:gravGradientStretching} and found to be both small enough in amplitude and likely out of band.
Nevertheless, a quake could be induced by this tidal strain if the stored energy due to the tide is released due to non-linear effects in the rock.
Parametrically, the energy stored in the tide is $\sim F_T \delta r \sim (G_{\text{N}}^2 M_p^2 R_{\alice}^7 \rho_{\alice}^2)/(d_p^6 E_{\alice})$. 
For our fiducial numbers, the stored tidal energy, arising from any object in the Solar System, in the entire asteroid is $\lesssim\text{mJ}$, with the largest energy arising from the tidal action of the Sun. 
This energy is considerably smaller than the heat deposited by the Sun on the asteroid due to the irradiance of $\sim \text{kW}/\text{m}^2$. 

The amplitude of thermal quakes can be estimated using lunar data. 
The measured amplitude of thermal lunar quakes range between $\sim 0.3$--$3$\,nm~\cite{Dunnebier:1974srg,1975LPSC....6.2863C}%
%%%%%%%%%%%%
\footnote{\label{ftnt:amplitudes}%
    Note that care must be taken in reading, e.g., \citeR{1975LPSC....6.2863C}: quakes are often referred to in that reference and some others of the same era by their amplitude \emph{as recorded on a compressed seismogram} (i.e., the size of the `needle' deflection on a trace), and not by their ground-movement amplitude, the latter of which is the physically relevant displacement. 
    The former is typically $\mathcal{O}(\text{mm})$, while the latter is $\mathcal{O}(\text{nm})$ for thermal moonquakes.
    } %
%%%%%%%%%%%%
(similar amplitudes were measured for quakes on Mars~\cite{INSIGHT:2020sdg}), smaller than the $\sim 0.1 \, \mu$m stability required for this mission. 
It is likely that thermal quakes on an asteroid will have a smaller amplitude than the thermal quakes on the Moon for a number of reasons. 
First, the lunar day is longer than the asteroid day by a factor of $\sim$ 100, permitting a larger energy to be stored in thermal stresses since the incident thermal power is approximately the same. 
Second, since the size of the asteroid $\sim 10$\,km is also a factor of $\sim 100$ smaller than the size of the Moon, the fundamental oscillation frequency of the asteroid is larger than that of the Moon by a factor of $\sim$ 100~\cite{WALKER2006142}. 
This larger oscillation frequency implies that, for a fixed energy, the amplitude of oscillations on the asteroid should likely be smaller than the amplitude of oscillations on the Moon. 
Third, it is likely that the amplitudes of such quakes on an asteroid will be lower than the corresponding quakes on the Moon because such quakes on the Moon are triggered by soil slippage in the lunar gravitational potential.
Since the gravitational acceleration on a $\sim 10$\,km asteroid is about a factor of $\sim 100$ smaller than the gravitational acceleration on the Moon, the gravitational potential energy available for triggering a quake on the asteroid due to soil/rock slipping is smaller than that available on the Moon.
Since the measured amplitudes on the Moon were already smaller than the amplitudes required for this mission, it is likely that thermal quakes are not an issue for the proposed measurement.
Nevertheless, without an \emph{in situ} measurement, the exact size of these quakes is difficult to robustly determine, especially since their intensity likely depends in detail on the properties of the asteroid.

This latter point is also especially true with regard to any sources of seismic activity on the asteroid that are cognate to the shallow moonquakes, the largest of the lunar seismic events~\cite{LAMMLEIN1977224,1979LPSC...10.2299N,1980LPSC...11.1847N,1982LPSC...13..117N,Watters:2019erh}.
On the Moon, these events appear to be essentially `tectonic' in origin: although the Moon lacks large-scale plate motion, they result from a secular accumulation of strain in the weak points or fractures in the rock that is then released suddenly by some trigger.
These events are very large on the Moon: they are estimated to be magnitude 4--5 on the Richter scale, compared to the $\lesssim 1.5$ magnitude deep quakes~\cite{LAMMLEIN1977224}.
There are however also very rare: only 28 such events were recorded on the Moon from 1969 to 1977~\cite{1982LPSC...13..117N} (and recent analyses indicate that they appear to be preferentially temporally clustered around times of peak compressional tidal stress~\cite{Watters:2019erh}).
Already, the average inter-quake period for shallow quakes on the Moon is $\sim 10^{7}\,\text{s} \gtrsim 1/f_{\textsc{gw}}$ for our GW band of interest, and we have no reason to suspect that such events would be more frequent on $\sim 10$\,km class asteroids (indeed, they may even be rarer if the size of the strained fault region cannot be as large as on the Moon, which is almost certainly the case).
Because large quakes would be measurable on the asteroid and, because they would presumably be as rare or rarer than on the Moon, one could simply analyze asteroid ranging data over disjoint temporal durations that do not bridge such large quake occurrences on either asteroid.

Moreover, the available tidal energy for stressing an asteroid is quite small; the available thermal energy that could stress rock in this fashion is of course larger than the available tidal energy, but thermal cycling can only differentially heat and cool at most the upper $\sim 10$--$30\,\text{m}$ of the asteroid rock over orbital timescales $\sim 10^8$--$10^9$\,s on which the solar irradiance (and hence equilibrium surface temperature) might vary by a reasonable fraction (and a much smaller depth is differentially heated and cooled on the rotational period).
It is difficult to see how this could lead to a significant secular strain accumulation in any of the larger fractures an asteroid may possess, which are presumably the ones that would need to build up large strain to lead to large seismic events.
Moreover, residual strain in the asteroid rock due to cooling of the asteroid to its thermal steady state has had a longer time to relax than on the Moon: 10\,km--class asteroids would have cooled anciently to their thermal steady states, whereas the Moon lost its heat of formation and contracted relatively more recently~\cite{Watters:2010etq}.
And finally, because the shallow moonquakes were found to occur on the Moon close to the surface, but not deeper inside the Moon, their origin appears intimately tied to the specific lunar internal structure; whether or not asteroids have similar-enough internal structure to support similar events is unknown.
For all these reasons, we would not expect tectonic-like events similar in origin to the shallow moonquakes to be a problem, but this is clearly a detailed question requiring \emph{in situ} measurement on specific target asteroids.

A key limitation of the above analysis is the fact that seismometry of the Moon was not performed at $\mu$Hz frequencies; the seismometers deployed during the various Apollo missions had peak sensitivities at frequencies~$\gtrsim 0.1\,\text{Hz}$~\cite{Lammlein:1974dbf,LunarSeismicReport}.
It is thus possible that there might be larger displacements on the asteroid at lower frequencies than implied by the lunar measurement data.
Seismic activity at these low frequencies cannot be described as conventional seismic waves since the fundamental mode of the asteroid is $\sim$ 100 mHz. 
Instead, such activity would correspond to slow plastic deformations and relaxation of the asteroid. 
Since the diurnal thermal stresses on the asteroid are at $\sim 0.1$\,mHz, there do not appear to be significant sources of noise that would drive such plastic deformations at $\sim \mu$Hz frequencies. 
It is however difficult to robustly estimate these effects, especially since they may depend upon the details of the asteroid. 
For example, a stony asteroid likely has smaller plastic deformations than a rubble pile. 
For all these reasons, it is important to perform \emph{in situ}  measurements of such low frequency seismic activity, as we discuss in \sectref{sect:exploratory}. 

The final source of seismic activity on the Moon is impacts on the Moon from meteorites. 
We can estimate that any collisional excitation of normal modes of an asteroid such as 433~Eros appears to be safe. 
Explicit finite element modeling of the response of 433~Eros to a surface explosion has been performed in \citeR{WALKER2006142}, motivated by the opportunities that seismic monitoring of artificial explosions hold for understanding asteroid internal structure (see generally \citeR[s]{Plescia:2017wrt,Plescia:2018wfb,Plescia:2016sfg,Schmerr:2018rtb,Walker:2006bqw,MURDOCH201789}).
In \citeR{WALKER2006142}, read in conjunction with the results of \citeR{WALKER20041564}, it was shown that if a 1\,kg charge of Composition C-4 explosive were detonated on the surface of 433~Eros, global seismic waves with an acceleration amplitude of $|\Delta a| \sim 10^{-10}\,\text{m/s}^2$ would be induced.
Although the results of \citeR{WALKER2006142} seem to indicate that this would ring up normal modes of the asteroid in the $f\sim 0.1\,\text{Hz}$ range, the lowest normal modes lie closer to $f_0\sim0.01\,\text{Hz}$ and we assume that all the ringing goes into such low-frequency modes as a worst-case scenario: this then leads to an estimate of the amplitude of the global seismic motion of $|\Delta r| \sim |\Delta a|/(2\pi f_0)^2 \sim 3\times 10^{-8}\,$m.
On a $L\sim 1\,$AU baseline, this would constitute a strain noise of $h_c \sim 2\times 10^{-19}$, which is just about safe at any frequency of interest to us.

It remains to estimate how this artificial explosive impulse translates to an impact on the asteroid surface.
The initial condition used in \citeR{WALKER2006142} for the downward momentum of the surface area under the 1\,kg C-4 explosion site was $p_z \sim 10^8\,\text{g\,cm/s}$; see also \citeR{WALKER20041564}.
The results of \citeR{WALKER20041564} indicate that a 100\,g copper impactor striking an asteroid with a speed somewhere between 5 and 7.5\,\text{km/s} would impart a similar downward momentum impulse to the asteroid surface.
Note that this makes sense on energetics grounds, since C-4 has an explosive energy content of approximately 6\,MJ/kg~\cite{ArmyExplosivesManual},%
%%%%%%%%%%%%
\footnote{\label{ftnt:kilocalJoule}%
    \citeR{ArmyExplosivesManual} quotes values of 1.59 and 1.40\,kcal/g for the heat of detonation of C-4 with liquid and gaseous water, respectively. 
    Recall also that $1\,\text{kcal} \equiv 4 184\,$J.
    } %
%%%%%%%%%%%%
while a 0.1\,kg impactor moving at 5\,km/s carries a kinetic energy of 1.3\,MJ; naturally, the directed motion of an impactor into an asteroid surface will be more efficient in causing a given downward surface perturbation of the asteroid than an unshaped surface charge explosion, so the fact that the impactor energy is slightly smaller than the explosive energy content makes sense on physical grounds.
It also makes sense on momentum-conservation grounds since a 100\,g impactor at 5\,km/s carries a linear momentum $p \sim 5\times 10^7\,\text{g cm/s}$. 

However, a 100\,g object moving at 5\,km/s does not necessarily match well with the fluxes of the kinds of objects in the inner Solar System, which are typically smaller and moving faster.
If we were instead to assume that a smaller impactor moving at a higher speed imparts the same downward momentum perturbation to the asteroid surface, then we can match onto a more realistic flux.
A 3\,g object moving at 30\,km/s carries approximately the same kinetic energy%
%%%%%%%%%%%%
\footnote{\label{ftnt:EorP}%
    It can be argued that one should find the smaller object that carries the same momentum at the higher speed, but this leads to a larger mass object ($\sim 17\,$g), which implies rarer impacts; assuming that the energy matches is thus a more conservative estimate for the present purposes.
    } %
%%%%%%%%%%%%
as a 100\,g object moving at 5\,km/s, so we will adopt a 3\,g object as the scale of object that would, with a high impact speed on the asteroid surface, cause global seismic perturbations at an amplitude at the borderline of being an extra non-negligible noise source.
However, on the basis of the flux shown in \figref{fig:IM}, an object of mass 3\,g impacts an asteroid the size of 433~Eros (or \Alice) only once every $\sim 10^7\,\text{s}$ (note that the smaller the object, the more common the impact, so considering the smallest object at the highest possible impact speeds is conservative).
This means that the potentially borderline troublesome impactor only collides with an asteroid the size of \Alice\ less than once per GW period, over essentially the entirety of the range of GW frequencies of interest to us.
Moreover, the seismic motion it rings up appears to be out of band since $f_0 \gg f_{\textsc{gw}}$ for our band.

If one focuses instead on the more common impacts of less massive objects, a similar picture emerges. 
The mass-weighted dust flux of interest peaks around 1--10\,$\mu$g per the lower panel of \figref{fig:IM}.
Over $T_{\textsc{gw}}\sim 10^5\,\text{s}$ (recall: our other noise sources reach $h_c \sim 10^{-19}$ at $f_{\textsc{gw}} \sim 10^{-5}\,$Hz), around 1.2\,g in total of such dust will pass through the cross-sectional area of an asteroid the size of \Alice.
At 30\,km/s impact speeds, this amount of dust impact would inject 0.5\,MJ of energy into the asteroid, which is of the same order of magnitude as the amount of energy imparted by the rarer single collisions.
Because the asteroid likely responds less violently to such small-mass impacts, and because this is slightly less energy input than the rarer collisions, this would appear to again be at most a borderline noise issue. 
For all of these reasons, we estimate that collision-induced seismic motion should not be a serious issue for a mission of this type.

It is interesting to note that the measured seismic activity~\cite{INSIGHT:2020sdg} on Mars has also largely followed the above expectations. 
In addition to the above, Marsquakes are also triggered by atmospheric activity.
This is not a concern on asteroids, since they lack an atmosphere.

%%%%%%%%%%%%%%%%%%%%%%%%%%%%%%%%%%%%%%%%%%%%%%%%%%%%%%
%%%%%%%%%%%%%%%%%%%%%%%%%%%%%%%%%%%%%%%%%%%%%%%%%%%%%%
\subsection{Discussion}
\label{sect:TMnoiseDiscussion}
%%%%%%%%%%%%%%%%%%%%%%%%%%%%%%%%%%%%%%%%%%%%%%%%%%%%%%
%%%%%%%%%%%%%%%%%%%%%%%%%%%%%%%%%%%%%%%%%%%%%%%%%%%%%%
We have shown that asteroids of radius $R_{\alice} \sim 8$\,km are sufficiently good test masses to permit the detection of gravitational waves with strains $\sim 10^{-17}$--$10^{-19}$ in the $\mu$Hz frequency band.
The main noise source that limits the use of $R_{\alice} \sim 8$\,km as test masses is the fluctuation in the center of mass position of the asteroid caused by the fluctuations in the solar intensity and wind at these frequencies. 
The displacement noise from these solar-origin fluctuations scales $\propto R_{\alice}^{-1}$. 

It is interesting to ask if the requirement of ${R_{\alice} \sim 8\,\text{km}}$ can be relaxed. 
While the displacement noise would increase for a smaller asteroid, it might be possible to subtract out this noise to some extent. 
For example, fluctuations of the solar intensity and wind can be directly measured to several digits independent of the distance measurement between the test masses. 
With these measured intensity fluctuations, the response of the asteroid could potentially be modeled out to the extent to which the properties of the asteroid (such as mass, albedo, etc.) are known. 

This would permit us to use smaller asteroids, potentially allowing the use of a larger number of asteroids as test bodies. 
This larger number of possible candidates may be beneficial in terms of optimizing the mission for technical and cost considerations, issues that we have not addressed in this paper. 
A smaller asteroid is also likely to rotate more rapidly; noise associated with the rotational frequency would thus be further away from the measurement band, possibly increasing the high-frequency reach of this mission. 

Assuming that solar-origin noise is modeled out to 1--2 significant figures, the noise source that most rapidly becomes a problem as the asteroid radius under consideration is decreased, is the position fluctuation caused by impacts; see \sectref{sect:collisions}. 
The displacement noise from collisions scales as $\propto R_{\alice}^{-3/2}$ (see footnote \ref{ftnt:scalingWithAsteroidSize}).
While this noise source is about a factor of $\sim 30$ below our overall enveloped noise projection for an asteroid with $R_{\alice} \sim 8$\,km (see \figref{fig:results}), it will equal or exceed that level for an asteroid with $R_{\alice} \sim 0.8$\,km, and therefore begin to limit access to interesting strain parameter space.
These impacts and their associated effects on the asteroid are harder to monitor and model than the effects of solar intensity and wind fluctuations on the asteroid and thus it is likely prudent to use asteroids with $R_{\alice} \gtrsim 0.8$\,km for this mission. 
There is thus a tradespace in asteroid radii between $1\,\text{km} \lesssim R_{\alice} \lesssim 10\,\text{km}$ where a combination of measurement and modeling could expand the range of asteroid targets and allow for additional optimization of the mission. 
These possibilities deserve further study. 

In addition to smaller asteroids, it may also be interesting to consider the Moon, Mars, and Mars's moons (Phobos and Deimos) as potential test masses. 
There are identifiable issues with each of these, but further study may show that these issues are not particularly limiting. 
The issues of concern are as follows.
The Moon has a rotational period that is in the frequency band of interest: noise associated with the rotational frequency could thus be problematic. 
Mars has an atmosphere and this atmosphere might generate noise, for example, atmospheric disturbances of the link system could be limiting. 
Phobos and Deimos are close to Mars and thus likely experience tidal quakes. 
Further, Phobos is believed to be a rubble pile~\cite{Hurford:2016nuq} and that might make it particularly sensitive to seismic disturbances.

%%%%%%%%%%%%%%%%%%%%%%%%%%%%%%%%%%%%%%%%%%%%%%%%%%%%%%%%%%%%%%%%%%%%%%%%%%%%%%%%%%%%%%%%%%
\section{Metrology noise sources}
\label{sect:linkNoise}
%%%%%%%%%%%%%%%%%%%%%%%%%%%%%%%%%%%%%%%%%%%%%%%%%%%%%%%%%%%%%%%%%%%%%%%%%%%%%%%%%%%%%%%%%%
Having demonstrated the utility of appropriately selected asteroids as TMs in the GW detector, we now turn to the separate question of how to link them together in order to measure the GW-induced strain across the baseline.

In its simplest incarnation, our mission concept is to perform a simple ranging measurement across a single baseline between two asteroids, akin the concept of Lunar Laser Ranging~\cite{Murphy:2013qya}.
The limitations on this method naturally separate into two classes of metrology noise sources: (1) the `link noise', or the accuracy with which an electromagnetic link system could possibly extract the ranging information given perfect timing measurements; and (2) the `clock noise', or the accuracy of the timing system with which one records the round-trip travel time for an electromagnetic signal traveling between the asteroid base stations.

In this section, we consider the link noise for both a conceptual laser ranging system and a conceptual radio ranging system.
We also consider the clock noise.
We defer to \sectref{sect:missionDesign} questions that may arise about the technical implementation of these systems, most of which are for future work; our purpose in this section (and indeed in this paper) is solely to establish the rough performance requirements on the link systems so as to enable our mission concept.
We do not design the link system in detail.

%%%%%%%%%%%%%%%%%%%%%%%%%%%%%%%%%%%%%%%%%%%%%%%%%%%%%%
%%%%%%%%%%%%%%%%%%%%%%%%%%%%%%%%%%%%%%%%%%%%%%%%%%%%%%
\subsection{Link noise: laser pulsing}
\label{sect:laserPulsing}
%%%%%%%%%%%%%%%%%%%%%%%%%%%%%%%%%%%%%%%%%%%%%%%%%%%%%%
%%%%%%%%%%%%%%%%%%%%%%%%%%%%%%%%%%%%%%%%%%%%%%%%%%%%%%

%%%%%%%%%%%%%%%%%%%%%
\subsubsection{Link concept, noise estimate, and parameters}
\label{sect:laserPulsingLink}
%%%%%%%%%%%%%%%%%%%%%
Consider that the base stations on \Alice\ and \Bob\ are equipped with a laser of wavelength $\lambda_{\text{laser}}$ and average power $\bar{P}_{\text{laser}}$ which is directed through a telescope of diameter $D_{\text{telescope}}$.
Conceptually, the base station at \Alice\ aligns its telescope to point to the instantaneous location%
%%%%%%%%%%%%
\footnote{\label{ftnt:leadLocationOfBob}%
    As we will see below, because \Bob\ is moving, this really means that the laser must be pointed to some position ahead of the current instantaneous location of \Bob, so that the pulse and \Bob\ will intersect after the $\sim 500\,$s light-travel time across a $\sim 1$\,AU baseline.
    } %
%%%%%%%%%%%%
of \Bob\ (perhaps using guide stars for pointing orientation), and sends out a train of laser pulses of duration $\Delta t$ with a repeat time of $\Delta t$ (i.e., the laser pulses on and off with frequency $1/\Delta t$).
This pulse train may include some digital encoding (e.g., a series of deleted pulses) to allow for unambiguous pulse-number identification.
We will assume that the system is engineered such that the base station at \Bob\ will receive on average one photon per pulse emitted by \Alice\ (see discussion below), a nominal time $L$ after it is emitted.
The \Bob\ base station then immediately returns a \emph{powered}%
%%%%%%%%%%%%
\footnote{\label{ftnt:whyPowered}%
    The return pulse must be powered; the passively reflected power scales as $L^{-4}$, and is much too small given the available laser power and baselines involved.
    See discussion in the main text.
    } %
%%%%%%%%%%%%
pulse back toward the location of \Alice.
The base station at \Alice\ will receive back on average one photon from this powered return pulse, a nominal time $2L$ after emission of the original outgoing pulse.

Of course, a GW will modulate these round-trip times accordingly, by a fractional amount $\sim h_c$ over the GW period.
Because the pulse receipt at each end of the baseline is only located with a single received photon on average, the uncertainty on individual round-trip pulse timing is of order $\Delta t$, up to an $\order{1}$ numerical factor arising from the exact pulse shape, and from accounting for the uncertainty on each leg of the round trip across the baseline.
However, over a GW period $T_{\textsc{gw}}$, $N_{\text{{pulses}}} \sim T_{\textsc{gw}}/ \Delta t$ such pulses can be sent and received,%
%%%%%%%%%%%%
\footnote{\label{ftnt:digitalCode}%
    As mentioned above in the text, to allow for completely unambiguous pulse identification even in the face of possible dropped pulses, it may be necessary to encode a digital signal in the pulse train: e.g., drop the $(r_i)$\up{th} pulse sent from \Alice\ toward \Bob, where $r_i$ is a uniformly distributed integer in the range $i < r_i /n \leq i + 1 $ for some fixed $n$ and $i = 0, \ldots, i_{\text{max}}$ where $i_{\text{max}} = (N_{\text{pulses}}/n) - 1$.
    This results in the loss of only a fraction of $1/n$ of the total possible pulses, so the estimates in the text are unaffected at the 10\% level as long as $n\gtrsim 10$.
    } %
%%%%%%%%%%%%
so the residual timing accuracy can be estimated as $\sigma_t^{\text{laser}} \sim \Delta t / \sqrt{ N_{\text{{pulses}}} } \sim (\Delta t)^{3/2} f^{1/2}_{\textsc{gw}}$.
This leads to a strain link noise of order
%%%%%%
\begin{align}
    h^{\text{laser}}_c &\sim \frac{\sigma^{\text{laser}}_t}{L} \times \max\lb[ 1, \pi f_{\textsc{gw}} L \rb] \\
    &\sim \frac{(\Delta t)^{3/2} f^{1/2}_{\textsc{gw}}}{L} \times \max\lb[ 1, \pi f_{\textsc{gw}} L \rb],
    \label{eq:LaserLinkNoise1}
\end{align}
%%%%%%
where we have also included a possible baseline penalty factor $\sim \omega_{\textsc{gw}}L/2$ that occurs because the GW strain response of the detector is suppressed once the GW wavelength is inside the baseline (note: we have ignored this factor until now in the paper as none of the noise estimates preceding this point are the relevant ones at high frequencies).
This estimate neglects any electronics delay noise, although as we will see, $\Delta t \sim \text{ns}$, so this should be manageable.

The pulse time $\Delta t$ in \eqref{eq:LaserLinkNoise1} is constrained by the requirement that at least one photon must be received at the far end of the baseline.
A collimated Gaussian laser beam with initial radial beam waist
$w_0 = D^{\text{emitter}}_{\text{telescope}}/2$ undergoes diffractive beam spreading beyond the Rayleigh range $z_{\text{Rayleigh}} = \pi w_0^2 / \lambda_{\text{laser}}$: the beam width spreads as $w(z) = w_0 \sqrt{ 1 + (z/z_{\text{Rayleigh}})^2}$~\cite{GaussianBeams}.
Assuming that $L\gg z_{\text{Rayleigh}}$, for an (average) emitted total power $\bar{P}_{\text{laser}}$, a receiver telescope of diameter $D^{\text{receiver}}_{\text{telescope}}$ at axial distance $L$ will receive an (average) total power%
%%%%%%%%%%%%
\footnote{\label{ftnt:gaussianBeams}%
    The received power in \eqref{eq:laserRecvdPower} assumes a Gaussian beam and on-axis reception.
    Gaussian beams have a transverse intensity profile that is a factor of 2 higher in the center of the beam than the value obtained by averaging the total beam power over the beam's cross-sectional area: i.e., $I_{\text{beam}}(r=0,z) \approx 2 P_{\text{beam}} / \lb( \pi [w(z)]^2 \rb)$.
    } %
%%%%%%%%%%%%
%%%%%%
\begin{align}
    P_{\text{received}}^{\text{laser}} \sim \frac{\pi^2(1-e^{-2})}{8} \lb[  \frac{  D^{\text{receiver}}_{\text{telescope}} D^{\text{emitter}}_{\text{telescope}} }{ L \lambda_{\text{laser}} } \rb]^2\bar{P}_{\text{laser}},
    \label{eq:laserRecvdPower}
\end{align}
%%%%%%
where the extra factor of $(1-e^{-2})$ arises from assuming that the laser power is effectively sent through an aperture of radius $w_0 = D_{\text{telescope}}^{\text{emitter}}/2$ when it is emitted by the telescope; but $\pi^2(1-e^{-2})/8\sim 1.07$, so we will drop this whole $\order{1}$ numerical factor in what follows.
Assuming that $D^{\text{receiver}}_{\text{telescope}} \sim D^{\text{emitter}}_{\text{telescope}} \sim D_{\text{telescope}}$, then numerically we have
%%%%%%
\begin{align}
    P_{\text{received}}^{\text{laser}} &\sim 200\,\text{pW} \times \lb( \frac{ \bar{P}_{\text{laser}} }{1\,\text{W}} \rb) \times \lb( \frac{D_{\text{telescope}}}{1.5\,\text{m}} \rb)^4 \nl \times \lb( \frac{\lambda_{\text{laser}}}{1064\,\text{nm}} \rb)^{-2} \times \lb( \frac{ L }{1\,\text{AU}} \rb)^{-2}.
    \label{eq:laserRecvdPowerNumerical}
\end{align}
%%%%%%
Moreover, the number of received laser photons of energy $E_{\gamma} = 2\pi / \lambda_{\text{laser}}$ assuming a pulse of duration $\Delta t$ is thus
%%%%%%
\begin{align}
    N_{\gamma}^{\text{received}} \sim \frac{1}{2\pi} \frac{  D^4_{\text{telescope}} }{ L^2 \lambda_{\text{laser}} }  \bar{P}_{\text{laser}} \Delta t.
\end{align}
%%%%%%
Fixing $N_{\gamma}^{\text{received}}=1$ gives
%%%%%%
\begin{align}
   \Delta t &\sim 2\pi \frac{ L^2 \lambda_{\text{laser}} }{  D^4_{\text{telescope}} \bar{P}_{\text{laser}} }\\
   &\sim 0.9\,\text{ns} \times \lb( \frac{ \lambda_{\text{laser}} }{ 1064\,\text{nm} } \rb) \times  \lb( \frac{ D_{\text{telescope}} }{ 1.5\,\text{m} } \rb)^{-4} \nl \times \lb( \frac{\bar{P}_{\text{laser}}}{1\,\text{W}} \rb)^{-1}\times \lb( \frac{L}{1\,\text{AU}} \rb)^{2}.
\end{align}
%%%%%%
Substituting this value of $\Delta t$ into \eqref{eq:LaserLinkNoise1} then yields the laser link noise estimate
%%%%%%
\begin{align}
    h_c^{\text{laser}}& \sim \lb(2\pi\rb)^{3/2} \frac{f^{1/2}_{\textsc{gw}}L^2 \lambda^{3/2}_{\text{laser}}}{  D^6_{\text{telescope}}  \bar{P}_{\text{laser}}^{3/2} }\times \max\lb[ 1, \pi f_{\textsc{gw}} L \rb]
    \label{eq:LaserLinkNoise2}\\[1ex]
    &\sim 1.8\times 10^{-19} \times \lb( \frac{f_{\textsc{gw}}}{10\,\mu\text{Hz} } \rb)^{1/2} \times \lb( \frac{ \lambda_{\text{laser}} }{ 1064\,\text{nm} } \rb)^{3/2} \nl \times  \lb( \frac{ D_{\text{telescope}} }{ 1.5\,\text{m} } \rb)^{-6} \times \lb( \frac{\bar{P}_{\text{laser}}}{1\,\text{W}} \rb)^{-3/2}\nl \times \lb( \frac{L}{1\,\text{AU}} \rb)^{2} \times \max\lb[ 1, \lb(\frac{L }{1\,\text{AU}} \times \frac{f_{\textsc{gw}}}{0.6\,\text{mHz}} \rb)\rb].
\end{align}
%%%%%%
Noise curves based on \eqref{eq:LaserLinkNoise2} are shown in \subfigref{fig:noise_contributions}{f} assuming (a) $\bar{P}_{\text{laser}}=1\,\text{W}$ and $D_{\text{telescope}} = 1.5\,\text{m}$, which gives $P_{\text{received}}^{\text{laser}} \approx 0.2\,\text{nW}$ and $\Delta t \sim 0.9\,\text{ns}$ for $N_{\gamma}^{\text{received}}\sim 1$ (solid gold line), and (b) $\bar{P}_{\text{laser}}=3\,\text{W}$ and $D_{\text{telescope}} = 1.5\,\text{m}$, which gives $P_{\text{received}}^{\text{laser}} \approx 0.6\,\text{nW}$ and $\Delta t \sim 0.31\,\text{ns}$ for $N_{\gamma}^{\text{received}}\sim 1$ (dashed gold line).
Note that the Rayleigh range for these parameters is $z_{\text{Rayleigh}}\sim 1.7\times 10^6\,\text{m} \sim 10^{-5}\,\text{AU}$.
Estimate (b) is slightly more aggressive in assuming a larger laser power, and the ability to modulate the laser above GHz frequencies to create sub-nanosecond pulses.

We note that, in the above estimates, we have taken an $L\sim 1$\,AU baseline.
As \figref{fig:baselineASDs} makes clear, typical asteroid baseline distances for some pairs of asteroids shown in \tabref{tab:asteroids} spend long periods around $L \sim 3$\,AU.
For all other noise source estimates we have thus far given in this paper, assuming the shorter baseline was the conservative choice (see comments in, e.g., \sectref{sect:TSI}).
However, for the link system, assuming a shorter baseline is not conservative: $h_c^{\text{laser}}$ in \eqref{eq:LaserLinkNoise2} scales up as $h_c^{\text{laser}} \propto L^2$. 
Our link noise estimates may therefore be slightly aggressive by a factor of $\sim 3^2 \sim 10$ for the given link system parameters if the link is to be employed during typical periods of asteroid separations of $L \sim 3\,$AU.
There are however lengthy periods of time ($\sim$ years) for some asteroid pairs (e.g., 433~Eros and 6618~Jimsimons) where the asteroids are within $L \lesssim 1\,$AU of each other, and a mission could in principle be planned to fly during such a period; that might however limit useful mission lifetime.
Alternatively, asteroids would need to be chosen more judiciously; in this vein, as we have discussed in \sectref{sect:TMnoiseDiscussion}, it would be useful to consider the tradespace available to optimize the mission between the larger noise on smaller asteroids and the larger number of such asteroids available~\cite{JPL-SBD}. 
As there are more smaller asteroids, it may be easier to find appropriate pairs for which $L\sim 1$\,AU separations are maintained over an $\mathcal{O}(1)$ fraction of a planned mission lifetime.
Of course, it would also be possible to still use the larger asteroids but instead consider a slightly larger telescope: since $h_c^{\text{laser}} \propto D_{\text{telescope}}^{-6}$, even a modest increase the diameter above $D_{\text{telescope}}\sim 1.5$\,m could remedy issues with the larger baseline distance, the trade-off being $D_{\text{telescope}}\propto L^{1/3}$ for fixed $h_c$.
We note here simply that if we employed a link-system telescope the size of the Hubble main mirror (which has a 2.4\,m diameter~\cite{HubbleWebsite}) that would already even slightly over-compensate for an $L\sim 3\,\text{AU}$ baseline; this would however likely increase engineering complications and cost.
One could also consider operating a laser with a shorter wavelength to boost the received power and increase strain sensitivity.
These are all detailed optimization questions best left to a future detailed mission study.

We are satisfied that these estimates demonstrate, at a conceptual level, that a relatively modest laser ranging link system consisting of a $\order{1\,\text{W}}$ laser (similar to the assumed power for the LISA metrology system~\cite{LISA_L3,LISAMetrology}) running at 1064\,\text{nm}, coupled with a telescope of $\order{1.5\,\text{m}}$ diameter (smaller than the Hubble main mirror) would suffice to enable access to almost the entirety of the frequency range where asteroids are sufficiently good TMs.

%%%%%%%%%%%%%%%%%%%%%
\subsubsection{Pulse background}
%%%%%%%%%%%%%%%%%%%%%
Note that the above protocol relies on being able to identify the arrival of one laser photon at the far end of the baseline.
One might therefore be concerned about potential backgrounds: both the reflection of the solar irradiance off the surface of the asteroid from whose base station the pulse is emitted, and the intrinsic thermal emission from the surface of that asteroid, are potential sources of photons at the laser frequency.

The receiving telescope has an angular resolution of $\Delta \theta \sim \lambda_{\text{laser}}/D_{\text{telescope}} \sim 7\times 10^{-7}\,$rad for our fiducial 1.5\,m telescope.
Suppose that the emitting telescope is closer than $\sim 100$\,km from the asteroid it is located nearest to, either because it is on the asteroid surface or because the orbit of the base station satellite is chosen to be low (see discussion in \sectref{sect:engineering}); e.g., NEAR--Shoemaker was initially in a slightly elliptical $\sim 300$\,km orbit around 433~Eros, but spent the majority of its science observation time in orbits with $35$\, and $50$\,km radii, as measured from the CoM of 433~Eros~\cite{Yeomans:2000eth,MILLER20023,NearShoemakerWebsite}.
In this case, the light from the emitting telescope and the emitted/reflected light from the asteroid surface will not be resolvable.
The flux of photons from the latter at the receiver must thus be estimated.

Suppose the laser receiver system has a band-pass filter with a width of $\Delta \lambda \sim 10\,$nm centered on $\lambda_{\text{laser}} = 1064\,$nm; such filters are commonly commercially available, and will admit all the incoming laser light, but only a fraction of the broad thermal distributions emitted/reflected from the asteroid surface.
Let us assume that the emitted/reflected light from the asteroid surface follows a perfect blackbody spectrum with a \emph{color} temperature $T$, such that the spectral radiance (power per unit solid angle per unit cross-sectional asteroid area per unit frequency) is
%%%%%%
\begin{align}
    \frac{dP}{dA_{\perp}\,d\nu\,d\Omega} = \frac{4\pi \nu^3}{e^{2\pi \nu/T}-1}.
\end{align}
%%%%%%
A fraction $\beta$ of the total spectrum falls in the frequency range $\nu_0-\Delta \nu/2 \leq \nu \leq \nu_0+\Delta \nu/2$, where 
%%%%%%
\begin{align}
    \beta &\approx 240 \frac{\nu_0^3\Delta \nu}{T^4} \frac{1}{e^{2\pi \nu_0/T}-1} \\
    &= 240 \frac{\Delta \lambda}{T^4\lambda_{\text{laser}}^5} \frac{1}{e^{2\pi/(\lambda_{\text{laser}} T)}-1},
\end{align}
%%%%%%
assuming $|\Delta\nu| = |\Delta \lambda| / \lambda_{\text{laser}}^2$.

Let us start with the estimate of the thermal emission. 
Because Wien's displacement law indicates that the Planck distribution that peaks at $\lambda_{\text{laser}} \sim 1064\,$nm has $T\sim 2.7\times 10^3\,\text{K}$, it follows that a conservative estimate here would utilize the highest possible asteroid thermal temperature; typical maximum asteroid temperatures are on the order of $T_{\alice}\sim 300\,\text{K}$ (see discussion in \sectref{sect:thermal}).
In this case, $\beta \sim 2\times 10^{-16}$.
Suppose the surface of \Alice\ emits as a perfect blackbody at temperature $T_{\alice}\sim 300\,\text{K}$; the Stephan-Boltzmann law then tells us that the total power radiated per unit solid angle by \Alice\ is $dP_{\alice}/d\Omega \sim \pi^2 R_{\alice}^2T_{\alice}^4/60 \sim 2\times 10^{29}\,\text{eV/s/sr}$.
It follows that the power received by the telescope at the distant end of the baseline, after being passed through the band-pass filter, would be
$P_{\text{r,\ therm}} \sim \beta\cdot [\pi (D_{\text{telescope}}/2)^2/L^2]\cdot dP_{\alice}/d\Omega \sim 2\times 10^{-9}\,\text{eV/s}$.
This corresponds to a number of photons received in a pulse time of $\Delta t \sim 1\,\text{ns}$ of $N_{\text{r,\ therm}}\sim 2\times 10^{-18}$.
This is clearly safe by an exceedingly large margin in comparison to the one laser photon received during the same pulse duration.

Consider now the reflected sunlight whose color temperature is $T\sim 6\times 10^3\,$K (approximately the solar surface temperature), assuming that \Alice\ is a perfect reflector of sunlight.
In this case, $\beta \sim 4\times 10^{-3}$.
A total solar power of $P_{\odot,\alice} \sim I_{\odot}\lb( r_{\oplus}/r_{\alice} \rb)^2 \pi R_{\alice}^2$ falls on the surface of \Alice; assume that it gets re-radiated out into a hemisphere, leading to a re-radiated power per unit solid angle of    $dP_{\odot,\alice}/d\Omega \sim(1/2) I_{\odot}\lb( r_{\oplus}/r_{\alice} \rb)^2 R_{\alice}^2 \sim  10^{29}\,\text{eV/s/sr}$; this of course should match well with the power assumed from the thermal re-radiation, up to $\mathcal{O}(1)$ factors we have not been careful with here, as the solar radiation sets the equilibrium temperature.
This leads to a power received by the telescope at the far end of the baseline, after the band-pass filter, of $P_{\text{r,} \odot,\alice} \sim \beta\cdot [\pi (D_{\text{telescope}}/2)^2/L^2]\cdot dP_{\odot,\alice}/d\Omega \sim 4.2\times 10^4\,\text{eV/s}$.
This corresponds to a number of photons received in a pulse time of $\Delta t \sim 1\,\text{ns}$ of $N_{\text{r,} \odot,\alice}\sim 4\times 10^{-5}$.
This is again clearly safe by a large margin in comparison to the one laser photon received during the same pulse duration.

Single-photon pulse detection is thus background-free assuming that appropriate band-pass filtering of the received light is performed (indeed, such filtering may not even be necessary on the basis of the discussion above), even if the laser emitter cannot be resolved from the asteroid surface.

%%%%%%%%%%%%%%%%%%%%%
\subsubsection{Angular stability}
%%%%%%%%%%%%%%%%%%%%%
Note that there are two pointing requirements for the optical system to achieve: (1) the receiver telescope must be within the beam width at the location of the receiver, and (2) the angular jitter in the beam pointing must not induce a noise on the distance measurement. We consider these in turn.

(1) 
At a distance of $L\sim 1\,\text{AU}$, the beam waist (radius) assuming $D_{\text{telescope}}\sim 1.5\,$m and $\lambda_{\text{laser}} \sim 1064\,\text{nm}$ is $w \sim 5\times 10^{-7}\,\text{AU}\sim 70\,\text{km}$, so the angular pointing requirement is $\Delta \theta \sim 3\times 10^{-5}\,\text{deg} \approx 0.1\,\text{arcsec}$.
This is easily achievable: for instance, Hubble achieves a pointing stability of 7\,\text{mas} over 24\,\text{hr} periods~\cite{BELY1992457,doi:10.2514/3.20280}, LISA aims for the DC mis-pointing error of $\lesssim 10^{-8}\,\text{rad}\sim 6\times 10^{-7}\,\text{deg}\sim2\,\text{mas}$, and the more precise angular tracking system discussed briefly in \sectref{sect:rotationalMotion} would also vastly exceed this pointing requirement if deployed.
Note also that this beam-waist estimate makes clear that the ranging has to be done by aiming the outgoing laser beam at a point some $v_{\alice} L/c \sim 10^4\,$km ahead of the location of the receiving asteroid at the moment a pulse is fired, such that the receiving asteroid will move into the path of the outgoing pulse; there is no additional technical requirement associated with this, as modeling of the asteroid position to the requisite accuracy over $\sim L/c\sim 500\,$s timescales is straightforward.

(2) 
Na\"ively, with planar wave-fronts, a jitter in the beam pointing by an angle $\delta \theta$ would give rise to an error on the distance measurement of order $\Delta L \sim L (\delta \theta)^2$, or a strain noise of order $h_c \sim (\delta\theta)^2$. 

However, that estimate fails to account for the curvature of the wavefronts of the beam beyond the Rayleigh range; these become quasi-spherical in the large-distance limit, which suppresses this noise greatly (see generally \citeR{Hogan:2010fz}).
The radius of curvature of the beam at an axial distance $z$ from the emitter (assumed to be the location where the beam waist is minimized) is given by~\cite{GaussianBeams}
%%%%%%
\begin{align}
    R(z) = z \lb[ 1 + \lb( \frac{z_\text{Rayleigh}}{z} \rb)^2 \rb].
\end{align}
%%%%%%
We make the approximation that you can use phase evolution of the beam as a proxy for the arrival time of a pulse since phase and group velocities agree in vacuum.
At an axial distance $z$ and transverse distance $r$ from the beam axis, the spatial phase of a Gaussian beam is given by~\cite{Hogan:2010fz,GaussianBeams}
%%%%%%
\begin{align}
    \Phi = kz + \frac{kr^2}{2R(z)} - \psi(z), 
\end{align}
%%%%%%
where $k=2\pi/\lambda_{\text{laser}}$ and $\psi(z)$ is the Gouy phase, which does not depend on wavenumber $k$ and is thus irrelevant for this argument.
Assume that the beam is mis-pointed by an angle $\delta \theta \ll 1$ so that the receiving asteroid, which is a distance $L$ from the emitter, lies at $z \sim L \lb[ 1 - (\delta\theta)^2/2 \rb]$ and $r \sim L \delta \theta$ with respect to the mis-pointed beam.
Since $L \gg z_{\text{Rayleigh}}$ and $\delta \theta \ll 1$, we have a phase error as compared to a perfectly pointed beam of
%%%%%%
\begin{align}
    \Delta \Phi \sim kL \lb[ \frac{1}{2} (\delta \theta)^2 \frac{z_{\text{Rayleigh}}^2}{L^2} + \mathcal{O}\lb( \frac{\lambda_{\text{laser}}z_{\text{Rayleigh}}}{L^2},\ldots\rb) \rb],
\end{align}
%%%%%%
where we omitted subdominant and higher-order terms.
The dominant term looks like a strain error of order $h_c \sim (\delta\theta)^2(z_{\text{Rayleigh}}/L)^2/2$, so we adopt that as the strain error arising from the beam arrival time.
Then taking $h_c \sim 10^{-19}$ at $f_{\textsc{gw}}\sim 10^{-5}\,$Hz gives $\delta\theta \sim 4\times 10^{-5}\,\text{rad}\sim 2.3\times 10^{-3}\,\text{deg} \sim 8\,\text{arcsec}$ over $T_{\textsc{gw}}\sim 10^5\,$s, which translates to a broadband angular pointing stability ASD requirement of $\sqrt{S[\delta \theta]} \sim \delta\theta / \sqrt{f_{\textsc{gw}}} \sim 10^{-2}\, \text{rad}/\sqrt{\text{Hz}}$.
By comparison, LISA demands a pointing stability requirement of order $10^{-8}\sqrt{1+\lb(3\,\text{mHz}/f\rb)^4 }\,\text{rad}/\sqrt{\text{Hz}}$~\cite{LISA_L3}, which gives $\sim 10^{-3}\,\text{rad}/\sqrt{\text{Hz}}$ at $f_{\textsc{gw}} \sim 10^{-5}\,\text{Hz}$; our requirement is thus less severe.
Moreover, comparing to (1) above, it is clear that, as long as the pulse is received, this jitter error is not relevant.

%%%%%%%%%%%%%%%%%%%%%%%%%%%%%%%%%%%%%%%%%%%%%%%%%%%%%%
%%%%%%%%%%%%%%%%%%%%%%%%%%%%%%%%%%%%%%%%%%%%%%%%%%%%%%
\subsection{Link noise: radio pulsing}
\label{sect:radioPulsing}
%%%%%%%%%%%%%%%%%%%%%%%%%%%%%%%%%%%%%%%%%%%%%%%%%%%%%%
%%%%%%%%%%%%%%%%%%%%%%%%%%%%%%%%%%%%%%%%%%%%%%%%%%%%%%
In addition to the optical-pulsing link system, we outline the technical requirements and reach achievable for a similar setup using instead a radio-frequency link.%
%%%%%%%%%%%%
\footnote{\label{ftnt:radioStabilityTransfer}%
    Note that in this setup, the radio frequency stability could be guaranteed by referencing it to the local space-qualified atomic clock on the same base station; see, e.g., \citeR{Nakamura:2020eav}.
    } %
%%%%%%%%%%%%

Consider a system that emits radio-frequency ($f_{\text{radio}}=1/\lambda_{\text{radio}}$) pulses of duration $\Delta t$ with power $\bar{P}_{\text{transmit}}$ using a radio dish of diameter $D^{\text{emitter}}_{\text{dish}} \sim D_{\text{dish}}$.
Assume these are received at the far end of the baseline with a dish of diameter $D^{\text{receiver}}_{\text{dish}}\sim D_{\text{dish}}$, leading to a received power of [cf.~\eqref{eq:laserRecvdPower} and the comment below that equation]
%%%%%%
\begin{align}
    P_{\text{received}} &\sim \lb[  \frac{  D^2_{\text{dish}} f_{\text{radio}} }{ L } \rb]^2\bar{P}_{\text{radio}} \\
    &\sim 0.6\,\text{pW} \times \lb( \frac{D_{\text{dish}}}{5\,\text{m}} \rb)^4  \times \lb( \frac{f_{\text{radio}}}{100\,\text{GHz}} \rb)^{2} \nl
    \times \lb( \frac{\bar{P}_{\text{radio}}}{200\,\text{W}} \rb).
\end{align}
%%%%%%
Further, assume that the thermal (Johnson--Nyquist) noise in the receiver,
%%%%%%
\begin{align}
    P_{\text{thermal}} \sim 4 T_{\text{noise}} \, {\Delta f}_\text{radio},
    \label{eq:JohnsonNyquist}
\end{align}
%%%%%%
is characterized by a noise temperature $T_{\text{noise}} \sim 100\,$K, where $\Delta f$ is the relevant bandwidth (see below).
Assume further than parameters are selected such that each pulse is received with signal-to-noise ratio (SNR) $\rho_\text{pulse} = \sqrt{P_{\text{received}}/P_{\text{noise}} }> 1$.
If pulses are sent and received with a duty cycle of order unity, then $N_{\text{pulses}} \approx \sqrt{T_{\textsc{gw}}/\Delta t } = 1/\sqrt{f_{\textsc{gw}}\Delta t}$ pulses are received in a GW period.
Then the strain sensitivity of the setup is 
%%%%%%
\begin{align}
    h^{\text{radio}}_c &\sim \frac{\sigma_{t\text{,radio}}}{L} \times \max\lb[ 1, \pi f_{\textsc{gw}} L \rb] \\
    &\sim \frac{\Delta t}{L \sqrt{\rho_{\text{pulse}}}\sqrt{N_{\text{pulses}}}} \times \max\lb[ 1, \pi f_{\textsc{gw}} L \rb],
    \label{eq:RadioLinkNoise0}
\end{align}
%%%%%%
subject to the requirement that $\rho_{\text{pulse}} > 1$.

Suppose that the intrinsic bandwidth of the transmitter/receiver $\Delta f_0 = f_{\text{radio}}/Q$ is sufficiently small that $(\Delta t)^{-1} \gtrsim \Delta f_0$; that is, that the pulse duration is no longer in duration than the inverse intrinsic bandwidth of the receiver.
Then, $\Delta f_\text{radio} \sim 1/\Delta t$ is the appropriate bandwidth to use in \eqref{eq:JohnsonNyquist}; in this limit, we have
%%%%%%
\begin{align}
    h^{\text{radio}}_c \sim \frac{\Delta t \sqrt{4T_{\text{noise}} f_{\textsc{gw}} }}{L \sqrt{P_{\text{received}}}} \times \max\lb[ 1, \pi f_{\textsc{gw}} L \rb],
    \label{eq:RadioLinkNoise1}
\end{align}
%%%%%%
subject to $\Delta t \lesssim Q/f_{\text{radio}}$ and $\rho_{\text{pulse}} > 1$.
Assuming still that $D^{\text{receiver}}_{\text{dish}}\sim D^{\text{emitter}}_{\text{dish}} \sim D_{\text{dish}}$, we have%
%%%%%%%%%%%%
\footnote{\label{ftnt:notindependentofL}%
    While the expression at \eqref{eq:RadioLinkNoise2} appears to give a noise estimate that is independent of $L$ (except for the baseline penalty factor), this is a mirage.
    \eqref{eq:RadioLinkNoise2} only applies subject to the requirement on $\Delta t$ imposed by \eqref{eq:radioPowerReq}, which places an $L$-dependent lower bound on $h_c^{\text{radio}}$.
} %
%%%%%%%%%%%%
%%%%%%
\begin{align}
    h^{\text{radio}}_c \sim \frac{\Delta t \sqrt{4T_{\text{noise}} f_{\textsc{gw}} }}{   D^2_{\text{dish}} f_{\text{radio}} \sqrt{\bar{P}_{\text{radio}}} } \times \max\lb[ 1, \pi f_{\textsc{gw}} L \rb],
    \label{eq:RadioLinkNoise2}
\end{align}
%%%%%%
subject to the requirements that $\Delta t \lesssim Q/f_{\text{radio}}$ and 
%%%%%%
\begin{align}
    \rho_{\text{pulse}} > 1 
    \Rightarrow  \lb[  \frac{  D^2_{\text{dish}} f_{\text{radio}} }{ L } \rb]^2 \frac{ \bar{P}_{\text{radio}} \Delta t}{4T_{\text{noise}}} > 1.
    \label{eq:radioPowerReq}
\end{align}
%%%%%%
Note that if $\rho_{\text{pulse}} = 1$ is kept saturated, then $h_c \propto (\Delta t)^{3/2}$, which is most clear from \eqref{eq:RadioLinkNoise0} [cf.~also \eqref{eq:LaserLinkNoise1}].

Suppose that $\bar{P}_{\text{radio}} = 200\,\text{W}$, $D_{\text{dish}} = 5\,$m, $f_{\text{radio}}=100\,$GHz, $Q\sim 10^3$, $T_{\text{noise}}=100\,$K, and $L\sim 1\,\text{AU}$.
Then we find that $\Delta t \gtrsim 9\,\text{ns}$ (note: $Q/f_{\text{radio}} \sim 10\,\text{ns} \sim \Delta t$), and $h_c \sim 2\times 10^{-18} \sqrt{f_{\textsc{gw}}/\mu\text{Hz}}$ for all frequencies below $ f_{\textsc{gw}} \lesssim (\pi L)^{-1} \sim 0.6\,\text{mHz}$.
Assuming the same parameters, we plot the strain sensitivity given by \eqref{eq:RadioLinkNoise2} in \subfigref{fig:noise_contributions}{f}.
Note that similar comments as those made at the end of \sectref{sect:laserPulsingLink} regarding the assumed baseline of $1\,\text{AU}$ in light of the results of \figref{fig:baselineASDs} also apply to this link noise estimate.

It is clear that this link system requires large dishes and power, and, at least with the numbers we have assumed, does not achieve as much sensitivity as the optical system.
However we do not attempt a true engineering study to assess the feasibility of either system and thus it is possible that the radio system may be easier to implement or will have other advantages.
It might also be possible to improve the sensitivity with an interferometric radio system rather than the pulsed system we have discussed.
This could effectively reduce the pulse length to the radio-frequency period (i.e., $\Delta t \sim 1/f_{\text{radio}}$).
This could significantly improve the sensitivity of this system.
However, the construction of the interferometer may entail extra complication and we leave a study of its feasibility for future work.

Note that for a radio link, the plasma in the interplanetary medium (IPM) could potentially cause an additional source of noise.
The index of refraction of the IPM is $n = 1-\Delta n$ where
%%%%%%
\begin{align}
    \Delta n &\sim \frac{1}{2} \frac{f_p^2}{f_{\text{radio}}^2} \\
    &\sim 2\times 10^{-14} \times \lb( \frac{f_{\text{radio}}}{100\,\text{GHz}} \rb)^{-2} \lb( \frac{f_p}{20\,\text{kHz}} \rb)^2,
    \label{eq:indexOfRefraction}
\end{align}
%%%%%%
where $f_p \sim \sqrt{n_e\alpha/(\pi m_e)} \sim 20 \, \text{kHz}$ is a typical IPM (electron) plasma frequency at distances around 1\,AU from the Sun ($n_e \sim 5\,\text{cm}^{-3}$).
This index of refraction shifts the group velocity of the radio wave by $\Delta v_g \sim - c \Delta n$, leading to a change in the measured round-trip travel time of order $\Delta t \sim 2L \Delta n$.
The shift to the strain measurement is thus $h_c \sim \Delta n$, where we have assumed a fixed value for $f_p$ during the measurement.
Of course, the IPM plasma is not homogeneous and shows density fluctuations (see \sectref{sect:solarWind}), so this may be an overestimate of the effect: the line-of-sight-integrated index of refraction would average down in that case by $\sim 1/\sqrt{ N_{\text{patches}}}$ where $N_{\text{patches}} \sim L/d$ with $d$ being the characteristic length scale over which the plasma shows $\mathcal{O}(1)$ density fluctuations.
Moreover, were the IPM plasma frequency constant in time, this would not be a noise source; this only becomes a noise source to the extent that the line-of-sight-averaged plasma frequency fluctuates in time.
However, we would expect in general to see $\mathcal{O}(1)$ fluctuations in the IPM number density on the relevant GW-period timescales since the solar wind traverses distances of $\sim \text{AU}$ on timescales of $\sim 3\times 10^5\,\text{s}$.
Therefore, we conclude that the index of refraction fluctuations of the IPM may present a noise source for the radio link by some orders of magnitude, but this could in principle be measured and subtracted out by making use of two or more different frequencies in the link system, owing to the frequency dependence of $\Delta n(f_{\text{radio}})$.
We note that for laser frequencies, this is not a relevant noise source since $f_{\text{laser}} \sim 3\times 10^{14}\,\text{Hz} \sim 10^3 f_{\text{radio}}$ (assuming a 1064\,nm laser), which suppresses these effects by a factor of $\sim 10^6$, to the level of at worst $h_c \sim 10^{-20}$.

%%%%%%%%%%%%%%%%%%%%%%%%%%%%%%%%%%%%%%%%%%%%%%%%%%%%%%
%%%%%%%%%%%%%%%%%%%%%%%%%%%%%%%%%%%%%%%%%%%%%%%%%%%%%%
\subsection{Clock noise}
\label{sect:clocknoise}
%%%%%%%%%%%%%%%%%%%%%%%%%%%%%%%%%%%%%%%%%%%%%%%%%%%%%%
%%%%%%%%%%%%%%%%%%%%%%%%%%%%%%%%%%%%%%%%%%%%%%%%%%%%%%
In this section, we consider the limitations imposed by clock noise on the strain measurement.

Consider making a series of round-trip timing measurements $\tau_n$ of a nominal baseline distance $L$ with a clock whose nominal frequency is $\hat{\nu}$, with each measurement initiated at time $t_n = n \Delta t$, where $\Delta t$ is the inter-measurement spacing.
Atomic clocks of the class we consider in this work however have intrinsic frequency drift; let the average clock frequency over the $i$th measurement be $\nu_i = \hat{\nu} + \delta \nu_i$, where $|\delta \nu_i| \ll \nu_i$.
The number of clock cycles (`ticks') in the time $\tau_i$ it takes to measure the nominal round-trip time is then $N_i = \tau_i \nu_i = 2L ( \hat{\nu} + \delta \nu_i)$.
Consider making two such measurements a time $T = \mathcal{N}\Delta t$ apart.
Then the difference in the number of ticks on the same nominal baseline distance owing to the clock frequency drift is $\Delta N = 2L ( \delta \nu_{\mathcal{N}} - \delta \nu_0 )$.
This has a typical size $\Delta N \sim 2 L \sigma_\nu(T)$, where $\sigma_\nu(T)$ is the typical amplitude of the clock frequency change measured over the time $T$.

On the other hand, the GW signal we are looking for is a change to the nominal proper baseline distance by an amount $|\Delta L_{\textsc{gw}}| \sim h_c L \times \text{min}\lb[ 1 , 1/ ( \pi f_{\textsc{gw}} L ) \rb]$, over a typical timescale $T \sim T_{\textsc{gw}}$.
This change in proper length changes the round-trip time, causing the number of clock ticks elapsed during its measurement to change by $|\Delta N_{\textsc{gw}}| = 2 |\Delta L_{\textsc{gw}}| \hat{\nu} \sim 2 h_c L\hat{\nu} \times \text{min}\lb[ 1 , 1/ ( \pi f_{\textsc{gw}} L ) \rb]$; we neglect the second-order-small term at $\mathcal{O}(h_c \delta \nu L)$.

Comparing the frequency-drift-induced clock tick noise to the signal size, we find that the limiting strain is
%%%%%%
\begin{align}
    h_c   \sim \frac{\sigma_\nu(T_{\textsc{gw}})}{\hat{\nu}}\times \text{max}\lb[ 1 ,  \pi f_{\textsc{gw}} L \rb].
\end{align}
%%%%%%
However, $\xi(t) \equiv \sigma_\nu(t) / \hat{\nu}$ is just the fractional frequency instability of the clock over a time period $t$, so we conclude that 
%%%%%%
\begin{align}
    h_c^{\text{clock}} \sim \xi(1/f_{\textsc{gw}})  \times \text{max}\lb[ 1 , \pi f_{\textsc{gw}} L \rb].
    \label{eq:hcClock}
\end{align}
%%%%%%

The current world-leading ground-based SrI optical lattice clocks at JILA achieve~\cite{Bothwell_2019}
%%%%%%
\begin{align}
\xi(t) \sim 5\times 10^{-17} \times \lb( \frac{t}{1\,\text{s}} \rb)^{-1/2}\quad [t\lesssim 10^4\,\text{s}],
\end{align}
%%%%%%
with a systematic noise floor bounded around the $\xi \sim 5\times 10^{-19}$ level~\cite{Bothwell_2019}.
Note that the transfer of an optical clock's fractional frequency stability to a clock running at microwave frequencies has also been demonstrated to not degrade frequency stability, to a level that is comparable to the world-leading optical clock performance itself~\cite{Nakamura:2020eav}.

While clocks of this class have not yet been space qualified, there is currently a nascent effort underway to do exactly that (see, e.g., \citeR[s]{HighPerformanceClocksInSpace,ColdAtomsInSpace,Voyage2050ESA,Kaltenbaek:2021dgs}), driven in part by the broad and intrinsic scientific utility of such clocks (e.g., \citeR[s]{Tino:2007az,Kolkowitz:2016wyg,Su:2017kng,Ebisuzaki:2018ujm,Tsai:2021lly}), and in part by technology cross-over with mid-band GW detectors based on atomic interferometry~\cite{Dimopoulos:2007cj,Dimopoulos:2008sv,Hogan:2010fz,Graham:2017pmn,Coleman:2018ozp,Tino:2019tkb,AEDGE:2019nxb} techniques that are envisaged to be deployed in space.
On the other hand, in recent decades, clock improvements have been exponential with passing time (see, e.g., \citeR{Safronova:2017xyt}), and we thus expect further improvements of clock technology into the future.

With those countervailing considerations in mind, we present an estimate of the clock-noise limitation that a space-qualified clock of approximately the same stability as the current JILA SrI clock would impose on our proposal.
Specifically, we assume a clock with a fractional frequency instability
%%%%%%
\begin{align}
\xi(t) \sim 5\times 10^{-19} \times \text{max}\lb[\, 1\, , \lb( \frac{t}{10^4\,\text{s}} \rb)^{-1/2}\, \rb],
\label{eq:FFS}
\end{align}
%%%%%%
where we have assumed that the systematic noise floor on the clock is similar to that of the JILA SrI clock, around $\xi \sim 5\times 10^{-19}$; see Fig.~2(c) of \citeR{Bothwell_2019}.
We also assume that this systematic noise floor does not increase above this level until an clock-integration time $t \gtrsim 10^6\,\text{s}$.

Employing \eqref{eq:FFS} in \eqref{eq:hcClock} yields the noise curve shown in \subfigref{fig:noise_contributions}{e}.
A space-qualified clock of this class would allow our proposal to access the majority of the parameter space in which asteroids serve as sufficiently good TMs.

%%%%%%%%%%%%%%%%%%%%%%%%%%%%%%%%%%%%%%%%%%%%%%%%%%%%%%%%%%%%%%%%%%%%%%%%%%%%%%%%%%%%%%%%%%
\section{Mission design considerations}
\label{sect:missionDesign}
%%%%%%%%%%%%%%%%%%%%%%%%%%%%%%%%%%%%%%%%%%%%%%%%%%%%%%%%%%%%%%%%%%%%%%%%%%%%%%%%%%%%%%%%%%
The main purpose of this paper is as a first exploration of the use of asteroids as test masses for gravitational-wave detection in the $\mu$Hz band.
In particular, we have estimated the acceleration noise of asteroids in this band and find it to be low enough that asteroids appear to be promising candidates as test masses in this band.
We have also demonstrated at a conceptual level that a link system whose main elements lie within the feasible set of improvements from current metrological sensitivities would be capable of exploiting these TMs within the $\mu$Hz band well enough to reach interesting levels of strain sensitivity.

A detailed technical design study for a mission that would realize this concept would of course be a natural next step.
That study itself lies beyond the scope of the present work, and would require additional engineering and spaceflight expertise. 
Nevertheless, in this section we take the opportunity to comment on some of the more pressing issues that we have identified and that such a technical design study will need to address.
We also clearly delineate where current gaps in humanity's knowledge base need to be closed with experimental data from future spaceflight missions in order to fully characterize uncertain aspects of the asteroid environment prior to undertaking a full mission of this type.
Additionally, we comment on areas where technological research and development work is needed to advance capabilities toward this mission's requirements.

%%%%%%%%%%%%%%%%%%%%%%%%%%%%%%%%%%%%%%%%%%%%%%%%%%%%%%
%%%%%%%%%%%%%%%%%%%%%%%%%%%%%%%%%%%%%%%%%%%%%%%%%%%%%%
\subsection{Engineering considerations}
\label{sect:engineering}
%%%%%%%%%%%%%%%%%%%%%%%%%%%%%%%%%%%%%%%%%%%%%%%%%%%%%%
%%%%%%%%%%%%%%%%%%%%%%%%%%%%%%%%%%%%%%%%%%%%%%%%%%%%%%
There are obviously many challenging mission design and engineering issues to consider in planning such a mission.
We will not attempt to discuss them all, but merely bring up a few that must be considered in the future.
As discussed above in \sectref{sect:concept}, there are broadly two possible ways to design such a mission: (1) each base station can be landed on the surface of its asteroid, or (2) each base station remains orbiting its asteroid with only retroreflectors/transponders landed on the asteroid surface itself. 
In either case, each base station would house a link system (e.g., laser or radio) to communicate with the other asteroid, and an accurate clock (e.g., an optical atomic clock). 
There are trade-offs to each approach and we do not attempt to choose which one is ultimately optimal.

Concept (1) has the advantage of requiring neither a secondary ranging system to the surface transponders, nor a daughter deployment subsystem to land the surface transponders on the asteroid.
As asteroids lack atmosphere, landing a sizeable base station that includes a large telescope or dish for the link system would not require atmospheric entry subsystems (heat shields, etc.); the escape speed from the surface of \Alice\ is a modest 10\,m/s, and previous asteroid soft landings have been achieved (e.g., NEAR--Shoemaker performed a soft landing on 433~Eros itself at mission conclusion, and survived an estimated 1.8\,m/s touch down~\cite{SiddiqiNASA}).
On the other hand, as we mentioned in \sectref{sect:concept}, a base station anchored to the asteroid would have to compensate for the asteroid rotation in pointing the link system, and perhaps even suffers from periods where line of sight between the base stations is physically blocked by the asteroid itself.
This concern could perhaps be somewhat alleviated by judicious choice of asteroid pairs and landing sites on each asteroid; this would clearly require detailed mission planning.
Investigation would be required as to whether or not a base station of the requisite structural rigidity and stability to achieve, e.g., the mechanical pointing of a $\sim 1.5 \, \text{m}$ telescope for the laser link system at the desired accuracy, can be deployed and anchored sufficiently rigidly to the surface of an asteroid. 
As asteroids typically have some surface debris or regolith (see, e.g., \citeR[s]{10.1038/s41550-019-0731-1,Richardson1526,RICHARDSON2005325}), this may require investigation of an anchoring system capable of reaching `bedrock' on the asteroid, and additional investigation of whether the asteroid might require some time to relax back to a quiescent state.
A radio link system might have an advantage here, as beam pointing is less of a concern for radio, and could potentially also be achieved via phased-array approaches instead of a mechanical steering system for the dish.
There may potentially also be some concerns about the effects of thermal cycling of the asteroid on the base station itself that would need study.

Concept (2) on the other hand has the advantage that the pointing and angular stability of the satellite are easier to control in free space using, e.g., reaction wheels (if necessary, inertial stability could even be referenced to distant stars via dedicated star trackers).
It also separates the base-station systems from potential asteroid surface thermal cycling effects, to the extent that these are a concern.
However, it would require additional subsystems to range to transponders/retroreflectors on the local asteroid, as well as a system to deploy those transponders/retroreflectors to the asteroid surface; this would need to be investigated in more detail, but is likely achievable.
This design would however alleviate some issues with anchoring a large base station to the asteroid surface, and mechanically steering its link system, if required.
Of course it will still be necessary for the link system to turn to remain continually pointed at the other asteroid.
Moreover, it could alleviate the duty cycle issue arising from line-of-sight concerns since the satellite may be able to orbit far enough from the local asteroid that it is not eclipsed (or only eclipsed rarely).
Additionally, being able to monitor from orbit and in real time the locations of multiple well-defined independent locations on the asteroid surface would allow both for better rotational characterization of the asteroid~\cite{FRENCH2020113537} if necessary, and would also be one way to monitor for plastic deformation of the asteroid surface over $\sim 10^6\,$s timescales (see next subsection).

For concept (2) it seems likely that we would want the satellite orbital frequency around the asteroid to lie outside the gravitational-wave frequency band of interest.
Of course the satellite is being referenced to the asteroid (which is the true test mass) and so the satellite position is not relevant to leading order.
Nevertheless, technical noise sources might lead to some enhanced noise at the satellite orbital frequency.
The orbital frequency can be raised above our band, to around the asteroid rotational frequency with an orbit near the asteroid.
For example, for \Alice~a 15\,km radius orbit has an orbital frequency $\sim 5 \times 10^{-5} \, \text{Hz}$, just above our band.
This will keep all related noise out of our band; however; this orbit brings the satellite close to the asteroid itself and so would likely still lead to some time spent in eclipse.
Alternatively, one could choose to keep the orbital frequency below our band.
For example, for \Alice~a 200\,km radius orbit has an orbital frequency $\sim \mu \text{Hz}$.
It can be checked that even at distances of several hundred km, mirrors or retroreflectors of $\sim$ few cm diameter and a laser power around 0.1\,W are sufficient to range the near asteroid with sufficient accuracy.
Thus it seems that having the base station orbit the asteroid may be achievable without introducing extra noise in our frequency band. 

On the link side, we have mentioned and given estimates for both laser and radio pulsing systems at a conceptual level.
The detailed engineering study of these systems would of course be required, and power and size requirements optimized.
In principle, it may also be possible to consider implementing instead either radio or laser interferometric systems, similar to the laser interferometric system envisaged for LISA~\cite{LISAMetrology}.
However, one challenge here that would require further work is that the asteroids have uncontrolled relative motion on the order of $\sim 10\,\text{km/s}$, which is much larger than the $\sim 5\,\text{m/s}$ LISA requirements~\cite{LISAMetrology,LISA_L3,LISA-Pre-Phase-A}.
Work would be required to understand if a working interferometric laser link system could be made to function under those conditions.
A radio interferometric system with the oscillators stabilized to the atomic clocks~\cite{Nakamura:2020eav} may also be an alternative worthy of study in future work.
It might also be possible to use an interferometric link system in a triangular (`LISA-like') configuration with three asteroids,%
%%%%%%%%%%%%
\footnote{\label{ftnt:notInFormation}%
    For the avoidance of all doubt, the asteroids as TMs would still be moving on their natural, unaltered orbits.
    We do not intend to suggest in any way the idea of orbital alterations to asteroids to put them in formation flight; that would clearly be an impossibility.
    } %
%%%%%%%%%%%%
which would then remove the need for accurate atomic clocks on each base station.
There are naturally trade-offs with each approach.
Implementing a triangular approach would relieve the need for a space-qualified optical atomic clock, but would add the complexity of an interferometric system with uncontrolled relative motion of the asteroids.
The interferometric link system in our case however would not need anywhere near as much strain sensitivity (in $h_c$) as in the case of LISA, as can be seen by comparing the high frequency behaviors of the sensitivity curve we hope to achieve, and that which LISA aims for; see \figref{fig:results}.
But of course the relative motion of the asteroids is orders of magnitude larger than the relative motion of the planned LISA satellites, which will impose a significant challenge on an interferometric link system.  
It is thus not trivial to evaluate the feasibility of the kind of interferometric link system we would need; nor, indeed, the feasibility of implementing the triangular configuration. 
Since our main goal here is the evaluation of asteroids as test masses, we leave consideration of this possibility for future work.

We also note that because the two asteroids in our concept will likely orbit the Sun with different periods, they will spend a fraction of the time separated by distances larger than the 1\,AU which we assumed for the link noise estimates in \sectref{sect:linkNoise} (see, e.g., \figref{fig:baselineASDs}).
For the link noises, a shorter baseline assumption is not conservative; cf.~the displacement noise estimates in \sectref{sect:noise}, where it is.  
Of course it may be possible to improve other parameters in the link noise system estimate to compensate for this.  
Or, at worst, we may have to give up some sensitivity during a fraction of the observing time while the asteroids are well separated.  
One may also choose to use slightly smaller asteroids than we considered in \tabref{tab:asteroids}.  
Since the number of inner-Solar-System asteroids increases rapidly at smaller sizes~\cite{JPL-SBD}, it would likely be possible to find asteroids orbiting at smaller radii which then could have an average distance between them around 1\,AU. 
We leave the exact choice of optimal asteroid pairs for future work.

%%%%%%%%%%%%%%%%%%%%%%%%%%%%%%%%%%%%%%%%%%%%%%%%%%%%%%
%%%%%%%%%%%%%%%%%%%%%%%%%%%%%%%%%%%%%%%%%%%%%%%%%%%%%%
\subsection{Exploratory paths}
\label{sect:exploratory}
%%%%%%%%%%%%%%%%%%%%%%%%%%%%%%%%%%%%%%%%%%%%%%%%%%%%%%
%%%%%%%%%%%%%%%%%%%%%%%%%%%%%%%%%%%%%%%%%%%%%%%%%%%%%%
In this subsection we comment on future work that would likely be useful to undertake before a full gravitational-wave mission.

It would be key to fully understanding the limitations imposed by the internal excitations of the asteroid itself on this mission concept to characterize the seismic environment of an asteroid.
These measurements should be included as a goal for asteroid-visiting missions to fly prior to full deployment of a mission of type we propose.
This is already well motivated from the utility that such studies would have on enlarging our knowledge of asteroid internal structure~\cite{Walker:2006bqw,Plescia:2016sfg,Plescia:2017wrt,MURDOCH201789,Plescia:2018wfb,Schmerr:2018rtb}.
We would strongly encourage a future asteroid mission to deploy a seismometer on an asteroid, as has recently been done on Mars~\cite{INSIGHT:2020sdg}, to characterize the higher frequency seismic environment (typically around $\sim$ Hz) and learn more about the internal structure and dynamics of the asteroid.
Seismic measurements of asteroids could also reveal the motion of boulders on the surface and other such asteroid dynamics, helping to pin down the relevant noise estimates further.
Moreover, we would need to also understand in more detail the plastic deformation/creep (or other longer-term changes) of asteroid surfaces over longer ($\sim 10^6\,$s) timescales, as this is a relevant noise for our mission concept. 
For the latter measurement, a dispersed collection of deployed transponders/retroreflectors placed on an asteroid surface that can be remotely monitored by, e.g., laser ranging from a near-orbiting parent satellite would be one way perform this characterization; this would also of course allow for rotational characterization which is already an existing goal of asteroid research (see, e.g., \citeR{FRENCH2020113537}).

Fortuitously, there has already been significant exploration of asteroids in the recent past, and a significant number of new missions are operating now or planned for the near future; see, e.g., \citeR[s]{NASA-asteroidmissionspage,SiddiqiNASA, Yeomans2085,NearShoemakerWebsite,Baker1327,HayabusaWebsite,TSUDA202042,Hayabusa2Website,Lauretta:2017erq,Smith672,OsirisRex,OririsRexWebsite1,OririsRexWebsite2,Bibring493,Taylor387,RosettaWebsite,RAYMAN2006605,DawnWebsite,LUCYwebsite}.  
This study of asteroids is motivated by multiple goals as disparate as planetary defense (i.e., avoiding collisions with Earth) and understanding the origin and evolution of the Solar System.  
Although motivated by these other goals, the multiple future missions to asteroids that are in progress or being planned are also very useful for the goal of gravitational-wave detection, as they appear likely to obtain more, useful data on the properties of asteroids.  
Moreover, the goal of using asteroids as test masses for gravitational-wave detection in turn undoubtedly further motivates future asteroid studies and missions.

The space qualification of optical atomic clocks in the $10^{-19}$ accuracy class is also a prerequisite for our mission to utilize, e.g., the pulse-timing link system approach.
Momentum is currently gathering behind a nascent research and development effort to achieve this goal; see, e.g., \citeR[s]{HighPerformanceClocksInSpace,ColdAtomsInSpace,Voyage2050ESA,Kaltenbaek:2021dgs,10.1117/12.2308164}.

%%%%%%%%%%%%%%%%%%%%%%%%%%%%%%%%%%%%%%%%%%%%%%%%%%%%%%%%%%%%%%%%%%%%%%%%%%%%%%%%%%%%%%%%%%
\section{Sensitivity projection}
\label{sect:sensitivity}
%%%%%%%%%%%%%%%%%%%%%%%%%%%%%%%%%%%%%%%%%%%%%%%%%%%%%%%%%%%%%%%%%%%%%%%%%%%%%%%%%%%%%%%%%%

%%%%%%%%%%%%%%%%%%%%%%%%%%%%%%%%%%%%%%%%%%%%%%%%%%%%%%%%%%%%%%%%%%%%%%%%%%%%%%%%
\begin{figure*}[!p]
    \centering
    \includegraphics[width=0.85\textwidth]{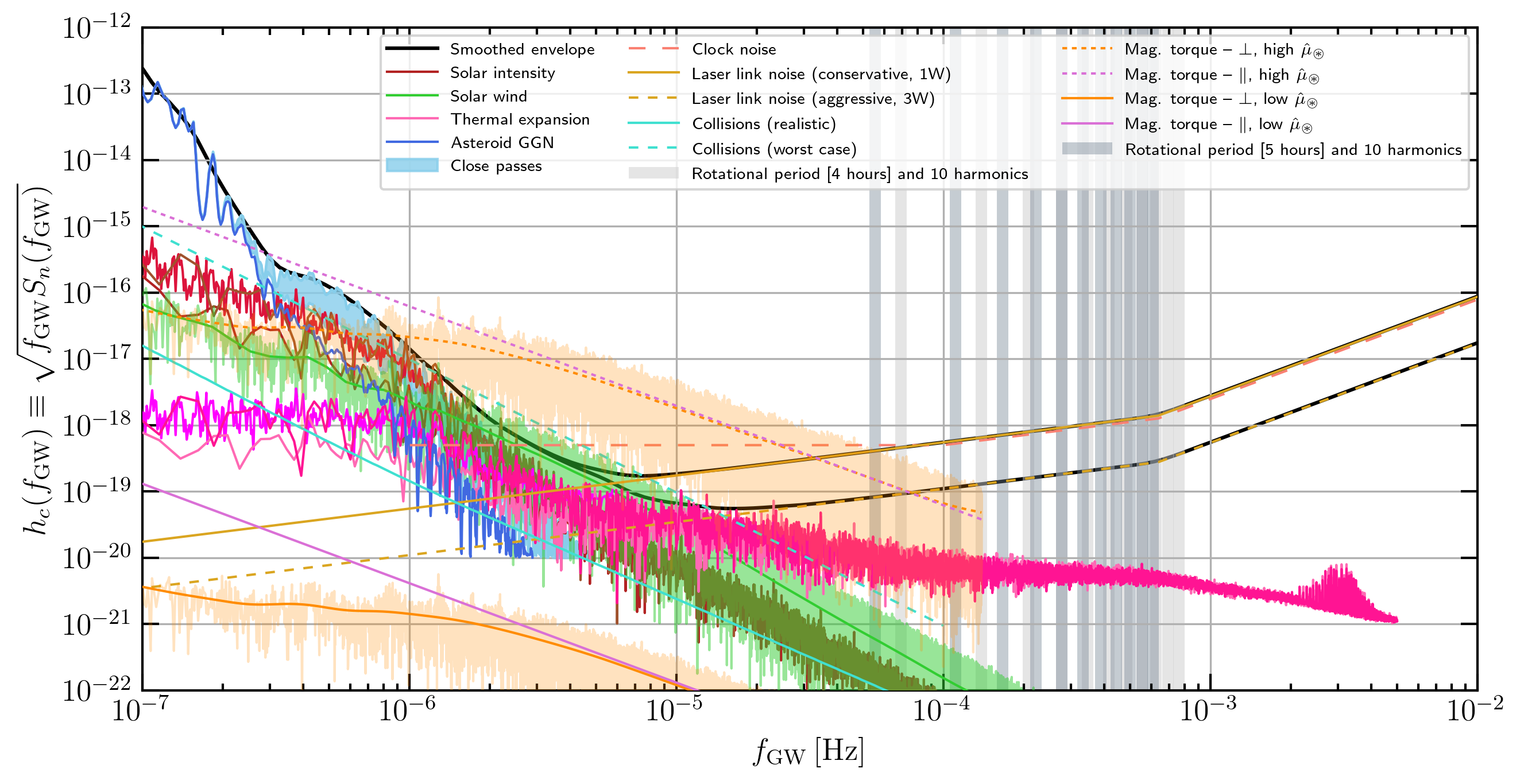}
    \caption{\label{fig:results_combo}%
        Combined plot of all contributing noise sources for our mission concept; see also \figref{fig:noise_contributions} for a clearer version of the individual noise contributions shown here and a more detailed explanation of the various lines and bands.
        The two smoothed envelope lines (solid black thick lines), computed using a log-frequency-space Gaussian smoothing kernel, take into account all noise sources, except for the following estimates (the reasons for their omission are discussed in the text): clock noise, upper bound on collisions, and high specific magnetic moment torquing.
        At high frequency, the enveloped lines track the conservative and aggressive laser ranging noise curves, respectively.
        }
\end{figure*}
%%%%%%%%%%%%%%%%%%%%%%%%%%%%%%%%%%%%%%%%%%%%%%%%%%%%%%%%%%%%%%%%%%%%%%%%%%%%%%%%

%%%%%%%%%%%%%%%%%%%%%%%%%%%%%%%%%%%%%%%%%%%%%%%%%%%%%%%%%%%%%%%%%%%%%%%%%%%%%%%%
\begin{figure*}[!p]
    \centering
    \includegraphics[width=0.85\textwidth]{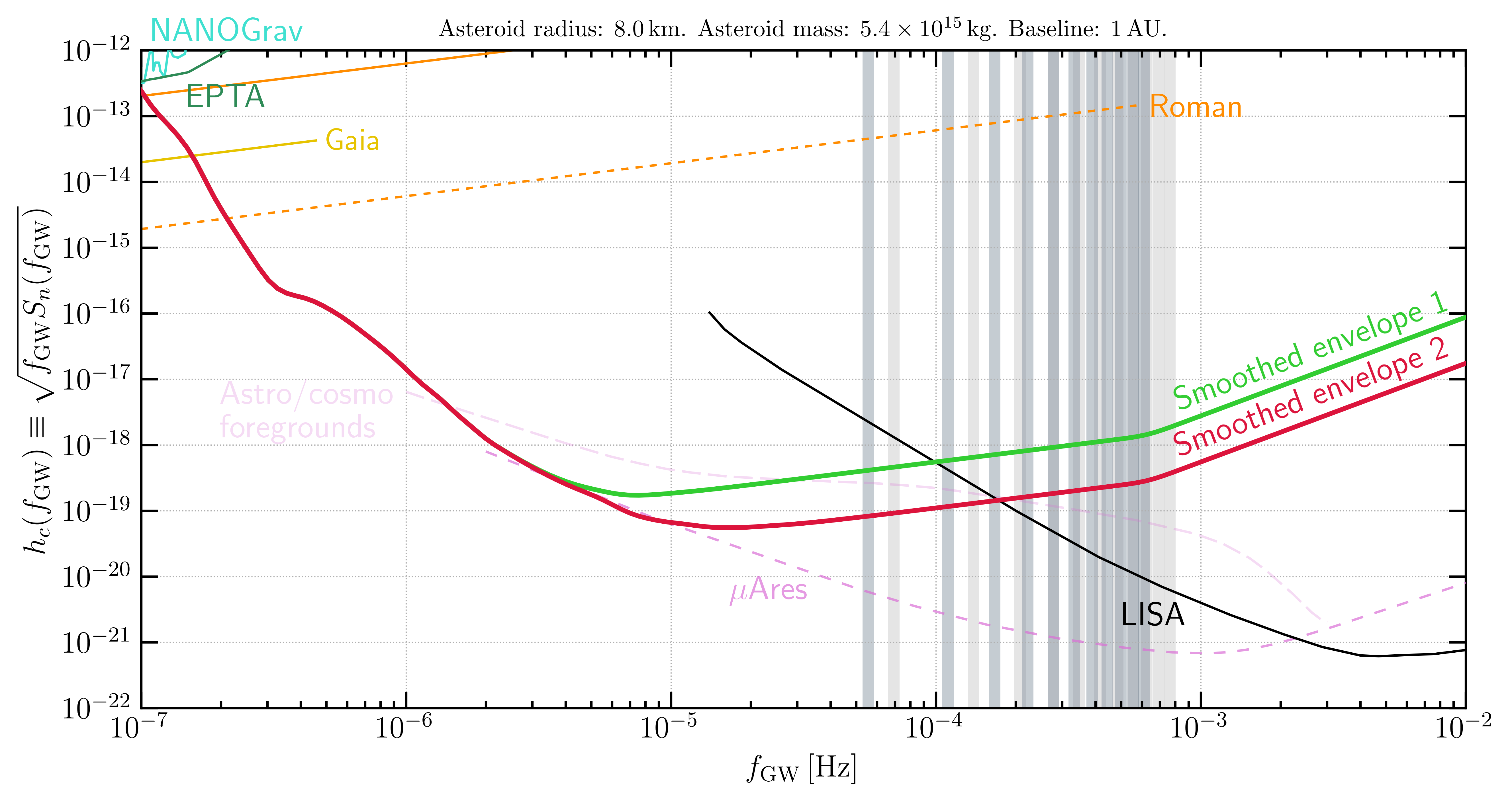}
    \caption{\label{fig:results}%
        Smoothed and enveloped projected characteristic-strain sensitivity reach $h_c$ for our mission concept as a function of the gravitational-wave period $f_{\textsc{gw}}$, assuming either conservative (thick green line, labeled 1) or aggressive (thick red line, labeled 2) laser link-noise at high frequency.
        Also shown for comparison are limits or projections for other existing and proposed facilities: existing NANOGrav limits~\cite{Aggarwal:2018mgp} (teal; converted to $h_c$ per the discussion in \sectref{sect:sensitivity}), existing EPTA limits~\cite{Babak:2015lua} (dark green; converted to $h_c$ per the discussion in \sectref{sect:sensitivity}), the LISA L3 design sensitivity (solid black)~\cite{LISA_L3}, the $\mu$Ares strawman mission concept projected sensitivity (darker, short-dashed purple, plotted above $\sim 2\,\mu\text{Hz}$; see text)~\cite{Sesana:2019vho}, and projected future astrometric GW detection reach using \emph{Gaia} or Roman Space Telescope survey data~\cite{Wang:2020pmf} (gold and orange, respectively; converted to $h_c$ per the discussion in \sectref{sect:sensitivity}).
        Also shown is the astrophysical/cosmological confusion noise estimate (lighter, long-dashed purple, plotted above $\sim \mu\text{Hz}$; see text) from \citeR{Sesana:2019vho}.
        Vertical shaded bands show the rotational frequencies of asteroids with rotational periods of 5\,hrs (darker blue-gray) or 4\,hrs (light gray), as well as the first ten harmonics of these rotational frequencies (the bands are $\pm 5\%$ wide), which limit the high-frequency reach of our concept. 
        }
\end{figure*}
%%%%%%%%%%%%%%%%%%%%%%%%%%%%%%%%%%%%%%%%%%%%%%%%%%%%%%%%%%%%%%%%%%%%%%%%%%%%%%%%

We combine all of the noise contributions shown in the various panels of \figref{fig:noise_contributions} into one master plot in \figref{fig:results_combo}, and include also as the solid black lines on that plot a smoothed%
%%%%%%%%%%%%
\footnote{\label{ftnt:smoothing}%
    The smoothing is performed over a Gaussian kernel in log-frequency space with a width (standard deviation parameter) of $\Delta \log_{10} f = 5\times 10^{-2}\log_{10}[\text{Hz}]$.
} %
%%%%%%%%%%%%
and enveloped combined noise curve taking into account the following relevant, dominant noise sources: the solar radiation pressure noise [\sectref{sect:TSI}; see also  \sectref{sect:TSIandWindTorque}], the solar wind noise [\sectref{sect:solarWind}; see also \sectref{sect:TSIandWindTorque}], asteroid GGN (including close passes) [\sectref{sect:GGN}], the thermal expansion noise [\sectref{sect:thermal}], and either of the two laser ranging system estimates [\sectref{sect:laserPulsing}] (we do not include a envelope based on the radio system we estimated [\sectref{sect:radioPulsing}]).
We have not included clock noise~[\sectref{sect:clocknoise}] in the enveloped results, as we expect future clock improvements; we have also assumed that the magnetic torquing on the asteroid is negligible, in line with the lower range of the estimates given [\sectref{sect:EMforcesTorque}] (this is an asteroid candidate selection criterion); and we assume the negligible `realistic' collisions noise estimates [\sectref{sect:collisions}; see also \sectref{sect:collisionsTorque}].
Additionally, we show in \figref{fig:results_combo} the rotational-period bands that will limit the high-frequency reach of this proposal [\sectref{sect:rotationalMotion}].

We transfer the enveloped and smoothed overall noise curves and rotational frequency bands from \figref{fig:results_combo} to \figref{fig:results}, which shows the sensitivity of our mission concept in the context of other existing missions or techniques that access this band, or the neighboring bands.

At the very lowest end of our frequency band of interest, for $f_{\textsc{gw}}\sim 0.01\,\mu$Hz, pulsar timing arrays begin to provide better sensitivity.
PTA limits are typically given on the instantaneous strain amplitude $h_0$ for a quasi-monochromatic source that would be detectable given the full PTA dataset~\cite{Aggarwal:2018mgp,Babak:2015lua}.
On the other hand, characteristic strain $h_c$ corresponds to the instantaneous strain amplitude of a monochromatic source that is detectable given \emph{one period of the GW signal}; see generally also \citeR{Hazboun:2019vhv} for detailed discussion.
The appropriate conversion between these is $h_c = h_0 \sqrt{N_{\text{GW periods}}} = h_0\sqrt{ f_{\textsc{gw}} T_{\text{obs}} }$ where $T_{\text{obs}}$ is the observation time for the array.
This conversion however assumes constant-in-time characteristic-strain performance for the array, which is not generally true since PTA networks have generally expanded to include more pulsars over time~\cite{Aggarwal:2018mgp,Babak:2015lua}; stated differently, the value computed as $h_c = h_0 \sqrt{ f_{\textsc{gw}} T_{\text{obs}} }$ gives the existing observation-time-averaged characteristic strain performance for the array.
It is expected that array performance is somewhat better during the best (typically, later) periods of the observation run, and that a burst signal occurring during such period would be more easily detected than this value of $h_c$ would credit. 
Nevertheless, in order to be able to make a direct comparison between our projected $h_c$ sensitivity and that of the PTA arrays, we convert the $h_0$ limits from \citeR[s]{Aggarwal:2018mgp,Babak:2015lua} to $h_c = h_0 \sqrt{ f_{\textsc{gw}} T_{\text{obs}} }$.
We plot in \figref{fig:results} values of $h_c$ for the PTAs, making use for these purposes of $T_{\text{obs}} = 11\,$yrs for the 11-year dataset results of NANOGrav~\cite{Aggarwal:2018mgp} and $T_{\text{obs}} = 18\,$yrs for the results of EPTA~\cite{Babak:2015lua} (longest timing baseline used in \citeR{Babak:2015lua}; see also \citeR{Desvignes:2016yex}).

Additionally, in the $\mu$Hz band, we present projected sensitivities of stellar astrometric GW detection making use of future \emph{Gaia} and/or Roman Space Telescope large-scale survey data taken from \citeR{Wang:2020pmf}.
The projections of \citeR{Wang:2020pmf} are however again presented as `the detectable instantaneous (time-domain) strain, $h$, of monochromatic GWs, assuming end-of-survey performance'~\cite{Wang:2020pmf}.
As such, they must also be converted to%
%%%%%%%%%%%%
\footnote{\label{ftnt:altConversion}%
    Note that this converts Eq.~(10) of \citeR{Wang:2020pmf} to $h_c \sim ( \Delta \theta  / \sqrt{N_{\text{s}}} ) \sqrt{f_{\textsc{gw}} \Delta t} \sim  \Delta \theta / \sqrt{N_{\text{s}} N_{\text{obs,\,}\textsc{gw}}} $, where $\Delta \theta$ is the single-star, single-measurement angular measurement accuracy; $N_\text{s}$ is the number of stars in the survey; $\Delta t$ is the time between observations of a single star; and $N_{\text{obs,\,}\textsc{gw}} = T_{\textsc{gw}}/\Delta t = 1/(f_{\textsc{gw}}\Delta t)$ is the number of observations of a single star in a GW period, which is the relevant timescale over which a strain of amplitude $h\sim h_c$ is obtained.} %
%%%%%%%%%%%%
$h_c = h \sqrt{ f_{\textsc{gw}} T_{\text{obs}} }$.
Following discussion in \citeR{Wang:2020pmf}, we take $T_{\text{obs}} = 5\,$yrs for the \emph{Gaia} results, and $T_{\text{obs}} = 432$\,days for the Roman Space Telescope results.
These converted $h_c$ curves are shown in \figref{fig:results}: the gold line shows the \emph{Gaia} projected sensitivity, and the solid and dashed orange lines show the Roman Space Telescope projected sensitivity under differing assumptions (see discussion in \citeR{Wang:2020pmf}).

At the upper end of our band, we run into the LISA L3 design sensitivity; we display in \figref{fig:results} the LISA $h_c$ sensitivity curve taken from Fig.~1 of \citeR{LISA_L3}.

Also shown in \figref{fig:results} is the sensitivity of the $\mu$Ares `strawman mission concept'~\cite{Sesana:2019vho}, which would adopt a LISA-style approach, but with a baseline two orders of magnitude larger and a TM acceleration isolation at $f_{\textsc{gw}}\sim \mu$Hz which is approximately three orders of magnitude improved over the extrapolated LISA Pathfinder~\cite{PhysRevLett.120.061101} results.%
%%%%%%%%%%%%
\footnote{\label{ftnt:LISAPathfinder}%
    LISA Pathfinder~\cite{PhysRevLett.120.061101} results show a rising acceleration noise ASD $\sqrt{S[a](f)} \sim f^{-1}$ below the LISA-optimized sensitivity band [$f\lesssim \text{(few)} \times 10^{-4}\,\text{Hz}$]; this degradation is assumed to be absent in the $\mu$Ares strawman concept estimate~\cite{Sesana:2019vho}.
    } %
%%%%%%%%%%%%
While these may be achievable goals, our mission concept is an interesting alternative optimization of mission parameters that would achieve approximately the same reach in the $\mu$Hz band.
Note that we cut off the low-frequency sensitivity projections for $\mu$Ares at $f_{\textsc{gw}}\sim 2\,\mu\text{Hz}$; the asteroid GGN noise source that we recently identified and estimated~\cite{Fedderke:2020yfy}, and included in our sensitivity projections, would also limit $\mu$Ares, as $\mu$Ares proposes to employ a TM-based approach with the entire baseline within the inner Solar System.

Finally, we show in \figref{fig:results} a foreground projection for unresolved astrophysical and cosmological GW sources in this band, also taken from \citeR{Sesana:2019vho}.
Our concept would just begin to be sensitive to these foregrounds were they at the level of that estimate.
However, these foreground projections are somewhat larger those shown in \citeR{LISA_L3} in the frequency range where both are available near $f_{\textsc{gw}} \sim$ mHz.
Depending on how they are estimated then, these foregrounds may thus not be of concern for our concept; nevertheless, to be conservative, we have shown the (larger) estimates from \citeR{Sesana:2019vho} in \figref{fig:results}.
We only plot these above $f_{\textsc{gw}}\sim\mu$Hz for similar reasons owing to asteroid GGN as discussed in the preceding paragraph.

Overall, our projected strain sensitivity results as shown in \figref{fig:results} indicate significant reach from our mission concept well into scientifically interesting levels of strain sensitivity~\cite{Sesana:2019vho} in the otherwise hard-to-access $\mu$Hz band.
This is especially true in the band\linebreak $f_{\textsc{gw}} \sim 1$--$10\,\mu$Hz, where we demonstrate possible reach 2--4 orders of magnitude better than that of other estimated techniques, with the exception of the $\mu$Ares strawman concept~\cite{Sesana:2019vho}.
This demonstrates the remarkable utility of appropriately selected asteroids as TMs in a GW detector.

%%%%%%%%%%%%%%%%%%%%%%%%%%%%%%%%%%%%%%%%%%%%%%%%%%%%%%%%%%%%%%%%%%%%%%%%%%%%%%%%%%%%%%%%%%
\section{Conclusion}
\label{sect:conclusion}
%%%%%%%%%%%%%%%%%%%%%%%%%%%%%%%%%%%%%%%%%%%%%%%%%%%%%%%%%%%%%%%%%%%%%%%%%%%%%%%%%%%%%%%%%%
The historic discoveries of LIGO/Virgo have opened the gravitational-wave spectrum. 
Given the unique nature of gravitational waves to probe the existence of every massive object in the universe, there is little doubt that they will play a central role in astronomy and the pursuit of fundamental physics. 
It is thus of great importance to investigate all possible technological options to discover gravitational waves over a wide range of frequencies. 
In this paper, we explored the possible use of $\sim 10$\,km-scale asteroids as test masses for gravitational-wave detection in the frequency band around $\mu$Hz and we have found that there is good reason to expect them to be viable test masses in this band. 

While these reasons are strong enough to warrant further investigation of this concept, it is desirable to make additional measurements (such as low frequency seismic monitoring) to establish the stability of the asteroid surface at these frequencies before launching a full-scale mission. 
Interestingly, given the strong motivation to visit asteroids~\cite{DARTwebsite},  there appears to be a symbiotic path that can be pursued in concert with other planned exploration of asteroids~\cite{Plescia:2017wrt,Plescia:2018wfb,Plescia:2016sfg,Schmerr:2018rtb,Walker:2006bqw,MURDOCH201789} to perform these measurements. 
Further, the metrological demands of this mission have been demonstrated in terrestrial atomic clocks. 
The possibility of this kind of mission adds to the science case for creating a space-qualified optical atomic clock, a technological goal that has many other science applications including the search for gravitational waves~\cite{Graham:2017pmn} and dark matter with atomic interferometers~\cite{Arvanitaki:2016fyj}.

While the detection of gravitational waves has been the primary focus of this paper, given the fact that asteroids are likely to be excellent test masses in this low-frequency range, it is interesting to explore their use for the detection of a variety of dark-matter candidates such as ultralight dark matter~\cite{Graham:2015ifn}, compact dark-matter blobs~\cite{Grabowska:2018lnd}, and primordial black holes~\cite{Seto:2004zu,Adams:2004pk}. 
It would also be interesting to consider the use of asteroids to probe violations of the equivalence principle and test gravitation in the Solar System, similar to the tests that have been performed with Lunar Laser Ranging.

%%%%%%%%%%%%%%%%%%%%%%%%%%%%%%%%%%%%%%%%%%%%%%%%%%%%%%%%%%%%%%%%%%%%%%%%%%%%%%%%%%%%%%%%%%
\acknowledgments
%%%%%%%%%%%%%%%%%%%%%%%%%%%%%%%%%%%%%%%%%%%%%%%%%%%%%%%%%%%%%%%%%%%%%%%%%%%%%%%%%%%%%%%%%%
We thank Sebastian Baum, Hartmut Grote, Jason Hogan, Leo Hollberg, David Hume, Mark Kasevich, Peter Michelson, Andrew Rivkin, James Thompson, and Jun Ye for useful discussions.  

We thank the ACE MAG instrument team and the ACE Science Center for providing the ACE data.  

S.R.~is supported in part by the U.S.~National Science Foundation (NSF) under Grant No.~PHY-1818899.   
This work was supported by the U.S.~Department of Energy (DOE), Office of Science, National Quantum Information Science Research Centers, Superconducting Quantum Materials and Systems Center (SQMS) under Contract No.~DE-AC02-07CH11359. 
S.R.~is also supported by the DOE under a QuantISED grant for MAGIS, and the Simons Investigator Award No.~827042.
This work was also supported by the Simons Investigator Award No.~824870, NSF Grant No.~PHY-2014215, DOE HEP QuantISED Award No.~100495, and the Gordon and Betty Moore Foundation Grant No.~GBMF7946.

The work of M.A.F.~was performed in part at the Aspen Center for Physics, which is supported by NSF Grant No.~PHY-1607611.

%%%%%%%%%%%%%%%%%%%%%%%%%%%%%%%%%%%%%%%%%%%%%%%%%%%%%%%%%%%%%%%%%%%%%%%%%%%%%%%%%%%%%%%%%%
%%%%%%%%%%%%%%%%%%%%%%%%%%%%%%%%%%%%%%%%%%%%%%%%%%%%%%%%%%%%%%%%%%%%%%%%%%%%%%%%%%%%%%%%%%
%%%%%%%%%%%%%%%%%%%%%%%%%%%%%%%%%%%%%%%%%%%%%%%%%%%%%%%%%%%%%%%%%%%%%%%%%%%%%%%%%%%%%%%%%%
\appendix
%%%%%%%%%%%%%%%%%%%%%%%%%%%%%%%%%%%%%%%%%%%%%%%%%%%%%%%%%%%%%%%%%%%%%%%%%%%%%%%%%%%%%%%%%%
%%%%%%%%%%%%%%%%%%%%%%%%%%%%%%%%%%%%%%%%%%%%%%%%%%%%%%%%%%%%%%%%%%%%%%%%%%%%%%%%%%%%%%%%%%
%%%%%%%%%%%%%%%%%%%%%%%%%%%%%%%%%%%%%%%%%%%%%%%%%%%%%%%%%%%%%%%%%%%%%%%%%%%%%%%%%%%%%%%%%%

%%%%%%%%%%%%%%%%%%%%%%%%%%%%%%%%%%%%%%%%%%%%%%%%%%%%%%%%%%%%%%%%%%%%%%%%%%%%%%%%%%%%%%%%%%
\section{Noise cross-power terms}
\label{app:crossPowerTerms}
%%%%%%%%%%%%%%%%%%%%%%%%%%%%%%%%%%%%%%%%%%%%%%%%%%%%%%%%%%%%%%%%%%%%%%%%%%%%%%%%%%%%%%%%%%
In this appendix, we consider whether the beat note between the rotational modulation of the area of an asteroid presented to the Sun, and the solar power fluctuation at roughly rotational frequencies, can induce low-frequency noise that would significantly modify the estimates provided in \sectref{sect:TSI}.

Consider the case where the rotational modulation of the effective cross-sectional area presented to the incoming solar radiation is narrowband in the sense that, given a total observation time $T$, it provides power in some single DFT frequency bin $f_r = r \Delta f \; (r\in\mathbb{Z};\ 0<r\leq N-1)$ where $\Delta f = 1/T$: that is, $\delta A_{\text{eff}}(t) = \delta A_{\text{eff}, 0} \cos( 2\pi f_r t + \phi )$.
In this case, the FFT of this function, $\widetilde{\delta A_{\text{eff}}}[n] \equiv \widetilde{\delta A_{\text{eff}}}(f_n)$ where $f_n = n\Delta f$, is (see conventions in Appendix C of \citeR{Fedderke:2020yfy})
%%%%%%
\begin{align}
    \widetilde{\delta A_{\text{eff}}}[n] = \frac{T}{2} \delta A_{\text{eff}, 0} \lb[ e^{i\phi} \delta_{n,r\mod N} + e^{-i\phi} \delta_{n,(N-r)\mod N} \rb],
    \label{eq:deltaAeff-FFT}
\end{align}
%%%%%%
with $\delta_{i,j}$ the Kronecker delta.

Let the FFT of the solar fractional power fluctuation be $\widetilde{\delta I_{\odot}}[n]\equiv\widetilde{\delta I_{\odot}}(f_n)$.

Consider the cross-term $\delta a(t) \supset C(t) \equiv \delta A_{\text{eff}}(t) \cdot \delta I_{\odot}(t)$ in \eqref{eq:aExp}. 
Because of the multiplication--convolution theorems of Fourier analysis, it follows that the FFT of $C(t)$, $\widetilde{C}[n] \equiv \widetilde{C}(f_n)$, is given by
%%%%%%
\begin{align}
    \widetilde{C}[n] = \frac{1}{T} \sum_{m=0}^{N-1} \widetilde{\delta A_{\text{eff}}}[m] \cdot \widetilde{\delta I_{\odot}}[(n-m)\mod N].
    \label{eq:C-FFT}
\end{align}
%%%%%%
Therefore,
%%%%%%
\begin{align}
    \widetilde{C}[n] &= \frac{1}{2}  \delta A_{\text{eff}, 0} \lb[ e^{i\phi} C_1[n] + e^{-i\phi}  C_2[n] \rb],
\end{align}
%%%%%%
where 
%%%%%%
\begin{align}
    C_1[n] &\equiv \widetilde{\delta I_{\odot}}[(n-(r\mod N))\mod N], \\
    C_2[n] &\equiv \widetilde{\delta I_{\odot}}[\lb(n-((N-r)\mod N)\rb)\mod N].
\end{align}
%%%%%%
But since $1\leq r\leq N-1$, we have $1\leq N-r\leq N-1$, so $\lb(n-((N-r)\mod N)\rb) = \lb(n+r-N\rb)$, and so 
%%%%%%
\begin{align}
    C_1[n] &\equiv \widetilde{\delta I_{\odot}}[(n-r)\mod N], \\
    C_2[n] &\equiv \widetilde{\delta I_{\odot}}[(n+r-N)\mod N].
\end{align}
%%%%%%

Now consider that we are looking for the low-frequency beat note at $f_n = n \Delta f$, such that $10^{-6}\,\text{Hz} \lesssim f_n  \lesssim 10^{-5}\,\text{Hz}$, whereas the asteroid rotational period is $f_r = r\Delta f \gtrsim (\text{few})\times 10^{-5}\,\text{Hz}$; therefore, $n\lesssim r$ (and possibly $n\ll r$). 
Further, let us assume that $N\gg 2r>r$, since $N$ sets the Nyquist sampling frequency $f_{\text{Nyq.}} = (N/2)\Delta f$, and we can in principle sample the acceleration much faster than the asteroid rotational rate.
In this case, we have $(n-r)\mod N \approx N-(r-n)$ and $(n+r-N)\mod N \approx r+n$, so that
%%%%%%
\begin{align}
    \widetilde{C}[n] &\approx \frac{1}{2}  \delta A_{\text{eff}, 0} \lb[ e^{i\phi} \widetilde{\delta I_{\odot}}[N-(r-n)] + e^{-i\phi} \widetilde{\delta I_{\odot}}[r+n] \rb].
\end{align}
%%%%%%
Because $\delta I_{\odot}(t)\in\mathbb{R}$, we have $\widetilde{\delta I_{\odot}}[N-k] = \widetilde{\delta I_{\odot}}[k]^*$, so
%%%%%%
\begin{align}
    \widetilde{C}[n] &\approx \frac{1}{2}  \delta A_{\text{eff}, 0} \lb[ e^{-i\phi} \widetilde{\delta I_{\odot}}[r+n] + e^{i\phi} \widetilde{\delta I_{\odot}}[r-n]^* \rb]\\
    &\approx  \frac{1}{2}  \delta A_{\text{eff}, 0} \lb[ e^{-i\phi} \widetilde{\delta I_{\odot}}[r] + \text{c.c.} \rb],
\end{align}
%%%%%%
where in the last line we have further assumed that $n\ll r$ (and that $\widetilde{\delta I_{\odot}}$ is reasonably smooth), and where $+\text{c.c.}$ denotes the addition of the complex conjugate of the previous term.

Therefore, it follows that 
%%%%%%
\begin{align}
    \widetilde{\delta a}[n] &\approx \widetilde{\delta I_{\odot}}[n] +   \delta A_{\text{eff}, 0} \Re\lb[ e^{-i\phi} \widetilde{\delta I_{\odot}}[r] \rb].
\end{align}
%%%%%%
From this form, it is clear that the additional cross-term is at most an $\order{1}$ correction to the direct solar fluctuation term, because (1) $\delta A_{\text{eff},0}$ is at most $\order{1}$; and (2) the solar fluctuation PSD falls as a function of increasing frequency~\cite{Froehlich:2004wed} very roughly as $|\widetilde{\delta I_{\odot}}| \propto \sqrt{S[\delta I_{\odot}]}\propto f^{-2}$ in the band between $10^{-6}$ and $4\times 10^{-6}\,\text{Hz}$, and as $|\widetilde{\delta I_{\odot}}| \propto \sqrt{S[\delta I_{\odot}]}\propto f^{-2/3}$ in the band between $4\times 10^{-6}$ and $10^{-4}\,\text{Hz}$ so that $|\widetilde{\delta I_{\odot}}[r]| \lesssim |\widetilde{\delta I_{\odot}}[n]|$.
As such, we ignore this extra term.

%%%%%%%%%%%%%%%%%%%%%%%%%%%%%%%%%%%%%%%%%%%%%%%%%%%%%%%%%%%%%%%%%%%%%%%%%%%%%%%%%%%%%%%%%%
%%%%%%%%%%%%%%%%%%%%%%%%%%%%%%%%%%%%%%%%%%%%%%%%%%%%%%%%%%%%%%%%%%%%%%%%%%%%%%%%%%%%%%%%%%
\bibliographystyle{JHEP}
\bibliography{references.bib}
%%%%%%%%%%%%%%%%%%%%%%%%%%%%%%%%%%%%%%%%%%%%%%%%%%%%%%%%%%%%%%%%%%%%%%%%%%%%%%%%%%%%%%%%%%
%%%%%%%%%%%%%%%%%%%%%%%%%%%%%%%%%%%%%%%%%%%%%%%%%%%%%%%%%%%%%%%%%%%%%%%%%%%%%%%%%%%%%%%%%%

%%%%%%%%%%%%%%%%%%%%%%%%%%%%%%%%%%%%%%%%%%%%%%%%%%%%%%%%%%%%%%%%%%%%%%%%%%%%%%%%%%%%%%%%%%
%%%%%%%%%%%%%%%%%%%%%%%%%%%%%%%%%%%%%%%%%%%%%%%%%%%%%%%%%%%%%%%%%%%%%%%%%%%%%%%%%%%%%%%%%%
%%%%%%%%%%%%%%%%%%%%%%%%%%%%%%%%%%%%%%%%%%%%%%%%%%%%%%%%%%%%%%%%%%%%%%%%%%%%%%%%%%%%%%%%%%
%%%%%%%%%%%%%%%%%%%%%%%%%%%%%%%%%%%%%%%%%%%%%%%%%%%%%%%%%%%%%%%%%%%%%%%%%%%%%%%%%%%%%%%%%%
\end{document}